\documentclass[12pt,aps,english,prd,onecolumn,nofootinbib,superscriptaddress]{revtex4-2}

\usepackage[utf8]{inputenc}
\usepackage[english]{babel}  
\usepackage{tikz}
\usepackage{tcolorbox}
\usepackage{physics}
\usepackage{mdframed}
\usepackage{fancybox}
\usepackage{textcomp}
\usepackage{amsmath,amssymb,amsfonts}
\usepackage{bm}
\usepackage{graphicx}
\usepackage{xcolor}               
\usepackage{color}                
\usepackage{textcomp}             
\usepackage{units}                
\usepackage[a4paper, margin=1.6cm]{geometry}
\usepackage{hyperref}             
\hypersetup{
    colorlinks=true,              
    breaklinks=true,              
    citecolor=blue,               
    linkcolor=[rgb]{0,0.5,0.9},   
    urlcolor=blue,                 
    filecolor=green               
}

\newcommand{\orcid}[1]{\href{https://orcid.org/#1}{\includegraphics[width=10pt]{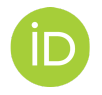}}}

\begin{document}

\title{Gauge-covariant Raychaudhuri dynamics for spin-nondegenerate Lorentz-violating congruences}

\author{A. A. Ara\'{u}jo Filho\orcid{0000-0002-8790-3944}}
\email{dilto@fisica.ufc.br}
\affiliation{Departamento de Física, Universidade Federal da Paraíba, Caixa Postal 5008, 58051--970, João Pessoa, Paraíba,  Brazil.}
\affiliation{Departamento de Física, Universidade Federal de Campina Grande Caixa Postal 10071, 58429-900 Campina Grande, Paraíba, Brazil.}
\affiliation{Center for Theoretical Physics, Khazar University, 41 Mehseti Street, Baku, AZ-1096, Azerbaijan.}
\author{A. F. Santos\orcid{0000-0002-2505-5273}}
\email{alesandroferreira@fisica.ufmt.br}
\affiliation{Programa de Pós-graduação em Física, Instituto de Física, Universidade Federal de Mato Grosso, Cuiabá, Brasil}
\author{J. A. A. S. Reis\orcid{0000-0002-2831-5317}}
\email{joao.reis@uesb.edu.br}
\affiliation{Departamento de Ciências Exatas e Naturais, Universidade Estadual do Sudoeste da Bahia, Campus Juvino Oliveira, Itapetinga -- BA, 45700-00, Brazil}
\author{L. Lisboa-Santos\orcid{0000-0003-4939-3856}}
\email{leticia.lisboa@discente.ufma.br}
\affiliation{Programa de Pós-Graduação em Física, Universidade Federal do Maranh\~{a}o, Campus Universit\'{a}rio do Bacanga, S\~{a}o Lu\'{i}s (MA), 65085-580, Brazil}
\author{V. B. Bezerra\orcid{0000-0001-7893-0265}}
\email{valdir@fisica.ufpb.br}
\affiliation{Departamento de Física, Universidade Federal da Paraíba, Caixa Postal 5008, 58051--970, João Pessoa, Paraíba,  Brazil.}

\date{\today}

\begin{abstract}
We investigate the Raychaudhuri dynamics of charged spin--nondegenerate Lorentz--violating particle congruences under minimal electromagnetic coupling. The coupling is introduced through the gauge--covariant momentum $P_{\mu}=\pi_{\mu}-qA_{\mu}$, so that the branch dispersion relation keeps its free functional form, while the electromagnetic field enters through the evolution of $P_{\mu}$. For a generic branch $\mathcal D^{(\pm)}(P)$, the tangent $k^{\mu}_{(\pm)}$ and the momentum Hessian $M^{\mu\nu}_{(\pm)}$ determine the covariant acceleration,
$a^{\mu}_{(\pm)}=-qM^{\mu\nu}_{(\pm)}F_{\nu\rho}k^{\rho}_{(\pm)}$.
As a consequence, the Raychaudhuri equation acquires the branch-dependent electromagnetic source
$-q\nabla_{\mu}\!\left(M^{\mu\nu}_{(\pm)}F_{\nu\rho}k^{\rho}_{(\pm)}\right)$.
We apply this construction to the $b_{\mu}$, $H_{\mu\nu}$, and $d_{\mu\nu}$ sectors, obtaining the corresponding branch tangents, Hessians, accelerations, and focusing equations. In flat spacetime, the electromagnetic field modifies the expansion through the divergence of the effective branch force. Therefore, uniform fields may bend the trajectories, whereas local focusing requires field gradients or, in the magnetic case, a coupling to an already deformed congruence. We also develop the analogous description for semiclassical quasiparticle beams, where the band Hessian plays the role of an effective electromagnetic response tensor. For anisotropic parabolic, Dirac--like, and Weyl--type dispersions, the same geometric structure relates electromagnetic textures to beam focusing. In two-branch systems, the opposite Hessians of the branches can produce focusing in one congruence and defocusing in the other, giving a quasiparticle realization of branch--dependent birefringence.
\end{abstract}

\maketitle

\tableofcontents


\section{Introduction}
\label{sec:introduction}

The Raychaudhuri equation is one of the most direct ways of describing the local evolution of a congruence. Its standard relativistic form separates the change of the expansion scalar into contributions from expansion, shear, vorticity, and curvature, without assuming a specific gravitational field equation \cite{Raychaudhuri1955,Ehlers1961,Sachs1961,HawkingEllis1973,Wald1984,Poisson2004,KarSengupta2007}. This kinematical character explains its central role in the analysis of gravitational focusing, singularity theorems, trapped surfaces, and causal properties of spacetime \cite{Penrose1965,Hawking1966,HawkingEllis1973,Tipler1978,Senovilla1998,GallowaySenovilla2010}. For nongeodesic flows, the divergence of the acceleration enters the same evolution equation, making the Raychaudhuri framework suitable for systems in which external forces, internal response tensors, or modified dispersion relations determine how neighboring trajectories focus or defocus \cite{Ehlers1961,Poisson2004,KarSengupta2007}.

{
During the last decades, the Raychaudhuri equation has also been extended beyond its original geodesic and Riemannian setting. Beyond its fundamental role in general relativity, the Raychaudhuri equation has found numerous applications in diverse areas of gravitational physics. Recent studies have explored several generalizations involving non-geodesic congruences, modified gravity theories, spacetime kinematics, and related topics, considerably extending the range of its applications. For representative recent developments, see Refs.~\cite{Chak1,Chak2,Chak3,Chak4,Chak5,Chak6,Chak7,Chak8,Chak9,Chak10,Chak11,Chak12,Chak13,Chak14,Chak15,Chak16}. In particular, generalized formulations have been developed for congruences in spacetimes with torsion and nonmetricity, where the kinematical evolution receives additional contributions from the affine structure of spacetime \cite{Kar:2006ms,Luz:2017, Iosifidis:2018}. Related analyses in modified gravity have used the Raychaudhuri equation to formulate focusing or defocusing conditions in settings such as nonlocal gravity, teleparallel gravity, and symmetric teleparallel models \cite{Conroy:2016, Beckering:2020, Geng:2026}. These works emphasize that expansion, shear, and vorticity provide a geometric diagnostic of whether a family of neighboring trajectories collapses, spreads out, or develops caustics under the influence of curvature, modified gravitational dynamics, or extra force terms\footnote{{
The present work is connected with this broader line of investigation, but it addresses a different source of generalized congruence dynamics. We do not modify the spacetime connection, nor do we derive the focusing condition from a modified gravitational field equation. Instead, the non--geodesic term is induced by minimal electromagnetic coupling to a spin--nondegenerate Lorentz--violating dispersion relation. 
}}.
}

Lorentz- and CPT-violating effective field theories provide a natural setting in which the geometry of a congruence is controlled not only by the spacetime metric, but also by background tensor coefficients. In the Standard-Model Extension, such coefficients parameterize deviations from exact local Lorentz invariance while preserving observer covariance \cite{KosteleckySamuel1989,ColladayKostelecky1997,ColladayKostelecky1998,Kostelecky2004}. The fermion sector contains spin-dependent coefficients such as $b_\mu$, $H_{\mu\nu}$, and $d_{\mu\nu}$, which split the particle dynamics into branches and modify the momentum--velocity relation \cite{ColladayKostelecky1998,KosteleckyRussell2011,BaileyKostelecky2006,KosteleckyTasson2011}. Since experimental searches constrain these effects in matter, antimatter, photons, gravity, and astrophysical propagation, the corresponding classical dynamics must be formulated in a way that keeps both covariance and branch dependence explicit \cite{Mattingly2005,Liberati2013,Bluhm2006,KosteleckyRussell2011}.

The classical limit of Lorentz-violating fermions is especially rich because the dispersion relation does not always lead to a single velocity field. For spin-nondegenerate sectors, the mass shell factorizes into branch functions $\mathcal D^{(\pm)}(p)$, and each branch defines its own tangent vector through the Hamiltonian flow \cite{KosteleckyRussell2010,Kostelecky2011Finsler,KosteleckyRussellTso2012}. This structure is closely related to pseudo-Finsler and bipartite Finsler geometries, where the line element depends on direction as well as position \cite{Randers1941,BaoChernShen2000,GibbonsGomisPope2007,CohenGlashow2006,PfeiferWohlfarth2012,JavaloyesSanchez2014}. In this formulation, the branch label selects a different momentum Hessian, and this Hessian fixes how a force changes the tangent field of the congruence.

Electromagnetic coupling gives a clean way of testing this branch-dependent response. Minimal coupling replaces the canonical momentum by the gauge--covariant momentum $P_\mu=\pi_\mu-qA_\mu$, leaving the functional form of the branch dispersion relation unchanged while making the external field enter through $F_{\mu\nu}$ \cite{Jackson1999,LandauLifshitzFields,Goldstein2002,Arnold1989}. For a branch Hamiltonian built from $\mathcal D^{(\pm)}(P)$, the tangent is determined by $k^\mu_{(\pm)}=(e/2)\partial \mathcal D^{(\pm)}/\partial P_\mu$, whereas the response of this tangent to changes in momentum is measured by
\begin{equation}
M^{\mu\nu}_{(\pm)}:=-\frac{\partial k^\mu_{(\pm)}}{\partial P_\nu}.
\end{equation}
The electromagnetic acceleration then takes the universal form
\begin{equation}
a^\mu_{(\pm)}=-qM^{\mu\nu}_{(\pm)}F_{\nu\rho}k^\rho_{(\pm)} ,
\end{equation}
so the Raychaudhuri equation acquires the source
\begin{equation}
-q\nabla_\mu\!\left(M^{\mu\nu}_{(\pm)}F_{\nu\rho}k^\rho_{(\pm)}\right).
\end{equation}
The field strength is common to both branches, but the tensor converting it into acceleration is branch dependent. This produces a branch-dependent electromagnetic focusing mechanism, in which different spin branches may undergo distinct focusing or
defocusing rates under the same external field configuration.

This mechanism defines a branch-dependent electromagnetic focusing
effect. Although the external electromagnetic field is the same for
both branches, the effective response tensor
$M^{\mu\nu}_{(\pm)}$
modifies how the field contributes to the local evolution of the
congruence expansion. As a consequence, two spin-nondegenerate
branches may experience different focusing or defocusing rates under
the same electromagnetic profile.

This point is important in flat spacetime, where curvature does not contribute to the focusing equation. In that case, a uniform electromagnetic field can bend individual trajectories, but a local change in the expansion requires a nonzero divergence of the effective branch force. Electric gradients contribute through projections of $\partial_\mu E_i$ weighted by $M^{\mu\nu}_{(\pm)}$ and $k^\mu_{(\pm)}$, while magnetic gradients contribute through the branch-dependent analogue of a magnetic lens. If the magnetic field is uniform, it can still modify the expansion when it couples to an already deformed congruence through shear, vorticity, or expansion. The distinction between trajectory bending and beam focusing is therefore essential: the central trajectory can be deflected even when the congruence width does not acquire a leading local source.

The same mathematical structure appears in semiclassical wave-packet dynamics. In a Bloch band with dispersion $\varepsilon_n(\mathbf{k})$, the group velocity is $v^i=\partial\varepsilon_n/\partial k_i$, and the band Hessian $M^{ij}=\partial^2\varepsilon_n/\partial k_i\partial k_j$ determines the electromagnetic acceleration of a wave packet \cite{Bloch1929,Peierls1933,AdamsBlount1959,AshcroftMermin1976,Kittel2005,Ziman1972}. Semiclassical dynamics in crystals already shows how electric and magnetic fields act through the local band geometry, with Berry phases and anomalous velocities playing a central role when such corrections are included \cite{Berry1984,KarplusLuttinger1954,Thouless1982,ChangNiu1995,ChangNiu1996,SundaramNiu1999,XiaoChangNiu2010,Vanderbilt2018,Haldane2004}. When Berry-curvature corrections are neglected, the Hessian alone acts as the effective susceptibility converting electromagnetic textures into acceleration and, after taking the divergence, into a Raychaudhuri-type source for the beam expansion.

Dirac- and Weyl-type systems give a direct condensed-matter analogue of the branch structure found in spin-nondegenerate Lorentz-violating particles. Relativistic linear dispersions originate from the Dirac and Weyl equations \cite{Dirac1928,Weyl1929}, while graphene and topological semimetals realize related low-energy quasiparticle spectra in solid-state systems \cite{Wallace1947,Novoselov2005,Zhang2005,CastroNeto2009,HasanKane2010,QiZhang2011}. In Weyl semimetals, the two chiral branches may carry opposite Hessians under suitable approximations, so the same electromagnetic texture can focus one beam and defocus the other \cite{Wan2011,BurkovBalents2011,Armitage2018,VafekVishwanath2014,Wehling2014,Goerbig2011}. This branch-selective behavior is the quasiparticle counterpart of birefringent propagation in Lorentz-violating systems.

In this work we develop a gauge--covariant Raychaudhuri formulation for charged spin-nondegenerate Lorentz-violating congruences. The construction is first derived for a generic branch dispersion relation $\mathcal D^{(\pm)}(P)$ and then applied to the $b_\mu$, $H_{\mu\nu}$, and $d_{\mu\nu}$ sectors. For each sector we compute the branch tangent, the momentum Hessian, the electromagnetic acceleration, and the corresponding focusing equation. We then analyze the flat-spacetime limit, identifying the conditions under which electric and magnetic field profiles act as focusing or defocusing sources. Finally, we formulate the analogous Raychaudhuri-type equation for semiclassical quasiparticle beams and apply it to anisotropic parabolic, Dirac-like, and Weyl-type dispersions. Throughout this work, the Lorentz-violating coefficients are treated
as fixed effective backgrounds within the validity regime of the
underlying effective field theory.

The paper is organized as follows. In Sec.~\ref{sec:em_coupling}, we introduce the minimal electromagnetic coupling and derive the branch-dependent Raychaudhuri equation in its general form. Secs.~\ref{sec:em_b_sector}, \ref{sec:em_H_sector}, and \ref{sec:em_d_sector} apply this construction to the $b_\mu$, $H_{\mu\nu}$, and $d_{\mu\nu}$ sectors, respectively. In Sec.~\ref{sec:unified_structure}, we identify the universal structure of electromagnetic forcing in spin-nondegenerate Lorentz-violating congruences. Sec.~\ref{sec:EM_focusing_flat} examines electromagnetic focusing in flat spacetime, while Sec.~\ref{sec:susceptibility} discusses the interpretation of the momentum Hessian as an effective electromagnetic response tensor tensor. In Sec.~\ref{sec:semiclassical}, we review the semiclassical dynamics of Bloch wave packets and formulate the associated congruence structure. Sec.~\ref{sec:raychaudhuri} derives the Raychaudhuri-type equation for quasiparticle beams. In Sec.~\ref{sec:susceptibility11} effective electromagnetic response tensor of band congruences are investigated. In Secs.~\ref{sec:parabolic}--\ref{sec:weyl}, we apply the formalism to concrete condensed-matter models, and Sec.~\ref{sec:fields} analyzes the focusing induced by electric and magnetic field gradients. Finally, Sec.~\ref{sec:discussion} presents the conclusion.


\section{Minimal electromagnetic coupling and branch-dependent Raychaudhuri equation}
\label{sec:em_coupling}

We now couple the type--2 Lorentz--violating particle \cite{Reis:2026rch} to an external electromagnetic field through the standard gauge one--form,
\begin{equation}
\tilde L_{\mathrm{em}}=-qA_\mu(x)\dot x^\mu .
\end{equation}
The branch Lagrangian is therefore
\begin{equation}
\tilde L^{(\pm)}_{\mathrm{tot}}
=
\tilde L^{(\pm)}_{\mathrm{LV}}
-qA_\mu(x)\dot x^\mu .
\end{equation}
Since this additional term is homogeneous of degree one in $\dot x^\mu$, the reparametrization symmetry of the worldline action is preserved. Moreover, under a gauge transformation $A_\mu\rightarrow A_\mu+\partial_\mu\Lambda$, the action changes only by the boundary term $-q\,d\Lambda/d\lambda$.

The canonical momentum is shifted by the electromagnetic potential,
\begin{equation}
\pi_\mu
:=
-\frac{\partial \tilde L^{(\pm)}_{\mathrm{tot}}}{\partial \dot x^\mu},
\qquad
P_\mu:=\pi_\mu-qA_\mu(x),
\end{equation}
where $P_\mu$ is the gauge--covariant momentum. Thus the electromagnetic coupling does not change the functional form of the branch Hamiltonian. It only promotes the free momentum to the gauge--covariant one,
\begin{equation}
\tilde H^{(\pm)}(x,\pi;\mathfrak e)
=
-\frac{\mathfrak e}{2}\,
\mathcal D^{(\pm)}(P),
\qquad
P_\mu=\pi_\mu-qA_\mu(x).
\end{equation}

Hamilton's equations then give
\begin{equation}
k^\mu_{(\pm)}
:=
\dot x^\mu
=
-\frac{\partial \tilde H^{(\pm)}}{\partial \pi_\mu}
=
\frac{\mathfrak e}{2}
\frac{\partial \mathcal D^{(\pm)}(P)}{\partial P_\mu},
\end{equation}
while the evolution of the gauge--covariant momentum becomes
\begin{equation}
\dot P_\mu
=
qF_{\mu\nu}k^\nu_{(\pm)},
\qquad
F_{\mu\nu}:=\partial_\mu A_\nu-\partial_\nu A_\mu .
\end{equation}
Therefore, the electromagnetic field acts through the usual antisymmetric field strength, whereas the Lorentz--violating structure remains contained in the branch dispersion relation $\mathcal D^{(\pm)}$.

It is useful to introduce the dispersion Hessian
\begin{equation}
H^{\mu\nu}_{(\pm)}
:=
\frac{\partial^2\mathcal D^{(\pm)}(P)}
{\partial P_\mu\partial P_\nu}.
\end{equation}
Since
\begin{equation}
k^\mu_{(\pm)}
=
\frac{\mathfrak e}{2}
\frac{\partial\mathcal D^{(\pm)}(P)}
{\partial P_\mu},
\end{equation}
the branch response tensor can be written as
\begin{equation}
M^{\mu\nu}_{(\pm)}
:=
-\frac{\partial k^\mu_{(\pm)}}{\partial P_\nu}
=
-\frac{\mathfrak e}{2}
H^{\mu\nu}_{(\pm)}.
\end{equation}

Then, one obtains the covariant branch acceleration
\begin{equation}
a^\mu_{(\pm)}
:=
k^\alpha_{(\pm)}\nabla_\alpha k^\mu_{(\pm)}
=
- q\,M^{\mu\nu}_{(\pm)}
F_{\nu\rho}k^\rho_{(\pm)} .
\end{equation}
Equivalently, in terms of the coordinate acceleration,
\begin{equation}
\frac{d k^\mu_{(\pm)}}{d\lambda}
=
-\Gamma^\mu{}_{\alpha\beta}
k^\alpha_{(\pm)}k^\beta_{(\pm)}
-
q\,M^{\mu\nu}_{(\pm)}
F_{\nu\rho}k^\rho_{(\pm)} .
\end{equation}

\subsection{Congruence geometry and transverse dimensionality}

Throughout this work, the congruence generator is identified with the
Hamiltonian tangent vector
\begin{equation}
k^\mu_{(\pm)}:=\dot x^\mu_{(\pm)} .
\end{equation}
The Raychaudhuri equation is written for the effective transverse
subspace orthogonal to the branch flow. In the situations considered
here, the physically relevant beam dynamics is effectively two-dimensional
in the transverse sector, so that the expansion contribution takes the form
\begin{equation}
-\frac12\theta_{(\pm)}^2 .
\end{equation}

More generally, for an $n$-dimensional transverse subspace, the
Raychaudhuri equation would contain the coefficient
$-1/n$. The present normalization therefore corresponds to
$n=2$, which is appropriate for the effective transverse evolution
of narrow beam congruences considered throughout this work.

{

The coefficient appearing in the quadratic expansion term is determined by the dimension of the subspace on which the congruence deformation is defined. For a congruence with tangent vector $k^\mu_{(\pm)}$, the evolution of neighboring trajectories is characterized only by directions orthogonal to the propagation.

In the present work, the physically relevant congruence corresponds to a collimated beam propagating along a preferred longitudinal direction. Only two independent transverse directions describe the local deformation of the beam cross section. The longitudinal component simply parametrizes the motion along the beam and does not contribute to the local change of the transverse area.

Accordingly, the deformation tensor is projected onto the effective transverse
subspace,
\begin{equation}
B^{\perp}_{\mu\nu} = h_\mu{}^{\alpha} h_\nu{}^{\beta} \nabla_\beta k_\alpha ,
\label{Bperp_projection}
\end{equation}
where $h_{\mu\nu}$ denotes the projector introduced below.
The trace of this projected tensor defines the expansion scalar,
\begin{equation}
\theta_{(\pm)} = h^{\mu\nu} B^{\perp}_{\mu\nu},
\label{theta_projection}
\end{equation}
while the shear and vorticity correspond to its traceless symmetric and antisymmetric parts, respectively.

For a transverse space of dimension $n_\perp$, the Raychaudhuri equation contains the universal coefficient $-1/n_\perp$ multiplying $\theta^2$. The present analysis adopts $n_\perp=2$, because the congruence represents a narrow beam whose observable deformation is entirely encoded in its two-dimensional transverse cross section. 

}

{
The choice $n_\perp=2$ changes only the numerical coefficient multiplying the quadratic expansion term. All branch-dependent contributions derived in the following sections remain unchanged, because they originate from the electromagnetic source $-q\nabla_\mu(M^{\mu\nu}_{(\pm)}F_{\nu\rho}k^\rho_{(\pm)})$, which is independent of the dimensionality of the transverse projection. The distinction between focusing and defocusing, the branch-selective response, and the physical interpretation of the momentum Hessian are unaffected by the choice of $n_\perp$. Different transverse dimensionalities merely modify the quantitative evolution of the expansion scalar through the self-focusing term. }

The branch-resolved Raychaudhuri equation takes the form
\begin{equation}
\frac{d\theta_{(\pm)}}{d\lambda}
=
-\frac12\theta_{(\pm)}^2
-\sigma^{(\pm)}_{\mu\nu}\sigma_{(\pm)}^{\mu\nu}
+\omega^{(\pm)}_{\mu\nu}\omega_{(\pm)}^{\mu\nu}
-R_{\mu\nu}k^\mu_{(\pm)}k^\nu_{(\pm)}
-q\,\nabla_\mu\!\left(
M^{\mu\nu}_{(\pm)}
F_{\nu\rho}
k^\rho_{(\pm)}
\right).
\label{eq:Ray_general}
\end{equation}
Here,
\begin{equation}
\theta_{(\pm)}
:=
\nabla_\mu k^\mu_{(\pm)}
\end{equation}
denotes the expansion scalar associated with the branch congruence,
measuring the local contraction or expansion rate of neighboring
trajectories. The covariant derivative of the congruence generator is
decomposed according to
\begin{equation}
\nabla_\nu k_\mu
=
\frac12\theta h_{\mu\nu}
+
\sigma_{\mu\nu}
+
\omega_{\mu\nu},
\end{equation}
where $h_{\mu\nu}$ is the effective transverse projector,
$\sigma_{\mu\nu}$ is the symmetric traceless shear tensor, and
$\omega_{\mu\nu}$ is the antisymmetric vorticity tensor.

Equation~\eqref{eq:Ray_general} separates the geometric and
electromagnetic contributions to the congruence evolution in a direct
way. The first four terms reproduce the standard kinematical structure
of the Raychaudhuri equation: expansion and shear favor focusing,
vorticity opposes it, and spacetime curvature contributes through the
Ricci projection
$R_{\mu\nu}k^\mu_{(\pm)}k^\nu_{(\pm)}$.
The last term represents the electromagnetic contribution induced by
the branch-dependent response tensor
$M^{\mu\nu}_{(\pm)}$.
Its sign is not fixed a priori, since it depends on the external field
configuration, on the local branch geometry, and on the modified
dispersion structure encoded in
$M^{\mu\nu}_{(\pm)}$.
Consequently, the same electromagnetic background may generate
different focusing or defocusing rates for the two branches, providing
a geometric realization of branch-dependent electromagnetic focusing.

For constant Lorentz--violating backgrounds, no additional force arises from gradients of the background coefficients. In this case, the non--geodesic contribution to the congruence is induced by the electromagnetic field, while the Lorentz--violating coefficients modify how this force is converted into acceleration through $k^\mu_{(\pm)}$ and $M^{\mu\nu}_{(\pm)}$. In the neutral limit $q\rightarrow0$, the last term vanishes and the free branch congruence is recovered. In the Lorentz--symmetric limit, $M^{\mu\nu}_{(\pm)}$ reduces to the usual mass--shell response tensor, and the standard Raychaudhuri equation for a charged congruence is obtained.


{\subsection{Relation with generalized and non--geodesic Raychaudhuri frameworks}}
\label{subsec:relation_generalized_raychaudhuri}

{
It is useful to clarify how the present equation is related to other generalized Raychaudhuri frameworks. For a generic non--geodesic congruence with tangent $k^{\mu}$, the Raychaudhuri equation contains the divergence of the acceleration,
\begin{equation}
\frac{\mathrm{d}\theta}{\mathrm{d}\lambda} = -\frac{1}{2}\theta^{2} -\sigma_{\mu\nu}\sigma^{\mu\nu} +\omega_{\mu\nu}\omega^{\mu\nu} -R_{\mu\nu}k^{\mu}k^{\nu} +\nabla_{\mu}a^{\mu},
\label{generic_nongeodesic_raychaudhuri}
\end{equation}
where $a^{\mu}=k^{\nu}\nabla_{\nu}k^{\mu}$. In this form, the acceleration is usually regarded as an already specified property of the congruence. This viewpoint is common in treatments of accelerated flows, forced congruences, and effective non--geodesic motion in modified gravity.

In generalized geometric settings, the extra terms in the Raychaudhuri equation often arise from modifications of the spacetime connection or from modified gravitational field equations. For example, in torsionful or metric--affine geometries, torsion and nonmetricity alter the decomposition of $\nabla_{\nu}k_{\mu}$ and contribute directly to the evolution of the kinematical scalars \cite{Luz:2017,Iosifidis:2018}. In modified gravity, the Raychaudhuri equation is frequently used to reformulate focusing or defocusing conditions in terms of effective curvature or effective energy--momentum contributions \cite{Conroy:2016,Beckering:2020,Geng:2026}. In these cases, the main question is how the gravitational or affine sector changes the sign and magnitude of the focusing terms.

Our construction differs in both origin and interpretation. The spacetime connection is kept Levi--Civita, and the Lorentz--violating backgrounds are treated as fixed effective coefficients. The new contribution comes from the electromagnetic evolution of the gauge--covariant momentum,
\begin{equation}
\dot{P}_{\mu}=qF_{\mu\nu}k^{\nu}_{(\pm)},
\label{comparison_momentum_evolution}
\end{equation}
combined with the branch-dependent momentum--velocity map. As a result, the acceleration is not an arbitrary non--geodesic input, but follows from
\begin{equation}
a^{\mu}_{(\pm)}
=
-qM^{\mu\nu}_{(\pm)}F_{\nu\rho}k^{\rho}_{(\pm)}.
\label{comparison_branch_acceleration}
\end{equation}
Therefore, the force divergence in the Raychaudhuri equation becomes
\begin{equation}
\nabla_{\mu}a^{\mu}_{(\pm)}
=
-q\nabla_{\mu}
\left(
M^{\mu\nu}_{(\pm)}F_{\nu\rho}k^{\rho}_{(\pm)}
\right).
\label{comparison_force_divergence}
\end{equation}
This expression has the same kinematical position as the acceleration-divergence term in the usual non--geodesic Raychaudhuri equation, but it carries additional physical information: the response tensor $M^{\mu\nu}_{(\pm)}$ is determined by the Lorentz--violating branch dispersion relation.

Accordingly, the present formalism may be viewed as a branch-resolved, gauge--covariant realization of non--geodesic Raychaudhuri dynamics. It keeps the standard geometric interpretation of expansion, shear, and vorticity, but it changes the mechanism that generates the acceleration. Instead of asking whether curvature, torsion, nonmetricity, or effective gravitational sources focus a congruence, we ask how a common electromagnetic field is converted into different focusing sources by the two spin--nondegenerate branches. This is why the same $F_{\mu\nu}$ can produce different values of $\theta_{(+)}$ and $\theta_{(-)}$, and may even focus one branch while defocusing the other.
}


\section{Minimally coupled $b_\mu$ sector}
\label{sec:em_b_sector}

We now specialize the electromagnetic construction to the spin--nondegenerate
$b_\mu$ sector. The Lorentz--violating background $b_\mu$ is kept constant,
whereas the electromagnetic potential $A_\mu(x)$ is allowed to depend on
spacetime.

\subsection{Gauge--covariant branch Hamiltonian}

For the free $b_\mu$ sector, the two branch dispersion functions are
\begin{equation}
\mathcal D_b^{(\pm)}(p)
=
p^2-b^2-m^2
\pm 2\sqrt{(b\cdot p)^2-b^2p^2}.
\end{equation}
After minimal coupling, the canonical momentum is replaced by the
gauge--covariant momentum
\begin{equation}
P_\mu:=\pi_\mu-qA_\mu(x).
\end{equation}
Thus the branch Hamiltonians take the form
\begin{equation}
\tilde H_b^{(\pm)}(x,\pi;\mathfrak e)
=
-\frac{\mathfrak e}{2}\,
\mathcal D_b^{(\pm)}(P),
\label{eq:Hb_em}
\end{equation}
with
\begin{equation}
\mathcal D_b^{(\pm)}(P)
=
P^2-b^2-m^2
\pm 2\sqrt{\Delta_b(P)},
\qquad
\Delta_b(P):=(b\cdot P)^2-b^2P^2 .
\label{eq:Db_em}
\end{equation}
The electromagnetic field therefore enters only through the replacement
$p_\mu\rightarrow P_\mu$. The branch structure of the free theory is not
changed by the minimal coupling, but the motion becomes sensitive to the
spacetime dependence of $A_\mu(x)$ through $P_\mu$.

\subsection{Branch tangent}

It is useful to introduce the $b$--Gram vector
\begin{equation}
Q^\mu(P):=(b\cdot P)b^\mu-b^2P^\mu .
\label{eq:Qb_em}
\end{equation}
Since
\begin{equation}
\frac{\partial}{\partial P_\mu}\sqrt{\Delta_b(P)}
=
\frac{Q^\mu(P)}{\sqrt{\Delta_b(P)}},
\end{equation}
Hamilton's equation gives
\begin{align}
k^\mu_{(\pm)}
:=
\dot x^\mu_{(\pm)}
&=
-\frac{\partial \tilde H_b^{(\pm)}}{\partial \pi_\mu}
=
\frac{\mathfrak e}{2}
\frac{\partial \mathcal D_b^{(\pm)}(P)}{\partial P_\nu}
\frac{\partial P_\nu}{\partial \pi_\mu}.
\end{align}
Using $\partial P_\nu/\partial\pi_\mu=\delta^\mu{}_\nu$, one obtains
\begin{equation}
k^\mu_{(\pm)}
=
\mathfrak e
\left(
P^\mu
\pm
\frac{Q^\mu(P)}{\sqrt{\Delta_b(P)}}
\right).
\label{eq:kb_em}
\end{equation}
The two signs correspond to the two spin--nondegenerate branches. If the
opposite convention is adopted for labeling the branches, one simply exchanges
$(+)\leftrightarrow(-)$ throughout the following expressions.

\subsection{Electromagnetic evolution of the gauge--covariant momentum}

The Hamilton equation for $\pi_\mu$ reads
\begin{equation}
\dot\pi_\mu
=
\frac{\partial \tilde H_b^{(\pm)}}{\partial x^\mu}.
\end{equation}
Since the spacetime dependence of \eqref{eq:Hb_em} appears through
$A_\mu(x)$, we have
\begin{align}
\dot\pi_\mu
&=
-\frac{\mathfrak e}{2}
\frac{\partial \mathcal D_b^{(\pm)}(P)}{\partial P_\nu}
\frac{\partial P_\nu}{\partial x^\mu}
\nonumber\\
&=
\frac{\mathfrak e q}{2}
\frac{\partial \mathcal D_b^{(\pm)}(P)}{\partial P_\nu}
\partial_\mu A_\nu
=
q\,k^\nu_{(\pm)}\partial_\mu A_\nu .
\end{align}
Therefore,
\begin{align}
\dot P_\mu
&=
\dot\pi_\mu
-
q\,\partial_\alpha A_\mu\,\dot x^\alpha
\nonumber\\
&=
q\,k^\alpha_{(\pm)}
\left(
\partial_\mu A_\alpha-\partial_\alpha A_\mu
\right),
\end{align}
or, equivalently,
\begin{equation}
\dot P_\mu
=
qF_{\mu\nu}k^\nu_{(\pm)},
\qquad
F_{\mu\nu}:=\partial_\mu A_\nu-\partial_\nu A_\mu .
\label{eq:Lorentz_force_b}
\end{equation}
This has the same covariant Lorentz--force structure as in the standard case.
The difference is that the tangent vector is now branch dependent and is
determined by the Lorentz--violating dispersion relation.

\subsection{Branch Hessian}

The response of the branch tangent to changes in the gauge--covariant momentum
is measured by the momentum Hessian
\begin{equation}
M^{\mu\nu}_{(\pm)}
:=
-\frac{\partial k^\mu_{(\pm)}}{\partial P_\nu}.
\label{eq:Mb_def}
\end{equation}
From \eqref{eq:kb_em}, one finds
\begin{equation}
M^{\mu\nu}_{(\pm)}
=
-\mathfrak e
\left[
g^{\mu\nu}
\pm
\frac{\partial}{\partial P_\nu}
\left(
\frac{Q^\mu(P)}{\sqrt{\Delta_b(P)}}
\right)
\right].
\end{equation}
Using
\begin{equation}
\frac{\partial Q^\mu}{\partial P_\nu}
=
b^\mu b^\nu-b^2g^{\mu\nu},
\qquad
\frac{\partial\Delta_b}{\partial P_\nu}=2Q^\nu,
\end{equation}
we obtain
\begin{align}
\frac{\partial}{\partial P_\nu}
\left(
\frac{Q^\mu}{\sqrt{\Delta_b}}
\right)
&=
\frac{b^\mu b^\nu-b^2g^{\mu\nu}}{\sqrt{\Delta_b}}
-
\frac{Q^\mu Q^\nu}{\Delta_b^{3/2}} .
\end{align}
Therefore,
\begin{equation}
M^{\mu\nu}_{(\pm)}
=
-\mathfrak e
\left[
g^{\mu\nu}
\pm
\left(
\frac{b^\mu b^\nu-b^2g^{\mu\nu}}{\sqrt{\Delta_b(P)}}
-
\frac{Q^\mu(P)Q^\nu(P)}{\Delta_b(P)^{3/2}}
\right)
\right].
\label{eq:Mb_explicit}
\end{equation}
The Hessian controls how the electromagnetic force is converted into branch
acceleration. Hence, even for the same electromagnetic field, the two branches
do not respond in the same way. The terms proportional to inverse powers of
$\Delta_b(P)$ also show that the splitting becomes stronger near the
spin--degeneracy surface $\Delta_b(P)=0$, where the branch description must be
treated with care.

\subsection{Branch acceleration}

For constant $b_\mu$, the branch tangent depends on spacetime through
$P_\mu(x)$. Along the worldline,
\begin{align}
\frac{d k^\mu_{(\pm)}}{d\lambda}
&=
\frac{\partial k^\mu_{(\pm)}}{\partial P_\nu}
\dot P_\nu
=
-M^{\mu\nu}_{(\pm)}\dot P_\nu .
\end{align}
Using \eqref{eq:Lorentz_force_b}, we obtain
\begin{equation}
\frac{d k^\mu_{(\pm)}}{d\lambda}
=
-q\,M^{\mu\nu}_{(\pm)}
F_{\nu\rho}k^\rho_{(\pm)} .
\end{equation}
In flat spacetime, this gives
\begin{equation}
a^\mu_{(\pm)}
=
-q\,M^{\mu\nu}_{(\pm)}
F_{\nu\rho}k^\rho_{(\pm)} .
\label{eq:ab_em_flat}
\end{equation}
In a curved background, the covariant acceleration is defined by
\begin{equation}
a^\mu_{(\pm)}
:=
k^\nu_{(\pm)}\nabla_\nu k^\mu_{(\pm)} ,
\end{equation}
and the electromagnetic contribution is written in the covariant form
\begin{equation}
a^\mu_{(\pm)}
=
-q\,M^{\mu\nu}_{(\pm)}
F_{\nu\rho}k^\rho_{(\pm)} ,
\label{eq:ab_em}
\end{equation}
provided the Lorentz--violating background is fixed and no gradients of
$b_\mu$ are included. In coordinate form, this corresponds to
\begin{equation}
\frac{d k^\mu_{(\pm)}}{d\lambda}
+
\Gamma^\mu{}_{\alpha\beta}
k^\alpha_{(\pm)}k^\beta_{(\pm)}
=
-q\,M^{\mu\nu}_{(\pm)}
F_{\nu\rho}k^\rho_{(\pm)} .
\end{equation}
Thus the electromagnetic field is the source of non--geodesic motion, while
the Lorentz--violating coefficients determine the branch-dependent response.

\subsection{Raychaudhuri equation}

The branch-resolved Raychaudhuri equation is
\begin{equation}
\frac{d\theta_{(\pm)}}{d\lambda}
=
-\frac12\theta_{(\pm)}^2
-\sigma^{(\pm)}_{\mu\nu}\sigma_{(\pm)}^{\mu\nu}
+\omega^{(\pm)}_{\mu\nu}\omega_{(\pm)}^{\mu\nu}
-R_{\mu\nu}k^\mu_{(\pm)}k^\nu_{(\pm)}
+\nabla_\mu a^\mu_{(\pm)} .
\label{eq:Ray_b_em_1}
\end{equation}
Substituting \eqref{eq:ab_em}, one obtains
\begin{equation}
\frac{d\theta_{(\pm)}}{d\lambda}
=
-\frac12\theta_{(\pm)}^2
-\sigma^{(\pm)}_{\mu\nu}\sigma_{(\pm)}^{\mu\nu}
+\omega^{(\pm)}_{\mu\nu}\omega_{(\pm)}^{\mu\nu}
-R_{\mu\nu}k^\mu_{(\pm)}k^\nu_{(\pm)}
-q\,\nabla_\mu\!\left(
M^{\mu\nu}_{(\pm)}F_{\nu\rho}k^\rho_{(\pm)}
\right).
\label{eq:Ray_b_em_2}
\end{equation}
In Minkowski spacetime, this reduces to
\begin{equation}
\frac{d\theta_{(\pm)}}{d\lambda}
=
-\frac12\theta_{(\pm)}^2
-\sigma^{(\pm)}_{\mu\nu}\sigma_{(\pm)}^{\mu\nu}
+\omega^{(\pm)}_{\mu\nu}\omega_{(\pm)}^{\mu\nu}
-q\,\partial_\mu\!\left(
M^{\mu\nu}_{(\pm)}F_{\nu\rho}k^\rho_{(\pm)}
\right).
\label{eq:Ray_b_em_flat}
\end{equation}

Equation \eqref{eq:Ray_b_em_2} shows how the focusing properties of the
charged branch congruence are modified. The first four terms have the usual
geometrical meaning: the expansion and shear terms favor focusing, vorticity
opposes focusing, and the Ricci contraction measures the effect of spacetime
curvature along the branch tangent. The last term is the electromagnetic
contribution. Its sign is not fixed in general, because it depends on the field
configuration, on the branch tangent, and on the Hessian
$M^{\mu\nu}_{(\pm)}$. Hence the same electromagnetic background can either
increase or decrease the focusing rate, depending on the branch and on the
orientation of $P_\mu$ relative to $b_\mu$.

For constant Lorentz--violating backgrounds, there is no force generated by
gradients of $b_\mu$. The non--geodesic part of the motion is therefore induced
by the electromagnetic field, but the Lorentz--violating coefficients remain
present through $k^\mu_{(\pm)}$ and $M^{\mu\nu}_{(\pm)}$. This is the main
physical effect of the spin--nondegenerate sector: the electromagnetic field is
common to both branches, while the response tensor is different for each one.
In the neutral limit $q\rightarrow0$, the electromagnetic contribution
vanishes and the free branch congruence is recovered.

\section{Minimally coupled $H_{\mu\nu}$ sector}
\label{sec:em_H_sector}

We now apply the same electromagnetic construction to the spin--nondegenerate. The Lorentz--violating background is
kept fixed, while the electromagnetic potential $A_\mu(x)$ carries the
spacetime dependence associated with the external field.

\subsection{Gauge--covariant branch Hamiltonian}

For the free $H_{\mu\nu}$ sector with $Y=0$, we define
\begin{equation}
X:=\frac14 H_{\alpha\beta}H^{\alpha\beta},
\qquad
K_{\mu\nu}:=(HH)_{\mu\nu}+2X\,g_{\mu\nu},
\qquad
(HH)_{\mu\nu}:=H_{\mu}{}^{\alpha}H_{\alpha\nu}.
\end{equation}
The quadratic form controlling the branch splitting is
\begin{equation}
\mathcal R_H(p):=p\cdot K\cdot p
=
p_\mu K^{\mu\nu}p_\nu .
\end{equation}
Thus the two free branch factors are
\begin{equation}
\mathcal D_H^{(\pm)}(p)
=
p^2+2X-m^2
\pm 2\sqrt{\mathcal R_H(p)} .
\end{equation}

After minimal coupling, the canonical momentum is replaced by the
gauge--covariant momentum
\begin{equation}
P_\mu:=\pi_\mu-qA_\mu(x).
\end{equation}
The branch Hamiltonians are therefore
\begin{equation}
\tilde H_H^{(\pm)}(x,\pi;\mathfrak e)
=
-\frac{\mathfrak e}{2}\,
\mathcal D_H^{(\pm)}(P),
\label{eq:HH_em}
\end{equation}
where
\begin{equation}
\mathcal D_H^{(\pm)}(P)
=
P^2+2X-m^2
\pm 2Q_H,
\qquad
Q_H:=\sqrt{\mathcal R_H(P)}
=
\sqrt{P\cdot K\cdot P}.
\label{eq:DH_em}
\end{equation}
Hence the electromagnetic field does not change the form of the branch
dispersion relation. It enters through the gauge--covariant replacement
$p_\mu\rightarrow P_\mu$.

\subsection{Branch tangent}

Let
\begin{equation}
S^\mu(P):=(KP)^\mu=K^{\mu\nu}P_\nu .
\end{equation}
Since
\begin{equation}
\frac{\partial Q_H}{\partial P_\mu}
=
\frac{S^\mu(P)}{Q_H},
\end{equation}
Hamilton's equation gives
\begin{align}
k^\mu_{(\pm)}
:=
\dot x^\mu_{(\pm)}
&=
-\frac{\partial \tilde H_H^{(\pm)}}{\partial \pi_\mu}
=
\frac{\mathfrak e}{2}
\frac{\partial\mathcal D_H^{(\pm)}(P)}{\partial P_\nu}
\frac{\partial P_\nu}{\partial \pi_\mu}.
\end{align}
Using $\partial P_\nu/\partial\pi_\mu=\delta^\mu{}_\nu$, we obtain
\begin{equation}
k^\mu_{(\pm)}
=
\mathfrak e
\left(
P^\mu
\pm
\frac{S^\mu(P)}{Q_H}
\right).
\label{eq:kH_em}
\end{equation}
The tensor $K_{\mu\nu}$ therefore determines how the two spin--nondegenerate
branches separate in momentum space. In contrast with the $b_\mu$ sector, the
anisotropy is not controlled by a single preferred direction, but by the
tensorial structure induced by $H_{\mu\nu}$.

\subsection{Electromagnetic evolution of the gauge--covariant momentum}

The Hamilton equation for $\pi_\mu$ is
\begin{equation}
\dot\pi_\mu
=
\frac{\partial\tilde H_H^{(\pm)}}{\partial x^\mu}.
\end{equation}
For the electromagnetic part, the spacetime dependence enters through
$A_\mu(x)$. Thus
\begin{align}
\dot\pi_\mu
&=
-\frac{\mathfrak e}{2}
\frac{\partial\mathcal D_H^{(\pm)}(P)}{\partial P_\nu}
\frac{\partial P_\nu}{\partial x^\mu}
\nonumber\\
&=
\frac{\mathfrak e q}{2}
\frac{\partial\mathcal D_H^{(\pm)}(P)}{\partial P_\nu}
\partial_\mu A_\nu
=
q\,k^\nu_{(\pm)}\partial_\mu A_\nu .
\end{align}
Consequently,
\begin{align}
\dot P_\mu
&=
\dot\pi_\mu
-
q\,\partial_\alpha A_\mu\,\dot x^\alpha
\nonumber\\
&=
q\,k^\alpha_{(\pm)}
\left(
\partial_\mu A_\alpha-\partial_\alpha A_\mu
\right),
\end{align}
which gives
\begin{equation}
\dot P_\mu
=
qF_{\mu\nu}k^\nu_{(\pm)},
\qquad
F_{\mu\nu}:=\partial_\mu A_\nu-\partial_\nu A_\mu .
\label{eq:Lorentz_force_H}
\end{equation}
Thus the electromagnetic interaction keeps the usual gauge--covariant
Lorentz--force form. The branch dependence appears through
$k^\mu_{(\pm)}$, which is fixed by the modified dispersion relation.

\subsection{Branch Hessian}

The branch response tensor is defined by
\begin{equation}
M^{\mu\nu}_{(\pm)}
:=
-\frac{\partial k^\mu_{(\pm)}}{\partial P_\nu}.
\label{eq:MH_def_em}
\end{equation}
From \eqref{eq:kH_em},
\begin{equation}
M^{\mu\nu}_{(\pm)}
=
-\mathfrak e
\left[
g^{\mu\nu}
\pm
\frac{\partial}{\partial P_\nu}
\left(
\frac{S^\mu(P)}{Q_H}
\right)
\right].
\end{equation}
Since
\begin{equation}
\frac{\partial S^\mu}{\partial P_\nu}=K^{\mu\nu},
\qquad
\frac{\partial\mathcal R_H}{\partial P_\nu}=2S^\nu,
\qquad
\frac{\partial Q_H}{\partial P_\nu}=\frac{S^\nu}{Q_H},
\end{equation}
we have
\begin{align}
\frac{\partial}{\partial P_\nu}
\left(
\frac{S^\mu}{Q_H}
\right)
&=
\frac{K^{\mu\nu}}{Q_H}
-
\frac{S^\mu S^\nu}{Q_H^3}.
\end{align}
Therefore,
\begin{equation}
M^{\mu\nu}_{(\pm)}
=
-\mathfrak e
\left[
g^{\mu\nu}
\pm
\left(
\frac{K^{\mu\nu}}{Q_H}
-
\frac{S^\mu(P)S^\nu(P)}{Q_H^3}
\right)
\right].
\label{eq:MH_em_explicit}
\end{equation}
This tensor controls how the electromagnetic force changes the branch tangent.
The terms involving inverse powers of $Q_H$ show that the response becomes more
sensitive near the effective degeneracy region $P\cdot K\cdot P\simeq0$,
where the two-branch description must be handled with care.

\subsection{Branch acceleration}

For a fixed $H_{\mu\nu}$ background, the electromagnetic variation of the
branch tangent follows from its dependence on $P_\mu$:
\begin{align}
\frac{d k^\mu_{(\pm)}}{d\lambda}
&=
\frac{\partial k^\mu_{(\pm)}}{\partial P_\nu}\dot P_\nu
=
-M^{\mu\nu}_{(\pm)}\dot P_\nu .
\end{align}
Using \eqref{eq:Lorentz_force_H}, one obtains
\begin{equation}
\frac{d k^\mu_{(\pm)}}{d\lambda}
=
-q\,M^{\mu\nu}_{(\pm)}
F_{\nu\rho}k^\rho_{(\pm)}
\end{equation}
in flat spacetime. In a curved background, the corresponding covariant
equation of motion is written as
\begin{equation}
a^\mu_{(\pm)}
:=
k^\nu_{(\pm)}\nabla_\nu k^\mu_{(\pm)}
=
-q\,M^{\mu\nu}_{(\pm)}
F_{\nu\rho}k^\rho_{(\pm)} .
\label{eq:aH_em}
\end{equation}
Equivalently, in coordinate form,
\begin{equation}
\frac{d k^\mu_{(\pm)}}{d\lambda}
+
\Gamma^\mu{}_{\alpha\beta}
k^\alpha_{(\pm)}k^\beta_{(\pm)}
=
-q\,M^{\mu\nu}_{(\pm)}
F_{\nu\rho}k^\rho_{(\pm)} .
\end{equation}
Thus the electromagnetic field provides the non--geodesic force, while
$H_{\mu\nu}$ fixes the branch-dependent response through $K_{\mu\nu}$ and
$M^{\mu\nu}_{(\pm)}$.

\subsection{Raychaudhuri equation}

The branch-resolved Raychaudhuri equation is
\begin{equation}
\frac{d\theta_{(\pm)}}{d\lambda}
=
-\frac12\theta_{(\pm)}^2
-\sigma^{(\pm)}_{\mu\nu}\sigma_{(\pm)}^{\mu\nu}
+\omega^{(\pm)}_{\mu\nu}\omega_{(\pm)}^{\mu\nu}
-R_{\mu\nu}k^\mu_{(\pm)}k^\nu_{(\pm)}
+\nabla_\mu a^\mu_{(\pm)} .
\label{eq:Ray_H_em_1}
\end{equation}
Substituting \eqref{eq:aH_em}, we find
\begin{equation}
\frac{d\theta_{(\pm)}}{d\lambda}
=
-\frac12\theta_{(\pm)}^2
-\sigma^{(\pm)}_{\mu\nu}\sigma_{(\pm)}^{\mu\nu}
+\omega^{(\pm)}_{\mu\nu}\omega_{(\pm)}^{\mu\nu}
-R_{\mu\nu}k^\mu_{(\pm)}k^\nu_{(\pm)}
-q\,\nabla_\mu\!\left(
M^{\mu\nu}_{(\pm)}F_{\nu\rho}k^\rho_{(\pm)}
\right).
\label{eq:Ray_H_em_2}
\end{equation}
In Minkowski spacetime this reduces to
\begin{equation}
\frac{d\theta_{(\pm)}}{d\lambda}
=
-\frac12\theta_{(\pm)}^2
-\sigma^{(\pm)}_{\mu\nu}\sigma_{(\pm)}^{\mu\nu}
+\omega^{(\pm)}_{\mu\nu}\omega_{(\pm)}^{\mu\nu}
-q\,\partial_\mu\!\left(
M^{\mu\nu}_{(\pm)}F_{\nu\rho}k^\rho_{(\pm)}
\right).
\label{eq:Ray_H_em_flat}
\end{equation}

Equation \eqref{eq:Ray_H_em_2} shows that the electromagnetic contribution to
the expansion is filtered by the branch tensor $M^{\mu\nu}_{(\pm)}$. Therefore,
two particles subjected to the same external field may follow different
congruence evolutions once they belong to different spin--nondegenerate
branches. In the $H_{\mu\nu}$ sector, this dependence is governed by
$K_{\mu\nu}$ and by the projection $S^\mu(P)=(KP)^\mu$, so the result depends
on how the gauge--covariant momentum is oriented relative to the tensorial
Lorentz--violating background. When $q=0$, the electromagnetic contribution is
removed. When the branch splitting is absent, the usual charged congruence is
recovered.

\section{Minimally coupled $d_{\mu\nu}$ sector}
\label{sec:em_d_sector}

We now apply the electromagnetic construction to the spin--nondegenerate
$d_{\mu\nu}$ sector. The Lorentz--violating background is kept fixed, while
the electromagnetic potential $A_\mu(x)$ is allowed to depend on spacetime.

\subsection{Gauge--covariant branch Hamiltonian}

For the free $d_{\mu\nu}$ sector, we define
\begin{equation}
\mathcal S_{\mu\nu}:=(d^2)_{\mu\nu}:=
d_{\mu}{}^{\alpha}d_{\alpha\nu},
\qquad
\mathcal R_d(p):=p\cdot \mathcal S\cdot p
=
p_\mu \mathcal S^{\mu\nu}p_\nu .
\end{equation}
The two branch factors are
\begin{equation}
\mathcal D_d^{(\pm)}(p)
=
p^2-m^2
\pm 2m\sqrt{\mathcal R_d(p)} .
\end{equation}
After minimal coupling, the canonical momentum is replaced by the
gauge--covariant momentum
\begin{equation}
P_\mu:=\pi_\mu-qA_\mu(x).
\end{equation}
Thus the branch Hamiltonians become
\begin{equation}
\tilde H_d^{(\pm)}(x,\pi;\mathfrak e)
=
-\frac{\mathfrak e}{2}\,
\mathcal D_d^{(\pm)}(P),
\label{eq:Hd_em}
\end{equation}
where
\begin{equation}
\mathcal D_d^{(\pm)}(P)
=
P^2-m^2
\pm 2mQ_d,
\qquad
Q_d:=\sqrt{\mathcal R_d(P)}
=
\sqrt{P\cdot \mathcal S\cdot P}.
\label{eq:Dd_em}
\end{equation}
The electromagnetic field therefore enters through the gauge--covariant
replacement $p_\mu\rightarrow P_\mu$, without changing the branch structure of
the free dispersion relation.

\subsection{Branch tangent}

Let
\begin{equation}
\mathcal S^\mu(P):=(\mathcal SP)^\mu
=
\mathcal S^{\mu\nu}P_\nu .
\end{equation}
Since
\begin{equation}
\frac{\partial Q_d}{\partial P_\mu}
=
\frac{\mathcal S^\mu(P)}{Q_d},
\end{equation}
Hamilton's equation gives
\begin{align}
k^\mu_{(\pm)}
:=
\dot x^\mu_{(\pm)}
&=
-\frac{\partial \tilde H_d^{(\pm)}}{\partial \pi_\mu}
=
\frac{\mathfrak e}{2}
\frac{\partial\mathcal D_d^{(\pm)}(P)}{\partial P_\nu}
\frac{\partial P_\nu}{\partial \pi_\mu}.
\end{align}
Using $\partial P_\nu/\partial\pi_\mu=\delta^\mu{}_\nu$, we find
\begin{equation}
k^\mu_{(\pm)}
=
\mathfrak e
\left(
P^\mu
\pm
m\,\frac{\mathcal S^\mu(P)}{Q_d}
\right).
\label{eq:kd_em}
\end{equation}
The two branches are therefore separated by the projection of the
gauge--covariant momentum along the symmetric tensor
$\mathcal S_{\mu\nu}=(d^2)_{\mu\nu}$. If the opposite branch convention is
used, one simply exchanges $(+)\leftrightarrow(-)$.

\subsection{Electromagnetic evolution of the gauge--covariant momentum}

The Hamilton equation for $\pi_\mu$ reads
\begin{equation}
\dot\pi_\mu
=
\frac{\partial \tilde H_d^{(\pm)}}{\partial x^\mu}.
\end{equation}
Since the spacetime dependence in \eqref{eq:Hd_em} comes from $A_\mu(x)$, we
obtain
\begin{align}
\dot\pi_\mu
&=
-\frac{\mathfrak e}{2}
\frac{\partial\mathcal D_d^{(\pm)}(P)}{\partial P_\nu}
\frac{\partial P_\nu}{\partial x^\mu}
\nonumber\\
&=
\frac{\mathfrak e q}{2}
\frac{\partial\mathcal D_d^{(\pm)}(P)}{\partial P_\nu}
\partial_\mu A_\nu
=
q\,k^\nu_{(\pm)}\partial_\mu A_\nu .
\end{align}
Hence
\begin{align}
\dot P_\mu
&=
\dot\pi_\mu
-
q\,\partial_\alpha A_\mu\,\dot x^\alpha
\nonumber\\
&=
q\,k^\alpha_{(\pm)}
\left(
\partial_\mu A_\alpha-\partial_\alpha A_\mu
\right),
\end{align}
which gives the gauge--covariant Lorentz--force law
\begin{equation}
\dot P_\mu
=
qF_{\mu\nu}k^\nu_{(\pm)},
\qquad
F_{\mu\nu}:=\partial_\mu A_\nu-\partial_\nu A_\mu .
\label{eq:Lorentz_force_d}
\end{equation}
Thus the electromagnetic force keeps its usual form, while the branch
dependence enters through $k^\mu_{(\pm)}$.

\subsection{Branch Hessian}

The response of the branch tangent to a change in the gauge--covariant momentum
is described by
\begin{equation}
M^{\mu\nu}_{(\pm)}
:=
-\frac{\partial k^\mu_{(\pm)}}{\partial P_\nu}.
\label{eq:Md_def_em}
\end{equation}
From \eqref{eq:kd_em}, one obtains
\begin{equation}
M^{\mu\nu}_{(\pm)}
=
-\mathfrak e
\left[
g^{\mu\nu}
\pm
m\,\frac{\partial}{\partial P_\nu}
\left(
\frac{\mathcal S^\mu(P)}{Q_d}
\right)
\right].
\end{equation}
Using
\begin{equation}
\frac{\partial\mathcal S^\mu}{\partial P_\nu}
=
\mathcal S^{\mu\nu},
\qquad
\frac{\partial\mathcal R_d}{\partial P_\nu}
=
2\mathcal S^\nu(P),
\qquad
\frac{\partial Q_d}{\partial P_\nu}
=
\frac{\mathcal S^\nu(P)}{Q_d},
\end{equation}
we have
\begin{align}
\frac{\partial}{\partial P_\nu}
\left(
\frac{\mathcal S^\mu(P)}{Q_d}
\right)
&=
\frac{\mathcal S^{\mu\nu}}{Q_d}
-
\frac{\mathcal S^\mu(P)\mathcal S^\nu(P)}{Q_d^3}.
\end{align}
Therefore,
\begin{equation}
M^{\mu\nu}_{(\pm)}
=
-\mathfrak e
\left[
g^{\mu\nu}
\pm
m\left(
\frac{\mathcal S^{\mu\nu}}{Q_d}
-
\frac{\mathcal S^\mu(P)\mathcal S^\nu(P)}{Q_d^3}
\right)
\right].
\label{eq:Md_em_explicit}
\end{equation}
This tensor determines how the electromagnetic force changes the tangent of
each branch. The inverse powers of $Q_d$ show that the response becomes more
sensitive near the effective degeneracy region $P\cdot\mathcal S\cdot P\simeq0$,
where the branch separation must be treated carefully.

\subsection{Branch acceleration}

For constant $d_{\mu\nu}$, the branch tangent depends on spacetime through the
gauge--covariant momentum. Along the worldline,
\begin{align}
\frac{d k^\mu_{(\pm)}}{d\lambda}
&=
\frac{\partial k^\mu_{(\pm)}}{\partial P_\nu}\dot P_\nu
=
-M^{\mu\nu}_{(\pm)}\dot P_\nu .
\end{align}
Using \eqref{eq:Lorentz_force_d}, one finds
\begin{equation}
\frac{d k^\mu_{(\pm)}}{d\lambda}
=
-q\,M^{\mu\nu}_{(\pm)}
F_{\nu\rho}k^\rho_{(\pm)}
\end{equation}
in flat spacetime. In a curved background, the same result is written in
covariant form as
\begin{equation}
a^\mu_{(\pm)}
:=
k^\nu_{(\pm)}\nabla_\nu k^\mu_{(\pm)}
=
-q\,M^{\mu\nu}_{(\pm)}
F_{\nu\rho}k^\rho_{(\pm)} .
\label{eq:ad_em}
\end{equation}
Equivalently,
\begin{equation}
\frac{d k^\mu_{(\pm)}}{d\lambda}
+
\Gamma^\mu{}_{\alpha\beta}
k^\alpha_{(\pm)}k^\beta_{(\pm)}
=
-q\,M^{\mu\nu}_{(\pm)}
F_{\nu\rho}k^\rho_{(\pm)} .
\end{equation}
Hence the electromagnetic field generates the non--geodesic part of the
motion, while $d_{\mu\nu}$ fixes the branch-dependent response through
$\mathcal S_{\mu\nu}$ and $M^{\mu\nu}_{(\pm)}$.

\subsection{Raychaudhuri equation}

The branch-resolved Raychaudhuri equation is
\begin{equation}
\frac{d\theta_{(\pm)}}{d\lambda}
=
-\frac12\theta_{(\pm)}^2
-\sigma^{(\pm)}_{\mu\nu}\sigma_{(\pm)}^{\mu\nu}
+\omega^{(\pm)}_{\mu\nu}\omega_{(\pm)}^{\mu\nu}
-R_{\mu\nu}k^\mu_{(\pm)}k^\nu_{(\pm)}
+\nabla_\mu a^\mu_{(\pm)} .
\label{eq:Ray_d_em_1}
\end{equation}
Substituting \eqref{eq:ad_em}, we obtain
\begin{equation}
\frac{d\theta_{(\pm)}}{d\lambda}
=
-\frac12\theta_{(\pm)}^2
-\sigma^{(\pm)}_{\mu\nu}\sigma_{(\pm)}^{\mu\nu}
+\omega^{(\pm)}_{\mu\nu}\omega_{(\pm)}^{\mu\nu}
-R_{\mu\nu}k^\mu_{(\pm)}k^\nu_{(\pm)}
-q\,\nabla_\mu\!\left(
M^{\mu\nu}_{(\pm)}F_{\nu\rho}k^\rho_{(\pm)}
\right).
\label{eq:Ray_d_em_2}
\end{equation}
In Minkowski spacetime this reduces to
\begin{equation}
\frac{d\theta_{(\pm)}}{d\lambda}
=
-\frac12\theta_{(\pm)}^2
-\sigma^{(\pm)}_{\mu\nu}\sigma_{(\pm)}^{\mu\nu}
+\omega^{(\pm)}_{\mu\nu}\omega_{(\pm)}^{\mu\nu}
-q\,\partial_\mu\!\left(
M^{\mu\nu}_{(\pm)}F_{\nu\rho}k^\rho_{(\pm)}
\right).
\label{eq:Ray_d_em_flat}
\end{equation}

Equation \eqref{eq:Ray_d_em_2} shows that the electromagnetic contribution to
the expansion is weighted by the branch response tensor
$M^{\mu\nu}_{(\pm)}$. Therefore, the two branches can have different
congruence evolutions even when they propagate in the same electromagnetic
background. In the $d_{\mu\nu}$ sector, this difference is controlled by
$\mathcal S_{\mu\nu}=(d^2)_{\mu\nu}$ and by the projection
$\mathcal S^\mu(P)=(\mathcal SP)^\mu$. The result then depends on the
orientation of the gauge--covariant momentum relative to the tensorial
Lorentz--violating background. When $q=0$, the electromagnetic correction is
removed. When the branch splitting is absent, the usual charged congruence is
recovered.

\section{Unified structure of electromagnetic forcing in Lorentz-violating congruences}
\label{sec:unified_structure}

The previous subsections show that the minimally coupled spin--nondegenerate
sectors have the same dynamical form. The explicit expressions for
$k^\mu_{(\pm)}$ and $M^{\mu\nu}_{(\pm)}$ depend on the chosen
Lorentz--violating background, but the electromagnetic coupling enters through
the same gauge--covariant mechanism.

\subsection{Universal Hamiltonian structure}

For all sectors considered above, minimal coupling shifts the canonical
momentum according to
\begin{equation}
P_\mu=\pi_\mu-qA_\mu(x).
\end{equation}
The branch Hamiltonian can then be written as
\begin{equation}
\tilde H^{(\pm)}(x,\pi;\mathfrak e)
=
-\frac{\mathfrak e}{2}\,
\mathcal D^{(\pm)}(P),
\end{equation}
where the specific form of $\mathcal D^{(\pm)}$ depends on the
Lorentz--violating sector. Hamilton's equation gives
\begin{equation}
k^\mu_{(\pm)}
:=
\dot x^\mu_{(\pm)}
=
\frac{\mathfrak e}{2}
\frac{\partial\mathcal D^{(\pm)}(P)}{\partial P_\mu}.
\end{equation}
The electromagnetic field changes the motion through the evolution of the
gauge--covariant momentum,
\begin{equation}
\dot P_\mu
=
qF_{\mu\nu}k^\nu_{(\pm)}.
\end{equation}
Thus the Lorentz--force law keeps the same covariant form in all
spin--nondegenerate sectors. The sector dependence is carried by the branch
dispersion relation and, consequently, by the tangent vector
$k^\mu_{(\pm)}$.

\subsection{Universal branch acceleration}

Since the Lorentz--violating backgrounds are kept constant, the branch tangent
depends on spacetime only through $P_\mu$,
\begin{equation}
k^\mu_{(\pm)}=k^\mu_{(\pm)}(P).
\end{equation}
Introducing the momentum Hessian
\begin{equation}
M^{\mu\nu}_{(\pm)}
:=
-\frac{\partial k^\mu_{(\pm)}}{\partial P_\nu},
\end{equation}
one obtains, in flat spacetime,
\begin{equation}
\frac{d k^\mu_{(\pm)}}{d\lambda}
=
-q\,M^{\mu\nu}_{(\pm)}
F_{\nu\rho}k^\rho_{(\pm)} .
\end{equation}
In a curved background, the same result is written in covariant form as
\begin{equation}
a^\mu_{(\pm)}
:=
k^\nu_{(\pm)}\nabla_\nu k^\mu_{(\pm)}
=
-q\,M^{\mu\nu}_{(\pm)}
F_{\nu\rho}k^\rho_{(\pm)} .
\end{equation}
Equivalently,
\begin{equation}
\frac{d k^\mu_{(\pm)}}{d\lambda}
+
\Gamma^\mu{}_{\alpha\beta}
k^\alpha_{(\pm)}k^\beta_{(\pm)}
=
-q\,M^{\mu\nu}_{(\pm)}
F_{\nu\rho}k^\rho_{(\pm)} .
\end{equation}
Therefore, the electromagnetic field generates the non--geodesic part of the
motion, whereas the Lorentz--violating coefficients determine the branch
response through $M^{\mu\nu}_{(\pm)}$.

The tensor
$M^{\mu\nu}_{(\pm)}$
plays the role of an effective geometric response operator converting
electromagnetic forcing into congruence acceleration. In this sense,
the branch-dependent focusing properties are controlled not directly
by the external field itself, but by how the modified dispersion
geometry filters the electromagnetic interaction.

\subsection{Universal Raychaudhuri equation}

The branch-resolved Raychaudhuri equation is
\begin{equation}
\frac{d\theta_{(\pm)}}{d\lambda}
=
-\frac12\theta_{(\pm)}^2
-\sigma^{(\pm)}_{\mu\nu}\sigma_{(\pm)}^{\mu\nu}
+\omega^{(\pm)}_{\mu\nu}\omega_{(\pm)}^{\mu\nu}
-R_{\mu\nu}k^\mu_{(\pm)}k^\nu_{(\pm)}
+\nabla_\mu a^\mu_{(\pm)} .
\end{equation}
Substituting the covariant acceleration gives the universal form
\begin{equation}
\frac{d\theta_{(\pm)}}{d\lambda}
=
-\frac12\theta_{(\pm)}^2
-\sigma^{(\pm)}_{\mu\nu}\sigma_{(\pm)}^{\mu\nu}
+\omega^{(\pm)}_{\mu\nu}\omega_{(\pm)}^{\mu\nu}
-R_{\mu\nu}k^\mu_{(\pm)}k^\nu_{(\pm)}
-q\,\nabla_\mu\!\left(
M^{\mu\nu}_{(\pm)}F_{\nu\rho}k^\rho_{(\pm)}
\right).
\end{equation}
In Minkowski spacetime, this reduces to
\begin{equation}
\frac{d\theta_{(\pm)}}{d\lambda}
=
-\frac12\theta_{(\pm)}^2
-\sigma^{(\pm)}_{\mu\nu}\sigma_{(\pm)}^{\mu\nu}
+\omega^{(\pm)}_{\mu\nu}\omega_{(\pm)}^{\mu\nu}
-q\,\partial_\mu\!\left(
M^{\mu\nu}_{(\pm)}F_{\nu\rho}k^\rho_{(\pm)}
\right).
\end{equation}

These equations summarize the common result for the $b_\mu$, $H_{\mu\nu}$,
and $d_{\mu\nu}$ sectors. Since the Lorentz--violating backgrounds are
constant, they do not generate extra force terms through their derivatives.
Their role is to change the branch tangent and the Hessian that converts the
Lorentz force into acceleration. Consequently, the same electromagnetic field
can affect the expansion differently in the two branches, because
$M^{\mu\nu}_{(+)}$ and $M^{\mu\nu}_{(-)}$ are not the same tensor.

The effect becomes more sensitive near the degeneracy regions of the branch
dispersion relation. In those regions, the square-root factors that appear in
the explicit expressions for $M^{\mu\nu}_{(\pm)}$ can make the branch response
large. Thus the electromagnetic term in the Raychaudhuri equation may become
the dominant contribution to the local focusing behavior, even when the
external field itself is fixed. When $q=0$, this contribution disappears. When
the Lorentz--violating branch splitting is removed, the usual charged
congruence is recovered.


{\subsection{Novel physical content and observable signatures}}
\label{subsec:novel_observables}

{
The result obtained above should be distinguished from a direct application of the standard Raychaudhuri equation to an arbitrary non--geodesic congruence. In the usual non--geodesic case, the acceleration $a^{\mu}$ is supplied as an external vector field, and the only new source in the focusing equation is $\nabla_{\mu}a^{\mu}$. Here, by contrast, $a^{\mu}_{(\pm)}$ is not an independent input. It is determined by the gauge--covariant Hamiltonian flow of each Lorentz--violating branch and by the corresponding momentum Hessian,
\begin{equation}
a^{\mu}_{(\pm)} = -qM^{\mu\nu}_{(\pm)}F_{\nu\rho}k^{\rho}_{(\pm)}.
\label{novel_acceleration_observable}
\end{equation}
In this way, the electromagnetic field is common to both branches, but the tensor that converts this field into real--space acceleration is not. The new physical information is contained in the branch map
\begin{equation}
F_{\mu\nu} \quad \longrightarrow \quad a^{\mu}_{(\pm)} \quad \longrightarrow \quad \frac{\mathrm{d}\theta_{(\pm)}}{\mathrm{d}\lambda}.
\label{branch_map}
\end{equation}
It is worth mentioning that this map is absent in the ordinary charged--congruence analysis, where there is no spin--nondegenerate Hessian separating the response into two branch sectors.

A useful observable measure of the effect is the branch focusing asymmetry
\begin{equation}
\Delta_{\theta} := \frac{\mathrm{d}\theta_{(+)}}{\mathrm{d}\lambda} - \frac{\mathrm{d}\theta_{(-)}}{\mathrm{d}\lambda}.
\label{theta_asymmetry_definition}
\end{equation}
Using the universal Raychaudhuri equation, the electromagnetic part of this quantity is
\begin{equation}
\left(\Delta_{\theta}\right)_{\rm em}
=
-q\nabla_{\mu}
\left[
M^{\mu\nu}_{(+)}F_{\nu\rho}k^{\rho}_{(+)}
-
M^{\mu\nu}_{(-)}F_{\nu\rho}k^{\rho}_{(-)}
\right].
\label{theta_asymmetry_em}
\end{equation}
This quantity vanishes in the Lorentz--symmetric single-branch limit, but it is generally nonzero when the spin--nondegenerate dispersion relation gives different Hessians and tangents for the two branches. Therefore, $\Delta_{\theta}$ isolates the genuinely branch--dependent part of the focusing dynamics.

The same information can be expressed in terms of beam observables. If $A_{(\pm)}$ denotes the infinitesimal transverse area of a narrow branch congruence, then
\begin{equation}
\theta_{(\pm)}
=
\frac{1}{A_{(\pm)}}\frac{\mathrm{d}A_{(\pm)}}{\mathrm{d}\lambda}.
\label{area_expansion_observable}
\end{equation}
Subsequently, the equations derived here predict branch--dependent changes in the beam cross section, the respective branch--dependent caustic positions, and different focal lengths for the two spin--nondegenerate components. In a localized electromagnetic profile, one may characterize the effect through the caustic parameter $\lambda_{c}^{(\pm)}$, defined by $A_{(\pm)}(\lambda_{c}^{(\pm)}) \rightarrow 0$, or, more weakly, through the relative beam-width asymmetry
\begin{equation}
\mathcal{A}_{W}
:=
\frac{W_{(+)}-W_{(-)}}{W_{(+)}+W_{(-)}},
\label{beam_width_asymmetry}
\end{equation}
where $W_{(\pm)}$ is the transverse width of each branch after crossing the electromagnetic region. These quantities are directly tied to the expansion scalar and therefore provide operational signatures of the branch--resolved Raychaudhuri dynamics.

The distinction between trajectory bending and congruence focusing is also physically important. A uniform electromagnetic field can deflect the central trajectory of a charged branch, but it does not necessarily change the local expansion of an initially uniform beam. Local focusing requires a nonzero divergence of the effective branch force,
\begin{equation}
\nabla_{\mu}
\left(
M^{\mu\nu}_{(\pm)}F_{\nu\rho}k^{\rho}_{(\pm)}
\right)
\neq 0,
\label{effective_force_divergence_condition}
\end{equation}
which may arise from electric or magnetic field gradients, from spatial variation of the branch data across the beam, or from the coupling of a magnetic field to an already deformed congruence. 

In addition, these observables make clear what is new compared with existing non--geodesic congruence analyses. The standard equation tells us how a prescribed acceleration field changes $\theta$. The present construction derives, from the Lorentz--violating dispersion relation itself, the branch--dependent response tensor that produces the acceleration. It therefore predicts branch--selective focusing, branch-dependent caustic formation, and birefringent beam deformation under the same external electromagnetic background. These effects survive even in flat spacetime, where the curvature term is absent, and are controlled by the Lorentz--violating Hessians $M^{\mu\nu}_{(\pm)}$ (instead of the spacetime curvature itself).
}

{
\subsection{Domain of validity near branch-degeneracy surfaces}}
\label{subsec:branch_degeneracy_validity}

{
The branch-resolved formulation developed above is valid only in regions of phase space where the two spin--nondegenerate branches remain separated. In the explicit sectors considered in this work, the square-root structures entering the branch dispersion relations define invariants that control the separation between the two branches. We denote any of these invariants collectively by $\mathcal{I}(P)$, so that the branch-degeneracy surface is characterized by
\begin{equation}
\mathcal{I}(P)=0 .
\label{degeneracy_invariant_general}
\end{equation}
The response tensors $M^{\mu\nu}_{(\pm)}$ contain inverse powers of this invariant because they are obtained by differentiating the branch tangent with respect to the gauge--covariant momentum. Therefore, the apparent enhancement of $M^{\mu\nu}_{(\pm)}$ near $\mathcal{I}(P)=0$ reflects the rapid variation of the branch eigenstructure in momentum space.

This behavior should not be interpreted as a prediction of an arbitrarily large physical focusing effect. Exactly at a branch-degeneracy surface, the labels $(+)$ and $(-)$ cease to define two independent smooth congruences. The branch Hessian then becomes singular because the single-branch parametrization is no longer a regular set of variables. In this sense, the divergence is partly a coordinate or parametrization singularity of the branch-resolved effective description. A physical treatment at or extremely close to degeneracy would require returning to the underlying multicomponent system, where the two branches can mix and must be evolved together.

The effective branch description is restricted to an adiabatic regime in which the branch separation remains large compared with the momentum change induced across the relevant region of the beam. A convenient way to state this condition is
\begin{equation}
\sqrt{\mathcal{I}(P)}
\gg
\delta P_{\rm em},
\label{adiabatic_branch_condition}
\end{equation}
where $\delta P_{\rm em}$ denotes the characteristic electromagnetic change of the gauge--covariant momentum over the length scale on which the congruence is analyzed. Equivalently, for a beam crossing a region of size $\ell$,
\begin{equation}
\delta P_{\rm em}
\sim
|qF|\,\ell ,
\label{momentum_variation_scale}
\end{equation}
so that the branch-resolved Raychaudhuri equation should be used only when the electromagnetic perturbation does not drive the system through the branch-degeneracy surface.

Within this domain, the large values of $M^{\mu\nu}_{(\pm)}$ near degeneracy have a controlled interpretation: they indicate enhanced sensitivity of the branch tangent to small changes in momentum. This enhancement can strengthen the branch-dependent focusing source, but only while the adiabatic separation of the branches is maintained. Once the condition \eqref{adiabatic_branch_condition} fails, the quantities $\theta_{(\pm)}$, $\sigma^{(\pm)}_{\mu\nu}$, $\omega^{(\pm)}_{\mu\nu}$, and $M^{\mu\nu}_{(\pm)}$ no longer describe two autonomous branch congruences.

A regularized description would require additional physical input beyond the single-branch effective equations used here. Possible regularizations include keeping the full spinor or multiband structure before diagonalizing into branches, introducing finite wave-packet widths in momentum space, or including branch-mixing effects near avoided crossings. Such procedures can smooth the apparent singularity, but they are model dependent and lie outside the branch-resolved approximation adopted in this work.

Accordingly, all focusing and defocusing conclusions derived in the following sections should be understood as statements within the nondegenerate region
\begin{equation}
\mathcal{I}(P)>0,
\qquad
\sqrt{\mathcal{I}(P)}
\gg
\delta P_{\rm em}.
\label{validity_region_final}
\end{equation}
In this regime, the Hessian is finite, the two branches define independent congruence flows, and the Raychaudhuri equation gives a meaningful branch-resolved evolution of the expansion. The formal divergence of $M^{\mu\nu}_{(\pm)}$ at $\mathcal{I}(P)=0$ is therefore best viewed as a boundary of validity of the effective branch description, not as a physical claim of infinite acceleration or infinite focusing.
}


\section{Electromagnetic focusing in flat spacetime}
\label{sec:EM_focusing_flat}

A particularly useful regime is flat spacetime,
\begin{equation}
g_{\mu\nu}=\eta_{\mu\nu},
\qquad
R_{\mu\nu}=0,
\qquad
\Gamma^\mu{}_{\alpha\beta}=0,
\end{equation}
with constant Lorentz--violating backgrounds and an external electromagnetic
field $F_{\mu\nu}(x)$. In this limit, curvature does not contribute to the
Raychaudhuri equation, and the non--geodesic part of the congruence is governed
by the electromagnetic force.

The branch-resolved Raychaudhuri equation becomes
\begin{equation}
\frac{d\theta_{(\pm)}}{d\lambda}
=
-\frac12\theta_{(\pm)}^2
-\sigma^{(\pm)}_{\mu\nu}\sigma_{(\pm)}^{\mu\nu}
+\omega^{(\pm)}_{\mu\nu}\omega_{(\pm)}^{\mu\nu}
-q\,\partial_\mu\!\left(
M^{\mu\nu}_{(\pm)}F_{\nu\rho}k^\rho_{(\pm)}
\right).
\label{eq:Ray_flat_EM}
\end{equation}
The electromagnetic contribution can be decomposed as
\begin{align}
\partial_\mu\!\left(
M^{\mu\nu}_{(\pm)}F_{\nu\rho}k^\rho_{(\pm)}
\right)
&=
(\partial_\mu M^{\mu\nu}_{(\pm)})
F_{\nu\rho}k^\rho_{(\pm)}
\nonumber\\
&\quad
+
M^{\mu\nu}_{(\pm)}
(\partial_\mu F_{\nu\rho})
k^\rho_{(\pm)}
+
M^{\mu\nu}_{(\pm)}
F_{\nu\rho}
\partial_\mu k^\rho_{(\pm)} .
\label{eq:split_force_term}
\end{align}
Since the Lorentz--violating backgrounds are constant, the spacetime
dependence of $M^{\mu\nu}_{(\pm)}$ and $k^\mu_{(\pm)}$ is induced through
the gauge--covariant momentum $P_\mu(x)$. Therefore, for a narrow beam whose
branch data vary slowly across the transverse section, the term containing the
explicit field gradient gives the leading local contribution,
\begin{equation}
\frac{d\theta_{(\pm)}}{d\lambda}
\approx
-\frac12\theta_{(\pm)}^2
-\sigma^{(\pm)}_{\mu\nu}\sigma_{(\pm)}^{\mu\nu}
+\omega^{(\pm)}_{\mu\nu}\omega_{(\pm)}^{\mu\nu}
-q\,M^{\mu\nu}_{(\pm)}
(\partial_\mu F_{\nu\rho})
k^\rho_{(\pm)} .
\label{eq:Ray_flat_EM_gradient}
\end{equation}
Thus, in this approximation, electromagnetic focusing is associated with
inhomogeneities of the external field. A uniform field can bend the individual
trajectories, but it does not produce a leading local source for the expansion
of an initially uniform branch congruence.

\subsection{Pure electric field}

Consider a purely electric configuration,
\begin{equation}
F_{0i}=E_i(\vec x),
\qquad
F_{ij}=0.
\end{equation}
Then
\begin{equation}
F_{\nu\rho}k^\rho
=
\big(E_i k^i,\,-E_i k^0\big),
\end{equation}
and the electromagnetic term in Eq.~\eqref{eq:Ray_flat_EM} becomes
\begin{align}
-q\,\partial_\mu\!\left(
M^{\mu\nu}_{(\pm)}F_{\nu\rho}k^\rho_{(\pm)}
\right)
&=
-q\,\partial_\mu
\left[
M^{\mu 0}_{(\pm)}E_i k^i_{(\pm)}
-
M^{\mu i}_{(\pm)}E_i k^0_{(\pm)}
\right].
\label{eq:EM_pureE_full}
\end{align}
If the variation of $M^{\mu\nu}_{(\pm)}$ and $k^\mu_{(\pm)}$ across the beam is
subleading, the Raychaudhuri equation reduces to
\begin{equation}
\frac{d\theta_{(\pm)}}{d\lambda}
\approx
-\frac12\theta_{(\pm)}^2
-\sigma^{(\pm)}_{\mu\nu}\sigma_{(\pm)}^{\mu\nu}
+\omega^{(\pm)}_{\mu\nu}\omega_{(\pm)}^{\mu\nu}
-q
\left[
M^{\mu 0}_{(\pm)}k^i_{(\pm)}
-
M^{\mu i}_{(\pm)}k^0_{(\pm)}
\right]
\partial_\mu E_i .
\label{eq:Ray_pureE}
\end{equation}
This form shows that the electric contribution is controlled by the gradient
of the electric field projected along the branch-dependent tensor
$M^{\mu\nu}_{(\pm)}$ and tangent $k^\mu_{(\pm)}$. Hence two branches crossing
the same electric profile need not develop the same expansion.

For a purely electrostatic field directed along the $x$-axis,
\begin{equation}
\vec E=(E_x(x),0,0),
\end{equation}
Eq.~\eqref{eq:Ray_pureE} gives
\begin{equation}
\frac{d\theta_{(\pm)}}{d\lambda}
\approx
-\frac12\theta_{(\pm)}^2
-\sigma^{(\pm)}_{\mu\nu}\sigma_{(\pm)}^{\mu\nu}
+\omega^{(\pm)}_{\mu\nu}\omega_{(\pm)}^{\mu\nu}
-q
\left[
M^{x0}_{(\pm)}k^x_{(\pm)}
-
M^{xx}_{(\pm)}k^0_{(\pm)}
\right]
\frac{dE_x}{dx}.
\label{eq:Ray_pureE_1D}
\end{equation}
The sign of
\begin{equation}
q
\left[
M^{x0}_{(\pm)}k^x_{(\pm)}
-
M^{xx}_{(\pm)}k^0_{(\pm)}
\right]
\frac{dE_x}{dx}
\end{equation}
then decides whether the electric profile contributes to focusing or
defocusing. The factor multiplying $dE_x/dx$ is branch dependent, so the two
spin--nondegenerate modes may react oppositely to the same field gradient.

\subsection{Pure magnetic field}

Now consider a purely magnetic configuration,
\begin{equation}
F_{0i}=0,
\qquad
F_{ij}=-\epsilon_{ijk}B^k(\vec x).
\end{equation}
Then
\begin{equation}
F_{i\rho}k^\rho
=
-\epsilon_{ijk}B^k k^j
=
-(\vec k\times\vec B)_i,
\qquad
F_{0\rho}k^\rho=0.
\end{equation}
Therefore,
\begin{equation}
-q\,\partial_\mu\!\left(
M^{\mu i}_{(\pm)}F_{i\rho}k^\rho_{(\pm)}
\right)
=
q\,\partial_\mu
\left[
M^{\mu i}_{(\pm)}
(\vec k_{(\pm)}\times\vec B)_i
\right].
\label{eq:EM_pureB_full}
\end{equation}
Under the same slowly varying beam approximation, this becomes
\begin{equation}
\frac{d\theta_{(\pm)}}{d\lambda}
\approx
-\frac12\theta_{(\pm)}^2
-\sigma^{(\pm)}_{\mu\nu}\sigma_{(\pm)}^{\mu\nu}
+\omega^{(\pm)}_{\mu\nu}\omega_{(\pm)}^{\mu\nu}
+
q\,M^{\mu i}_{(\pm)}
\epsilon_{ijk}k^j_{(\pm)}
\partial_\mu B^k .
\label{eq:Ray_pureB}
\end{equation}
Thus the magnetic contribution is controlled by gradients of the magnetic
field transverse to the branch motion. A homogeneous magnetic field changes
the direction of the trajectory, whereas a nonuniform magnetic field can change
the local expansion of the congruence.

For a magnetic field directed along the $z$-axis,
\begin{equation}
\vec B=(0,0,B_z(x,y)),
\end{equation}
we have
\begin{equation}
\vec k\times\vec B
=
\big(k^yB_z,\,-k^xB_z,\,0\big).
\end{equation}
The full magnetic contribution can then be written as
\begin{equation}
\frac{d\theta_{(\pm)}}{d\lambda}
=
-\frac12\theta_{(\pm)}^2
-\sigma^{(\pm)}_{\mu\nu}\sigma_{(\pm)}^{\mu\nu}
+\omega^{(\pm)}_{\mu\nu}\omega_{(\pm)}^{\mu\nu}
+
q\,\partial_\mu
\left[
M^{\mu x}_{(\pm)}k^y_{(\pm)}B_z
-
M^{\mu y}_{(\pm)}k^x_{(\pm)}B_z
\right].
\label{eq:Ray_pureB_z}
\end{equation}
In the leading gradient approximation, this reduces to
\begin{equation}
\frac{d\theta_{(\pm)}}{d\lambda}
\approx
-\frac12\theta_{(\pm)}^2
-\sigma^{(\pm)}_{\mu\nu}\sigma_{(\pm)}^{\mu\nu}
+\omega^{(\pm)}_{\mu\nu}\omega_{(\pm)}^{\mu\nu}
+
q
\left[
M^{\mu x}_{(\pm)}k^y_{(\pm)}
-
M^{\mu y}_{(\pm)}k^x_{(\pm)}
\right]
\partial_\mu B_z .
\label{eq:Ray_pureB_z_gradient}
\end{equation}
This expression makes clear that a magnetic profile can act as a lens for the
charged congruence. The effective lensing strength depends on the magnetic
gradient and on the branch tensor multiplying it.

\subsection{Gaussian electromagnetic lens}

A simple realization of branch-dependent electromagnetic focusing
is obtained from a localized electrostatic profile acting as an
effective electromagnetic lens. Consider a narrow congruence
propagating predominantly along the $z$-direction in flat spacetime,
with an external electric field of Gaussian form,
\begin{equation}
\vec E(x)
=
E_0\,e^{-x^2/\sigma^2}\,\hat x .
\label{eq:Gaussian_field}
\end{equation}
The field is localized around $x=0$ and produces a transverse
electric gradient,
\begin{equation}
\frac{dE_x}{dx}
=
-\frac{2x}{\sigma^2}
E_0 e^{-x^2/\sigma^2}.
\label{eq:Gaussian_gradient}
\end{equation}

In the slowly varying beam approximation, the branch-resolved
Raychaudhuri equation reduces to
\begin{equation}
\frac{d\theta_{(\pm)}}{d\lambda}
\approx
-\frac12\theta_{(\pm)}^2
-\sigma^{(\pm)}_{\mu\nu}\sigma_{(\pm)}^{\mu\nu}
+\omega^{(\pm)}_{\mu\nu}\omega_{(\pm)}^{\mu\nu}
-
\mathcal F_{(\pm)}(x),
\label{eq:Gaussian_Ray}
\end{equation}
where the effective focusing profile is
\begin{equation}
\mathcal F_{(\pm)}(x)
:=
q
\left[
M^{x0}_{(\pm)}k^x_{(\pm)}
-
M^{xx}_{(\pm)}k^0_{(\pm)}
\right]
\frac{dE_x}{dx}.
\label{eq:Gaussian_focus}
\end{equation}

The Gaussian profile therefore acts as an effective electromagnetic
lens whose focusing properties depend explicitly on the branch
response tensor $M^{\mu\nu}_{(\pm)}$. Since the sign of
$\mathcal F_{(\pm)}$
depends on the branch structure, the same electromagnetic profile
may produce focusing in one branch and defocusing in the other.

The strongest effect occurs near the region
\begin{equation}
|x|
\sim
\sigma ,
\end{equation}
where the electric gradient is maximal. Near the beam center,
\begin{equation}
x\rightarrow0,
\end{equation}
the field becomes approximately uniform and the local focusing
source tends to vanish,
\begin{equation}
\frac{dE_x}{dx}\rightarrow0.
\end{equation}
This behavior reflects the general result that electromagnetic
focusing is controlled by field inhomogeneities rather than by
the field amplitude itself.

The resulting dynamics is analogous to the action of an optical
lens on neighboring light rays. However, in the present case,
the effective optical response is branch dependent. Consequently,
the electromagnetic profile acts as a birefringent congruence lens,
splitting the evolution of the two spin-nondegenerate branches. This mechanism provides a geometric realization of
branch-dependent electromagnetic lensing in Lorentz-violating
congruence dynamics.

\subsection{General criterion for electromagnetic focusing}

For hypersurface--orthogonal congruences,
$\omega^{(\pm)}_{\mu\nu}=0$, the electromagnetic term favors focusing whenever
\begin{equation}
q\,\partial_\mu\!\left(
M^{\mu\nu}_{(\pm)}F_{\nu\rho}k^\rho_{(\pm)}
\right)\ge 0.
\label{eq:EM_focus_condition}
\end{equation}
Under this condition, the electromagnetic contribution enters the
Raychaudhuri equation with the same sign as the shear term and decreases
$\theta_{(\pm)}$ along the flow. If the inequality is reversed, the external
field contributes to defocusing.

This criterion shows that the relevant quantity is not only the magnitude of
$F_{\mu\nu}$, but the divergence of the effective branch force. The Hessian
$M^{\mu\nu}_{(\pm)}$ fixes how each spin--nondegenerate branch converts a given
electric or magnetic inhomogeneity into a change of expansion. Near degeneracy
regions of the dispersion relation, the explicit square-root factors entering
$M^{\mu\nu}_{(\pm)}$ can amplify this response. For this reason, nonuniform
electromagnetic backgrounds provide the cleanest setting in which the
branch-dependent focusing effects can appear in flat spacetime.

{
The flat spacetime examples above show how the formal branch--resolved Raychaudhuri equation can be translated into observable quantities. A localized electromagnetic profile changes the transverse area and width of a charged beam through the branch--dependent source $-q\partial_{\mu}(M^{\mu\nu}_{(\pm)}F_{\nu\rho}k^{\rho}_{(\pm)})$. Therefore, the relevant signatures are the relative beam narrowing or broadening of the two branches, the displacement of their caustic positions, and the sign of the focusing asymmetry $\Delta_{\theta}$. These effects are not equivalent to ordinary trajectory bending, because they concern the deformation of a family of neighboring trajectories rather than the motion of a single representative particle.
}

{\subsection{Representative focusing evolution and scale estimates}
\label{subsec:representative_focusing_estimates}}

{
The previous discussion identified the sign of the electromagnetic source that controls focusing or defocusing. We now give a simple quantitative estimate of the corresponding expansion evolution. To avoid introducing model-dependent field configurations beyond the scope of the present work, we use the Gaussian electric lens discussed above and evaluate the source at a representative transverse position $x=x_{\ast}$ inside the lens. For the Gaussian profile, the magnitude of the electric gradient is maximal at
\begin{equation}
|x_{\ast}|=\frac{\sigma}{\sqrt{2}},
\label{xstar_gradient_maximum}
\end{equation}
so this point provides a useful estimate of the strongest local focusing effect.

For each Lorentz--violating sector $s=b,H,d$, define the local branch source
\begin{equation}
\mathcal{F}^{(s)}_{(\pm)} := q\,\mathcal{C}^{(s)}_{(\pm)} \left. \frac{\mathrm{d}E_x}{\mathrm{d}x} \right|_{x=x_{\ast}}, \qquad \mathcal{C}^{(s)}_{(\pm)} := \left[ M^{x0}_{(s,\pm)}k^{x}_{(s,\pm)} - M^{xx}_{(s,\pm)}k^{0}_{(s,\pm)} \right]_{P=P_0}.
\label{sector_source_definition}
\end{equation}
Here $P_0$ denotes the central gauge--covariant momentum of the beam. The coefficient
$\mathcal{C}^{(s)}_{(\pm)}$ contains the sector dependence through the corresponding Hessian and tangent. The same electromagnetic lens is compared across the $b_\mu$, $H_{\mu\nu}$, and $d_{\mu\nu}$ sectors only by changing $\mathcal{C}^{(s)}_{(\pm)}$.

Neglecting shear and vorticity for an initially collimated beam, and treating the source as locally constant across the thin lens, the expansion equation becomes
\begin{equation}
\frac{\mathrm{d}\theta^{(s)}_{(\pm)}}{\mathrm{d}\lambda} = -\frac{1}{2} \left(\theta^{(s)}_{(\pm)}\right)^2 - \mathcal{F}^{(s)}_{(\pm)}.
\label{reduced_quantitative_raychaudhuri}
\end{equation}
For an initially parallel beam, $\theta^{(s)}_{(\pm)}(0)=0$, the solution is
\begin{equation}
\theta^{(s)}_{(\pm)}(\lambda) = -\sqrt{2\mathcal{F}^{(s)}_{(\pm)}} \tan \left[ \sqrt{\frac{\mathcal{F}^{(s)}_{(\pm)}}{2}}\, \lambda \right], \qquad \mathcal{F}^{(s)}_{(\pm)}>0.
\label{theta_solution_positive_source}
\end{equation}
The corresponding focusing parameter, defined by the divergence of $\theta^{(s)}_{(\pm)}$, is
\begin{equation}
\lambda^{(s)}_{c,(\pm)} = \frac{\pi}{\sqrt{2\mathcal{F}^{(s)}_{(\pm)}}}.
\label{focusing_time_positive_source}
\end{equation}
When $\mathcal{F}^{(s)}_{(\pm)}<0$, the electromagnetic source acts instead as a defocusing term. In that case,
\begin{equation}
\theta^{(s)}_{(\pm)}(\lambda) = \sqrt{2|\mathcal{F}^{(s)}_{(\pm)}|} \tanh \left[ \sqrt{\frac{|\mathcal{F}^{(s)}_{(\pm)}|}{2}}\, \lambda \right], \qquad \mathcal{F}^{(s)}_{(\pm)}<0,
\label{theta_solution_negative_source}
\end{equation}
and no caustic is produced within this reduced approximation.

It is convenient to introduce dimensionless variables
\begin{equation}
\tau:=\frac{\lambda}{L}, \qquad \Theta^{(s)}_{(\pm)}:=L\,\theta^{(s)}_{(\pm)}, \qquad \Phi^{(s)}_{(\pm)}:=L^2\mathcal{F}^{(s)}_{(\pm)},
\label{dimensionless_focusing_variables}
\end{equation}
where $L$ is the characteristic longitudinal size of the electromagnetic lens. Equation
\eqref{reduced_quantitative_raychaudhuri} then becomes
\begin{equation}
\frac{\mathrm{d}\Theta^{(s)}_{(\pm)}}{\mathrm{d}\tau} = -\frac{1}{2} \left(\Theta^{(s)}_{(\pm)}\right)^2 - \Phi^{(s)}_{(\pm)}.
\label{dimensionless_quantitative_raychaudhuri}
\end{equation}
For $\Phi^{(s)}_{(\pm)}>0$, the dimensionless focusing time is
\begin{equation}
\tau^{(s)}_{c,(\pm)} = \frac{\pi}{\sqrt{2\Phi^{(s)}_{(\pm)}}}.
\label{dimensionless_focusing_time}
\end{equation}

A representative comparison between the three Lorentz--violating sectors is shown in Table~\ref{tab:representative_focusing_times}. The entries should be understood as normalized source strengths, not as universal SME bounds. They are chosen only to display the different possible behaviors of the branch-resolved Raychaudhuri equation.
\begin{table}[h]
\centering
{
\begin{tabular}{c c c c}
\hline\hline
Sector/branch & $\Phi^{(s)}_{(\pm)}$ & Behavior & $\tau^{(s)}_{c,(\pm)}$ \\
\hline
$b_\mu,(+)$        & $0.05$  & focusing   & $9.93$ \\
$b_\mu,(-)$        & $-0.02$ & defocusing & -- \\
$H_{\mu\nu},(+)$   & $0.10$  & focusing   & $7.02$ \\
$H_{\mu\nu},(-)$   & $0.04$  & focusing   & $11.11$ \\
$d_{\mu\nu},(+)$   & $0.20$  & focusing   & $4.97$ \\
$d_{\mu\nu},(-)$   & $-0.20$ & defocusing & -- \\
\hline\hline
\end{tabular}
}
\caption{{
Representative dimensionless focusing sources and focusing times for branch-resolved electromagnetic focusing. Positive values of $\Phi^{(s)}_{(\pm)}$ produce focusing, while negative values produce defocusing in the reduced approximation. The values are normalized benchmarks used to illustrate the dependence of the expansion evolution on the sector-dependent Hessian response.
}}
\label{tab:representative_focusing_times}
\end{table}

The same data can be displayed by plotting $\Theta^{(s)}_{(\pm)}(\tau)$ from Eq.~\eqref{dimensionless_quantitative_raychaudhuri}. Positive sources drive $\Theta$ to $-\infty$ at finite $\tau_c$, while negative sources make $\Theta$ approach a positive defocusing branch. Such a plot gives a direct visualization of the focusing time and of the branch-dependent sign of the electromagnetic response.

To connect these estimates with experimental constraints, one may write the sector-dependent source as a Lorentz-symmetric contribution plus a small branch correction,
\begin{equation}
\Phi^{(s)}_{(\pm)}
=
\Phi_0
\left[
1+\chi^{(s)}_{(\pm)}
\right],
\qquad
|\chi^{(s)}_{(\pm)}|
\lesssim
\epsilon^{(s)}_{\rm SME}.
\label{sme_compatible_source_scaling}
\end{equation}
Here $\epsilon^{(s)}_{\rm SME}$ denotes the appropriate dimensionless combination of SME coefficients for the chosen particle species, reference frame, and sector. The corresponding values should be taken from the current SME data tables \cite{Kostelecky:2008ts}. For small $\chi^{(s)}_{(\pm)}$, the focusing time changes according to
\begin{equation}
\frac{\lambda^{(s)}_{c,(\pm)}-\lambda^{(0)}_c}{\lambda^{(0)}_c}
\simeq
-\frac{1}{2}\chi^{(s)}_{(\pm)}.
\label{focusing_time_sme_shift}
\end{equation}
Therefore, ordinary-particle realizations constrained by current SME bounds are expected to produce very small fractional shifts in the focusing time, unless the system is tuned close to a controlled nondegenerate enhancement region. By contrast, the condensed-matter analogues discussed below can realize much larger effective Hessian contrasts, because the branch response is governed by the band geometry rather than by fundamental SME coefficients.
}


\section{effective electromagnetic response tensor of Spacetime Congruences}
\label{sec:susceptibility}

The form of the Raychaudhuri equation obtained above allows a direct physical
interpretation of the Lorentz--violating dynamics. The electromagnetic
contribution is controlled by the momentum Hessian
\begin{equation}
M^{\mu\nu}_{(\pm)}
=
-\frac{\partial k^\mu_{(\pm)}}{\partial P_\nu},
\end{equation}
which measures how the branch tangent changes under variations of the
gauge--covariant momentum.

For constant Lorentz--violating backgrounds, the covariant acceleration is
\begin{equation}
a^\mu_{(\pm)}
:=
k^\nu_{(\pm)}\nabla_\nu k^\mu_{(\pm)}
=
-q\,M^{\mu\nu}_{(\pm)}
F_{\nu\rho}k^\rho_{(\pm)} .
\end{equation}
Equivalently, in coordinate form,
\begin{equation}
\frac{d k^\mu_{(\pm)}}{d\lambda}
+
\Gamma^\mu{}_{\alpha\beta}
k^\alpha_{(\pm)}k^\beta_{(\pm)}
=
-q\,M^{\mu\nu}_{(\pm)}
F_{\nu\rho}k^\rho_{(\pm)} .
\end{equation}
Thus the electromagnetic field does not couple to the congruence only through
the metric structure. Its effect is filtered by $M^{\mu\nu}_{(\pm)}$, which is
fixed by the branch dispersion relation. The Lorentz--violating coefficients
therefore determine how the same external field is converted into acceleration
for each branch.

This interpretation is especially clear in flat spacetime, where
\begin{equation}
\frac{d\theta_{(\pm)}}{d\lambda}
=
-\frac12\theta_{(\pm)}^2
-\sigma^{(\pm)}_{\mu\nu}\sigma_{(\pm)}^{\mu\nu}
+\omega^{(\pm)}_{\mu\nu}\omega_{(\pm)}^{\mu\nu}
-q\,\partial_\mu\!\left(
M^{\mu\nu}_{(\pm)}F_{\nu\rho}k^\rho_{(\pm)}
\right).
\end{equation}
Defining
\begin{equation}
\mathcal F^\mu_{(\pm)}
:=
M^{\mu\nu}_{(\pm)}F_{\nu\rho}k^\rho_{(\pm)},
\end{equation}
the electromagnetic term becomes $-q\,\partial_\mu\mathcal F^\mu_{(\pm)}$.
Hence the focusing behavior is governed by the divergence of the branch force.
For a hypersurface--orthogonal congruence, a positive value of
$q\,\partial_\mu\mathcal F^\mu_{(\pm)}$ contributes to the decrease of
$\theta_{(\pm)}$, while the opposite sign favors defocusing.

In this sense, constant Lorentz--violating backgrounds do not generate an
independent force. Their role is to reshape the electromagnetic response of
the congruence. Different sectors produce different Hessians, and the two
branches of the same sector correspond to different response channels. This is
why the same electromagnetic profile may focus one branch more strongly than
the other, or even produce opposite behavior depending on the orientation of
$P_\mu$ relative to the background tensor.

The result can be summarized by the chain
\begin{equation}
\mathcal D^{(\pm)}(P)
\;\longrightarrow\;
k^\mu_{(\pm)}
=
\frac{\mathfrak e}{2}
\frac{\partial\mathcal D^{(\pm)}}{\partial P_\mu}
\;\longrightarrow\;
M^{\mu\nu}_{(\pm)}
=
-\frac{\partial k^\mu_{(\pm)}}{\partial P_\nu}
\;\longrightarrow\;
\nabla_\mu\!\left(
M^{\mu\nu}_{(\pm)}F_{\nu\rho}k^\rho_{(\pm)}
\right).
\end{equation}
The first step fixes the branch tangent, the second fixes the electromagnetic
response tensor, and the last one gives the source term that enters the
Raychaudhuri equation. Near degeneracy regions of the dispersion relation, the
Hessian may become large, so even a moderate electromagnetic inhomogeneity can
produce a stronger branch-dependent change in the expansion.


{
\subsection{Relativistic-to-band correspondence and scope of the analogue}}
\label{subsec:relativistic_band_dictionary}

{
Before moving to semiclassical quasiparticle beams, it is useful to make explicit the precise correspondence between the relativistic spin--nondegenerate construction and the band-theory construction used below. The condensed-matter part of the paper should not be read as an independent extension unrelated to the previous sections. It is an analogue application of the same kinematical mechanism: a branch dispersion relation fixes a velocity field, the Hessian of this dispersion fixes the response to electromagnetic forcing, and the divergence of the resulting acceleration enters a Raychaudhuri-type equation for the expansion of a congruence.

The correspondence can be summarized as follows:
\begin{equation}
\begin{array}{ccl}
\text{Relativistic branch dispersion} 
&:&
D^{(\pm)}(P)=0,
\\[2mm]
\text{Band dispersion}
&:&
\varepsilon_n(\mathbf{k}),
\\[2mm]
\text{Gauge--covariant momentum}
&:&
P_\mu=\pi_\mu-qA_\mu,
\\[2mm]
\text{Crystal momentum}
&:&
k_i,
\\[2mm]
\text{Branch tangent}
&:&
k^\mu_{(\pm)}
=
\dfrac{e}{2}\,
\dfrac{\partial D^{(\pm)}}{\partial P_\mu},
\\[3mm]
\text{Group velocity}
&:&
u^i_n
=
\dfrac{\partial \varepsilon_n}{\partial k_i},
\\[3mm]
\text{SME response Hessian}
&:&
M^{\mu\nu}_{(\pm)}
=
-\dfrac{\partial k^\mu_{(\pm)}}{\partial P_\nu},
\\[3mm]
\text{Band Hessian}
&:&
M^{ij}_n
=
\dfrac{\partial u^i_n}{\partial k_j}
=
\dfrac{\partial^2\varepsilon_n}{\partial k_i\partial k_j},
\\[3mm]
\text{Relativistic electromagnetic source}
&:&
-q\nabla_\mu
\left(
M^{\mu\nu}_{(\pm)}F_{\nu\rho}k^\rho_{(\pm)}
\right),
\\[3mm]
\text{Band electromagnetic source}
&:&
\partial_i
\left[
qM^{ij}_nE_j
+
qM^{ij}_n\epsilon_{j\ell m}u^\ell_nB^m
\right].
\end{array}
\label{relativistic_band_dictionary}
\end{equation}

This dictionary shows which conclusions transfer directly. In both systems, the Hessian is the object that converts electromagnetic forcing in momentum space into acceleration in real space. The electric and magnetic gradients enter the expansion equation only after being filtered by the appropriate response tensor. Also, if the dispersion has two branches with different Hessians, the same electromagnetic profile can generate different focusing rates for the two congruences. This is the common geometric mechanism behind branch-dependent focusing in the SME sectors and quasiparticle birefringence in band systems.

The relativistic theory is formulated covariantly in spacetime and uses the gauge--covariant momentum $P_\mu$, whereas the quasiparticle theory is a semiclassical real-space projection of a phase-space flow in $(\mathbf{x},\mathbf{k})$. The band Hessian is not an SME coefficient, and the quasiparticle branches are not fundamental Lorentz-violating spin branches. What transfers is the Raychaudhuri mechanism:
\begin{equation}
\text{branch dispersion}
\;\longrightarrow\;
\text{branch velocity}
\;\longrightarrow\;
\text{Hessian response}
\;\longrightarrow\;
\text{expansion source}.
\label{common_mechanism_chain}
\end{equation}

}

\section{Semiclassical wave-packet dynamics and congruences}
\label{sec:semiclassical}

In this section we develop the kinematical framework used below. We start from
the semiclassical equations of motion for a wave packet in a single Bloch band
and then reinterpret the resulting real--space flow as a congruence of
quasiparticle trajectories.

\subsection{Semiclassical equations of motion}

Let $\varepsilon_n(\mathbf{k})$ denote the dispersion relation of the $n$-th
band. In the semiclassical approximation, and neglecting Berry-curvature
corrections at this stage, the center of the wave packet obeys
\begin{subequations}
\label{eq:semiclassical_eqs}
\begin{align}
\dot{x}^i
&=
\frac{\partial \varepsilon_n(\mathbf{k})}{\partial k_i},
\label{eq:xdot_band}
\\[1ex]
\dot{k}_i
&=
qE_i(\mathbf{x},t)
+
q\,\epsilon_{ijk}\dot{x}^jB^k(\mathbf{x},t).
\label{eq:kdot_band}
\end{align}
\end{subequations}
Here $q$ is the quasiparticle charge, while $E_i$ and $B^i$ are the electric
and magnetic fields. Repeated spatial indices are summed over.

Equation \eqref{eq:xdot_band} shows that the velocity field is fixed by the
band dispersion. For simplicity, we suppress the band label and define
\begin{equation}
v^i(\mathbf{k})
:=
\frac{\partial \varepsilon(\mathbf{k})}{\partial k_i}.
\label{eq:def_v}
\end{equation}
Thus
\begin{equation}
\dot{x}^i=v^i(\mathbf{k}).
\end{equation}
The second derivative of the dispersion defines the band Hessian
\begin{equation}
\mathcal M^{ij}(\mathbf{k})
:=
\frac{\partial v^i}{\partial k_j}
=
\frac{\partial^2\varepsilon(\mathbf{k})}
{\partial k_i\partial k_j}.
\label{eq:def_M}
\end{equation}
This tensor gives the local response of the group velocity to changes in the
crystal momentum.

\subsection{Acceleration in terms of the band Hessian}

The acceleration follows by differentiating \eqref{eq:def_v} along the
trajectory,
\begin{equation}
\ddot{x}^i
=
\frac{d v^i}{dt}
=
\frac{\partial v^i}{\partial k_j}\dot{k}_j .
\end{equation}
Using \eqref{eq:def_M}, this becomes
\begin{equation}
\ddot{x}^i
=
\mathcal M^{ij}(\mathbf{k})\dot{k}_j .
\label{eq:accel_M}
\end{equation}
Substitution of \eqref{eq:kdot_band} gives
\begin{equation}
\ddot{x}^i
=
q\,\mathcal M^{ij}E_j
+
q\,\mathcal M^{ij}\epsilon_{j\ell m}\dot{x}^{\ell}B^m .
\label{eq:accel_EM}
\end{equation}
Thus the electromagnetic field accelerates the wave packet through the
band-dependent tensor $\mathcal M^{ij}$. This is the condensed--matter
counterpart of the response tensors that appeared in the spin--nondegenerate
Lorentz--violating sectors.

For an isotropic parabolic band,
\begin{equation}
\varepsilon(\mathbf{k})
=
\frac{\mathbf{k}^2}{2m^\ast},
\end{equation}
one obtains
\begin{equation}
v^i=\frac{k_i}{m^\ast},
\qquad
\mathcal M^{ij}=\frac{\delta^{ij}}{m^\ast}.
\end{equation}
Then \eqref{eq:accel_EM} reduces to the usual effective--mass equation,
\begin{equation}
m^\ast \ddot{x}^i
=
qE^i
+
q\,\epsilon^{i}{}_{\ell m}\dot{x}^{\ell}B^m .
\end{equation}
For a generic anisotropic band, $\mathcal M^{ij}$ replaces $1/m^\ast$ and
determines the direction and magnitude of the electromagnetic response.

\subsection{Congruence interpretation}

We now regard the semiclassical flow as a congruence of quasiparticle
trajectories. Let the family of wave-packet paths be parametrized by
\begin{equation}
x^i=x^i(t,s^A),
\qquad
A=1,2,
\label{eq:congruence_family}
\end{equation}
where $t$ is the evolution parameter and $s^A$ label neighboring trajectories
inside the beam. The tangent field of the congruence is
\begin{equation}
u^i(t,s^A)
:=
\frac{\partial x^i}{\partial t}
=
v^i(\mathbf{k}(t,s^A)).
\label{eq:def_u}
\end{equation}
The local deformation of the congruence is described by
\begin{equation}
B_{ij}:=\partial_j u_i,
\label{eq:def_B}
\end{equation}
where the Euclidean metric $\delta_{ij}$ is used to lower indices.

The deformation tensor is decomposed as
\begin{equation}
B_{ij}
=
\frac{1}{d}\theta\,\delta_{ij}
+\sigma_{ij}
+\omega_{ij},
\label{eq:B_decomp}
\end{equation}
with
\begin{subequations}
\label{eq:optical_scalars}
\begin{align}
\theta
&:=
\partial_i u^i,
\label{eq:def_theta}
\\[1ex]
\sigma_{ij}
&:=
\frac12
\left(
\partial_j u_i+\partial_i u_j
\right)
-\frac{1}{d}\theta\,\delta_{ij},
\label{eq:def_sigma}
\\[1ex]
\omega_{ij}
&:=
\frac12
\left(
\partial_j u_i-\partial_i u_j
\right).
\label{eq:def_omega}
\end{align}
\end{subequations}
The scalar $\theta$ measures the local expansion or contraction of the beam,
$\sigma_{ij}$ measures shape distortion at fixed volume, and $\omega_{ij}$
measures local rotation.

The semiclassical equations define a flow in the reduced phase space
$(\mathbf{x},\mathbf{k})$. After projection onto real space, neighboring
wave-packet trajectories form a congruence whose deformation is governed by
the same kinematical quantities appearing in \eqref{eq:B_decomp}. This is the
basis for the Raychaudhuri-type description used below.

\begin{figure}[h]
\centering
\begin{tikzpicture}[scale=1.15]

\draw[->] (-3.5,0) -- (3.8,0) node[right] {\small $x$};
\draw[->] (0,-2.8) -- (0,3.2) node[above] {\small $k$};

\draw[->,gray] (-3,2.2) -- (-2.5,1.8);
\draw[->,gray] (-2,2.0) -- (-1.4,1.5);
\draw[->,gray] (-1,1.8) -- (-0.3,1.2);
\draw[->,gray] (0.5,1.4) -- (1.1,0.8);
\draw[->,gray] (1.5,1.0) -- (2.0,0.5);
\draw[->,gray] (2.4,0.7) -- (2.9,0.3);

\draw[->,gray] (-3,1.0) -- (-2.3,0.7);
\draw[->,gray] (-2,0.8) -- (-1.2,0.5);
\draw[->,gray] (-1,0.6) -- (-0.2,0.3);
\draw[->,gray] (0.6,0.3) -- (1.3,0.1);
\draw[->,gray] (1.7,0.2) -- (2.3,0.0);
\draw[->,gray] (2.6,0.1) -- (3.1,-0.1);

\draw[->,gray] (-3,-0.4) -- (-2.2,-0.3);
\draw[->,gray] (-2,-0.2) -- (-1.2,-0.1);
\draw[->,gray] (-1,0.0) -- (-0.2,0.0);
\draw[->,gray] (0.6,-0.1) -- (1.3,-0.2);
\draw[->,gray] (1.7,-0.3) -- (2.3,-0.5);
\draw[->,gray] (2.6,-0.6) -- (3.1,-0.9);

\draw[->,gray] (-3,-1.6) -- (-2.4,-1.2);
\draw[->,gray] (-2,-1.4) -- (-1.3,-0.9);
\draw[->,gray] (-1,-1.1) -- (-0.2,-0.5);
\draw[->,gray] (0.5,-0.7) -- (1.1,-0.2);
\draw[->,gray] (1.5,-0.3) -- (2.0,0.1);
\draw[->,gray] (2.4,0.0) -- (2.9,0.3);

\draw[very thick,blue]
(-3,1.7) .. controls (-2.2,1.3) and (-1.4,0.9) .. (-0.4,0.4)
            .. controls (0.6,0.0) and (1.6,-0.2) .. (3,-0.6);

\draw[very thick,blue]
(-3,1.2) .. controls (-2.2,0.9) and (-1.4,0.6) .. (-0.4,0.2)
            .. controls (0.6,-0.1) and (1.6,-0.4) .. (3,-1.0);

\draw[dashed] (-2.2,1.15) -- (-2.2,0.85);
\node at (-2.55,1.2) {\small $\delta \xi_{\rm in}$};

\draw[dashed] (2.2,-0.55) -- (2.2,-0.95);
\node at (2.65,-1.1) {\small $\delta \xi_{\rm out}$};

\node at (-1.8,2.6) {\small phase-space congruence};
\node at (1.4,1.9) {\small local flow};
\node at (2.2,-1.6) {\small focusing/defocusing};

\end{tikzpicture}

\caption{
Reduced phase-space flow of a quasiparticle congruence in the $(x,k)$ plane.
The arrows represent the local semiclassical flow generated by
$\dot x=\partial\varepsilon/\partial k$ and by the electromagnetic evolution of
$k$. Two nearby trajectories define a beam whose local separation changes along
the flow. The corresponding Raychaudhuri-type equation describes this
deformation through the expansion, shear, vorticity, and the forcing term
controlled by the band Hessian.
}
\label{fig:phase_space_flow}
\end{figure}
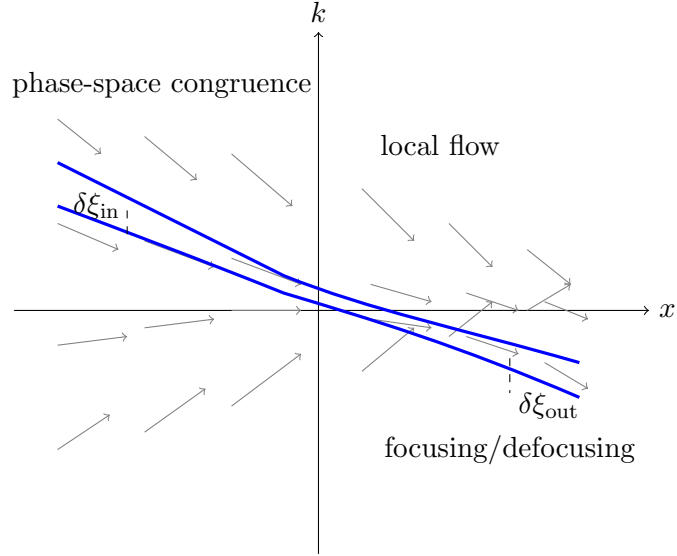

\subsection{Dependence of the deformation tensor on the band Hessian}

Since $u^i=v^i(\mathbf{k})$, the spatial gradient of the velocity is related
to gradients of the crystal momentum by
\begin{equation}
\partial_j u^i
=
\frac{\partial v^i}{\partial k_m}\partial_j k_m
=
\mathcal M^{im}\partial_j k_m .
\label{eq:B_M_gradk}
\end{equation}
Therefore,
\begin{equation}
B^i{}_{j}
=
\mathcal M^{im}(\mathbf{k})\,\partial_j k_m .
\label{eq:B_in_terms_M}
\end{equation}
This relation shows that real--space deformation is produced by the
combination of the band response tensor $\mathcal M^{ij}$ with the spatial
variation of the momentum field. Thus an initially narrow beam can expand,
contract, shear, or rotate depending both on the local curvature of the band
and on how $\mathbf{k}$ varies across the beam. For a parabolic isotropic band,
this dependence reduces to the usual effective--mass scaling. For a generic
band, however, the anisotropy of $\mathcal M^{ij}$ can make different
directions of the beam deform at different rates.

\subsection{Preview of the geometric dynamics}

The next step is to determine the evolution equation for the expansion
$\theta$. This requires differentiating $B_{ij}$ along the flow and using the
acceleration law \eqref{eq:accel_EM}. With the convective derivative
\begin{equation}
\frac{D}{Dt}
:=
\frac{\partial}{\partial t}
+
u^i\partial_i ,
\end{equation}
the expected Raychaudhuri-type equation is
\begin{equation}
\frac{D\theta}{Dt}
=
-\frac{1}{d}\theta^2
-\sigma_{ij}\sigma^{ij}
+\omega_{ij}\omega^{ij}
+
\partial_i a^i,
\label{eq:preview_ray}
\end{equation}
where
\begin{equation}
a^i:=\ddot{x}^i .
\end{equation}
Using \eqref{eq:accel_EM}, the forcing term $\partial_i a^i$ contains the
band Hessian and the gradients of the applied electromagnetic field. Hence the
same mechanism that determines the acceleration also controls the focusing
properties of the quasiparticle beam. This gives the condensed--matter
analogue of the branch-dependent electromagnetic forcing obtained in the
relativistic Lorentz--violating problem.


\section{Raychaudhuri-type equation for quasiparticle beams}
\label{sec:raychaudhuri}

In this section we derive the evolution equation for the expansion scalar of a
quasiparticle congruence. The derivation follows the same logic as the
Raychaudhuri equation, but it is adapted to the nonrelativistic semiclassical
flow introduced in Sec.~\ref{sec:semiclassical}.

\subsection{Convective derivative and kinematics}

Let $u^i(x,t)$ be the velocity field of the congruence. The derivative along
the flow is the convective derivative
\begin{equation}
\frac{D}{Dt}
:=
\partial_t+u^k\partial_k .
\label{eq:convective}
\end{equation}
The acceleration field is then
\begin{equation}
a^i
:=
\frac{D u^i}{Dt}.
\label{eq:def_accel}
\end{equation}

To avoid confusion with the magnetic field, we denote the deformation tensor by
$\mathcal B_{ij}$,
\begin{equation}
\mathcal B_{ij}:=\partial_j u_i .
\label{eq:def_deformation_tensor}
\end{equation}
We now compute its evolution along the flow,
\begin{equation}
\frac{D}{Dt}\mathcal B_{ij}
=
\frac{D}{Dt}\left(\partial_j u_i\right).
\label{eq:DB_start}
\end{equation}
Using the definition of the convective derivative, one has
\begin{align}
\frac{D}{Dt}\left(\partial_j u_i\right)
&=
\partial_t\left(\partial_j u_i\right)
+
u^k\partial_k\left(\partial_j u_i\right).
\end{align}
The first term is simply
\begin{equation}
\partial_t\left(\partial_j u_i\right)
=
\partial_j\left(\partial_t u_i\right).
\end{equation}
For the second one,
\begin{equation}
u^k\partial_k\left(\partial_j u_i\right)
=
\partial_j\left(u^k\partial_k u_i\right)
-
\left(\partial_j u^k\right)
\left(\partial_k u_i\right).
\end{equation}
Combining these identities, we find
\begin{align}
\frac{D}{Dt}\left(\partial_j u_i\right)
&=
\partial_j\left(
\partial_t u_i+u^k\partial_k u_i
\right)
-
\left(\partial_j u^k\right)
\left(\partial_k u_i\right).
\end{align}
Since
\begin{equation}
\partial_t u_i+u^k\partial_k u_i=a_i,
\end{equation}
the deformation tensor satisfies
\begin{equation}
\frac{D}{Dt}\mathcal B_{ij}
=
\partial_j a_i
-
\mathcal B_{ik}\mathcal B^k{}_{j}.
\label{eq:DB_final}
\end{equation}
This is the nonrelativistic deformation equation for the quasiparticle
congruence.

\subsection{Trace and evolution of the expansion}

The expansion scalar is defined by
\begin{equation}
\theta:=\partial_i u^i=\mathcal B^i{}_{i}.
\end{equation}
Taking the trace of Eq.~\eqref{eq:DB_final}, we obtain
\begin{equation}
\frac{D\theta}{Dt}
=
\partial_i a^i
-
\mathcal B_{ik}\mathcal B^{ki}.
\label{eq:theta_pre}
\end{equation}

We decompose $\mathcal B_{ij}$ into its trace, symmetric traceless, and
antisymmetric parts,
\begin{equation}
\mathcal B_{ij}
=
\frac{1}{d}\theta\,\delta_{ij}
+\sigma_{ij}
+\omega_{ij},
\label{eq:B_decomp_ray}
\end{equation}
where $d$ is the spatial dimension. Therefore,
\begin{align}
\mathcal B_{ik}\mathcal B^{ki}
&=
\left(
\frac{1}{d}\theta\,\delta_{ik}
+\sigma_{ik}
+\omega_{ik}
\right)
\left(
\frac{1}{d}\theta\,\delta^{ki}
+\sigma^{ki}
+\omega^{ki}
\right).
\end{align}
The trace part gives $\theta^2/d$, while the shear contribution is
$\sigma_{ij}\sigma^{ij}$. For the antisymmetric part,
\begin{equation}
\omega_{ik}\omega^{ki}
=
-\omega_{ij}\omega^{ij}.
\end{equation}
All mixed contractions vanish because $\sigma_{ij}$ is symmetric and
traceless, whereas $\omega_{ij}$ is antisymmetric. Hence
\begin{equation}
\mathcal B_{ik}\mathcal B^{ki}
=
\frac{1}{d}\theta^2
+
\sigma_{ij}\sigma^{ij}
-
\omega_{ij}\omega^{ij}.
\label{eq:B_contraction}
\end{equation}
Substituting this result into Eq.~\eqref{eq:theta_pre}, we obtain
\begin{equation}
\frac{D\theta}{Dt}
=
-\frac{1}{d}\theta^2
-\sigma_{ij}\sigma^{ij}
+\omega_{ij}\omega^{ij}
+
\partial_i a^i .
\label{eq:raychaudhuri_general}
\end{equation}
This equation describes the evolution of the local expansion of the
quasiparticle beam. The first two terms decrease $\theta$, the vorticity term
acts with the opposite sign, and the last term contains the external forcing
produced by the semiclassical acceleration.

\subsection{Electromagnetic forcing term}

The semiclassical acceleration is
\begin{equation}
a^i
=
q\,\mathcal M^{ij}E_j
+
q\,\mathcal M^{ij}\epsilon_{j\ell m}u^\ell B^m .
\label{eq:accel_repeat}
\end{equation}
The force contribution to Eq.~\eqref{eq:raychaudhuri_general} is therefore
$\partial_i a^i$.

For the electric part, one obtains
\begin{align}
\partial_i\left(q\,\mathcal M^{ij}E_j\right)
&=
q\,(\partial_i\mathcal M^{ij})E_j
+
q\,\mathcal M^{ij}\partial_i E_j .
\label{eq:div_E_term}
\end{align}
The first term appears when the band Hessian changes across the congruence,
whereas the second term is controlled by the electric-field gradient. Thus an
electric field with nonzero spatial variation can act as a local focusing or
defocusing source, with the strength weighted by $\mathcal M^{ij}$.

For the magnetic part, we find
\begin{align}
\partial_i
\left(
q\,\mathcal M^{ij}\epsilon_{j\ell m}u^\ell B^m
\right)
&=
q\,(\partial_i\mathcal M^{ij})
\epsilon_{j\ell m}u^\ell B^m
\nonumber\\
&\quad
+
q\,\mathcal M^{ij}\epsilon_{j\ell m}
(\partial_i u^\ell)B^m
\nonumber\\
&\quad
+
q\,\mathcal M^{ij}\epsilon_{j\ell m}
u^\ell \partial_i B^m .
\label{eq:div_B_term}
\end{align}
Using
\begin{equation}
\partial_i u^\ell=\mathcal B^\ell{}_{i},
\end{equation}
the middle contribution becomes
\begin{equation}
q\,\mathcal M^{ij}\epsilon_{j\ell m}
\mathcal B^\ell{}_{i}B^m .
\end{equation}
This term is worth keeping. Even if the magnetic field is uniform, it can
couple to an already deformed beam through $\mathcal B^\ell{}_{i}$. In
contrast, if the beam is initially uniform and the band Hessian is treated as
constant across it, then the leading magnetic focusing source comes from
gradients of $B^m$.

\subsection{Raychaudhuri equation with electromagnetic fields}

Combining the previous contributions, the expansion obeys
\begin{align}
\frac{D\theta}{Dt}
&=
-\frac{1}{d}\theta^2
-\sigma_{ij}\sigma^{ij}
+\omega_{ij}\omega^{ij}
\nonumber\\
&\quad
+
q\,(\partial_i\mathcal M^{ij})E_j
+
q\,\mathcal M^{ij}\partial_i E_j
\nonumber\\
&\quad
+
q\,(\partial_i\mathcal M^{ij})
\epsilon_{j\ell m}u^\ell B^m
\nonumber\\
&\quad
+
q\,\mathcal M^{ij}\epsilon_{j\ell m}
\mathcal B^\ell{}_{i}B^m
\nonumber\\
&\quad
+
q\,\mathcal M^{ij}\epsilon_{j\ell m}
u^\ell \partial_i B^m .
\label{eq:raychaudhuri_full}
\end{align}
This is the Raychaudhuri-type equation for a semiclassical quasiparticle
congruence in the absence of Berry-curvature corrections. It shows that the
band Hessian does not only determine the acceleration of a single wave packet;
it also controls how electromagnetic fields modify the expansion of a beam of
nearby trajectories.

The electric field contributes through $\partial_iE_j$ and through possible
spatial variation of $\mathcal M^{ij}$. The magnetic field contributes through
three channels: variation of the Hessian, coupling to the deformation tensor,
and the gradient $\partial_iB^m$. Therefore, a uniform electric field does not
produce a leading focusing source when $\mathcal M^{ij}$ is constant across the
beam. A uniform magnetic field is slightly different: it may still affect the
evolution of $\theta$ if the congruence already has shear, rotation, or
expansion through $\mathcal B^\ell{}_{i}$.

\subsection{Homogeneous Hessian approximation}

A useful limit is obtained when the Hessian is approximately constant across
the beam,
\begin{equation}
\partial_i\mathcal M^{ij}\simeq0.
\end{equation}
This may occur for an isotropic parabolic band or for a narrow wave packet
restricted to a small region of momentum space. Equation
\eqref{eq:raychaudhuri_full} then reduces to
\begin{align}
\frac{D\theta}{Dt}
&=
-\frac{1}{d}\theta^2
-\sigma_{ij}\sigma^{ij}
+\omega_{ij}\omega^{ij}
\nonumber\\
&\quad
+
q\,\mathcal M^{ij}\partial_i E_j
+
q\,\mathcal M^{ij}\epsilon_{j\ell m}
\mathcal B^\ell{}_{i}B^m
\nonumber\\
&\quad
+
q\,\mathcal M^{ij}\epsilon_{j\ell m}
u^\ell \partial_i B^m .
\label{eq:ray_simple}
\end{align}
In this form, the role of the band Hessian is explicit. It converts electric
and magnetic inhomogeneities into changes of the beam expansion, while the term
proportional to $\mathcal B^\ell{}_{i}B^m$ describes the action of a magnetic
field on an already deforming congruence. For a parabolic isotropic band,
$\mathcal M^{ij}=\delta^{ij}/m^\ast$, and the usual effective-mass response is
recovered. For anisotropic bands, different spatial directions are weighted
differently, so the same electromagnetic profile can focus the beam more
strongly along one direction than another.


\section{effective electromagnetic response tensor of Band Congruences}
\label{sec:susceptibility11}

In this section we discuss the physical meaning of the tensor
\begin{equation}
\mathcal M^{ij}
=
\frac{\partial^2\varepsilon}{\partial k_i\partial k_j},
\end{equation}
which appears in the Raychaudhuri-type equation
\eqref{eq:raychaudhuri_full}. This tensor determines how the band dispersion
converts electromagnetic forcing into acceleration and, consequently, into
focusing or defocusing of a quasiparticle congruence.

\subsection{Response structure of the acceleration}

From Eq.~\eqref{eq:accel_EM}, the acceleration of a quasiparticle is
\begin{equation}
a^i
=
q\,\mathcal M^{ij}E_j
+
q\,\mathcal M^{ij}\epsilon_{j\ell m}u^\ell B^m .
\label{eq:accel_sus}
\end{equation}
Therefore, the electromagnetic field does not determine the acceleration by
itself. Its effect is weighted by the band Hessian $\mathcal M^{ij}$, which
measures the variation of the group velocity with respect to the crystal
momentum. In this sense, the wave packet responds to the external field through
the local curvature of the band.

This point is important for anisotropic dispersions. Even if the external field
has a simple direction, the acceleration need not be parallel to it, because
$\mathcal M^{ij}$ mixes the field components according to the geometry of the
band. Thus the same electromagnetic configuration may produce different
accelerations in different regions of momentum space.

\subsection{Relation to the effective mass tensor}

In the conventions used here, the inverse effective mass tensor is
\begin{equation}
\left(m^{-1}\right)^{ij}
=
\frac{\partial^2\varepsilon}{\partial k_i\partial k_j}.
\label{eq:mass_tensor}
\end{equation}
Comparing this expression with Eq.~\eqref{eq:def_M}, we identify
\begin{equation}
\mathcal M^{ij}
=
\left(m^{-1}\right)^{ij}.
\end{equation}
Thus the tensor entering the Raychaudhuri equation is the inverse effective
mass tensor. The difference is in its role: in the present formulation,
$\mathcal M^{ij}$ does not only determine the acceleration of a single wave
packet, but also controls the collective deformation of a beam of nearby
trajectories.

For an isotropic parabolic band,
\begin{equation}
\mathcal M^{ij}
=
\frac{1}{m^\ast}\delta^{ij},
\end{equation}
all spatial directions respond in the same way. For a generic band, however,
the eigenvalues and eigenvectors of $\mathcal M^{ij}$ select preferred
directions for the response. This is the origin of anisotropic focusing in the
congruence description.

\subsection{Susceptibility interpretation}

The electric part of the acceleration can be written as
\begin{equation}
a^i_{\rm E}
=
q\,\mathcal M^{ij}E_j .
\end{equation}
This has the form of a linear response law, with $\mathcal M^{ij}$ acting as
the tensor that maps the applied field into the acceleration. For this reason,
$\mathcal M^{ij}$ may be viewed as an effective electromagnetic response tensor
of the quasiparticle motion.

The same tensor also appears in the forcing term of the Raychaudhuri equation.
In the homogeneous Hessian approximation, Eq.~\eqref{eq:ray_simple} contains
the contributions
\begin{equation}
\partial_i a^i
\supset
q\,\mathcal M^{ij}\partial_iE_j
+
q\,\mathcal M^{ij}\epsilon_{j\ell m}u^\ell\partial_iB^m
+
q\,\mathcal M^{ij}\epsilon_{j\ell m}
\mathcal B^\ell{}_{i}B^m .
\label{eq:focus_sus}
\end{equation}
The first two terms show that electric and magnetic field gradients can act as
sources for the expansion. The last term shows that a magnetic field can also
couple to an already deforming congruence through the deformation tensor
$\mathcal B^\ell{}_{i}$. Hence uniform fields are not all equivalent: a uniform
electric field does not give a leading focusing source when $\mathcal M^{ij}$
is constant, whereas a uniform magnetic field can still affect the expansion
if the beam already carries shear, rotation, or expansion.

This interpretation is close to the role of constitutive tensors in anisotropic
media. The comparison should be understood structurally: $\mathcal M^{ij}$ is
not a dielectric tensor, but it plays an analogous mathematical role by
selecting how the external field is converted into motion. In the present
case, the ``medium'' is the band structure itself.

For multiband systems, each band carries its own Hessian
$\mathcal M^{ij}_{(n)}$. Therefore, different branches of the dispersion can
respond differently to the same electromagnetic field. At the level of the
congruence, this produces branch-dependent focusing. This behavior is the
condensed-matter analogue of the branch-dependent response found in the
spin--nondegenerate Lorentz--violating sectors.

The tensor $\mathcal M^{ij}$ also has a geometric meaning. Since it is the
Hessian of the dispersion relation, it defines the local quadratic variation of
the energy in momentum space. Regions with large eigenvalues of
$\mathcal M^{ij}$ are more sensitive to electromagnetic gradients, while nearly
flat directions of the band respond more weakly. Therefore, the focusing
properties of the quasiparticle beam are directly tied to the local shape of
the dispersion surface.


\section{Anisotropic parabolic band}
\label{sec:parabolic}

We now apply the general formalism to the simplest nontrivial band structure:
an anisotropic parabolic band. This case is useful as a benchmark because the
band Hessian is constant and coincides with the inverse effective mass tensor.
Thus the geometric formulation can be compared directly with the standard
effective--mass description of electron dynamics in solids.

\subsection{Band structure and basic kinematics}

Consider the dispersion relation
\begin{equation}
\varepsilon(\mathbf{k})
=
\frac{k_x^2}{2m_x}
+
\frac{k_y^2}{2m_y}
+
\frac{k_z^2}{2m_z},
\label{eq:parabolic_dispersion}
\end{equation}
where $m_x$, $m_y$, and $m_z$ are the direction-dependent effective masses.
The group velocity is obtained from
\begin{equation}
v^i(\mathbf{k})
=
\frac{\partial\varepsilon}{\partial k_i}.
\end{equation}
Therefore,
\begin{subequations}
\label{eq:parabolic_velocity}
\begin{align}
v_x &= \frac{k_x}{m_x},
\\
v_y &= \frac{k_y}{m_y},
\\
v_z &= \frac{k_z}{m_z}.
\end{align}
\end{subequations}
The Hessian of the dispersion is
\begin{equation}
\mathcal M^{ij}
=
\frac{\partial^2\varepsilon}{\partial k_i\partial k_j}.
\end{equation}
For the band in Eq.~\eqref{eq:parabolic_dispersion}, this gives
\begin{equation}
\mathcal M^{ij}
=
\begin{pmatrix}
\dfrac{1}{m_x} & 0 & 0\\[1.2ex]
0 & \dfrac{1}{m_y} & 0\\[1.2ex]
0 & 0 & \dfrac{1}{m_z}
\end{pmatrix}.
\label{eq:parabolic_M}
\end{equation}
Hence $\mathcal M^{ij}$ is constant in momentum space and plays the role of
the inverse effective mass tensor. The anisotropy of the band is therefore
encoded only in the three weights $1/m_x$, $1/m_y$, and $1/m_z$.

Using the semiclassical equations
\begin{subequations}
\begin{align}
\dot{x}^i &= v^i(\mathbf{k}),
\\
\dot{k}_i &= qE_i+q\,\epsilon_{ijk}\dot{x}^jB^k,
\end{align}
\end{subequations}
the acceleration is
\begin{equation}
\ddot{x}^i
=
\mathcal M^{ij}\dot{k}_j.
\end{equation}
Since the Hessian is diagonal, one obtains
\begin{subequations}
\label{eq:parabolic_accel_components}
\begin{align}
\ddot{x}
&=
\frac{q}{m_x}
\left[
E_x+\left(\dot{\mathbf{x}}\times\mathbf{B}\right)_x
\right],
\\[1ex]
\ddot{y}
&=
\frac{q}{m_y}
\left[
E_y+\left(\dot{\mathbf{x}}\times\mathbf{B}\right)_y
\right],
\\[1ex]
\ddot{z}
&=
\frac{q}{m_z}
\left[
E_z+\left(\dot{\mathbf{x}}\times\mathbf{B}\right)_z
\right].
\end{align}
\end{subequations}
These equations show that the electromagnetic force is converted into
acceleration with a different strength along each principal direction of the
band. A lighter effective mass gives a larger response, while a heavier mass
suppresses the acceleration in that direction.

Because $\mathcal M^{ij}$ is constant, all terms involving
$\partial_\ell\mathcal M^{ij}$ vanish. The Raychaudhuri-type equation then
reduces to
\begin{align}
\frac{D\theta}{Dt}
&=
-\frac{1}{d}\theta^2
-\sigma_{ij}\sigma^{ij}
+\omega_{ij}\omega^{ij}
+
q\,\mathcal M^{ij}\partial_iE_j
\nonumber\\
&\quad
+
q\,\mathcal M^{ij}\epsilon_{j\ell m}u^\ell\partial_iB^m
+
q\,\mathcal M^{ij}\epsilon_{j\ell m}\mathcal B^\ell{}_{i}B^m .
\label{eq:ray_parabolic_general}
\end{align}
In matrix notation, the electric part can be written as
\begin{equation}
q\,\mathcal M^{ij}\partial_iE_j
=
q\,{\rm Tr}\!\left(\mathcal M\cdot\nabla\mathbf{E}\right).
\end{equation}
Thus the parabolic anisotropic band provides a clean setting in which the
focusing of the beam is weighted direction by direction by the inverse masses.

\subsection{Pure electric field}

Let us first consider a purely electric configuration,
\begin{equation}
\mathbf{B}=0.
\end{equation}
Equation \eqref{eq:ray_parabolic_general} becomes
\begin{equation}
\frac{D\theta}{Dt}
=
-\frac{1}{d}\theta^2
-\sigma_{ij}\sigma^{ij}
+\omega_{ij}\omega^{ij}
+
q\,\mathcal M^{ij}\partial_iE_j .
\label{eq:ray_parabolic_E}
\end{equation}
Using Eq.~\eqref{eq:parabolic_M}, the electric source is
\begin{equation}
q\,\mathcal M^{ij}\partial_iE_j
=
q\left(
\frac{1}{m_x}\partial_xE_x
+
\frac{1}{m_y}\partial_yE_y
+
\frac{1}{m_z}\partial_zE_z
\right).
\label{eq:electric_source_parabolic}
\end{equation}
Therefore,
\begin{equation}
\mathcal S_E
=
q\left(
\frac{1}{m_x}\partial_xE_x
+
\frac{1}{m_y}\partial_yE_y
+
\frac{1}{m_z}\partial_zE_z
\right).
\label{eq:SE_parabolic}
\end{equation}
For an isotropic band, $m_x=m_y=m_z=m^\ast$, this reduces to
\begin{equation}
\mathcal S_E
=
\frac{q}{m^\ast}\nabla\cdot\mathbf{E}.
\end{equation}
For an anisotropic band, however, each component of the electric gradient is
weighted by a different inverse mass. Hence the same electric texture may focus
the congruence more efficiently along a light-mass direction than along a
heavy-mass direction. A uniform electric field, for which
$\partial_iE_j=0$, does not contribute to the local expansion in this
approximation.

\subsection{One-dimensional electrostatic lens}

As a simple electric lens, consider
\begin{equation}
\mathbf{E}=\left(E_x(x),0,0\right).
\end{equation}
Then Eq.~\eqref{eq:ray_parabolic_E} gives
\begin{equation}
\frac{D\theta}{Dt}
=
-\frac{1}{d}\theta^2
-\sigma_{ij}\sigma^{ij}
+\omega_{ij}\omega^{ij}
+
\frac{q}{m_x}\frac{dE_x}{dx}.
\label{eq:ray_parabolic_1D}
\end{equation}
The sign of
\begin{equation}
\frac{q}{m_x}\frac{dE_x}{dx}
\end{equation}
determines whether the electric profile contributes to focusing or
defocusing. For electrons, $q<0$. Thus a positive gradient $dE_x/dx>0$
decreases $\theta$ and favors contraction of the beam, whereas a negative
gradient contributes with the opposite sign. The magnitude of this effect is
set by $1/m_x$, so the same electric lens is more efficient for a lighter
effective mass along the $x$-direction.

\subsection{Pure magnetic field}

We now set
\begin{equation}
\mathbf{E}=0
\end{equation}
and keep only the magnetic field. Equation
\eqref{eq:ray_parabolic_general} becomes
\begin{align}
\frac{D\theta}{Dt}
&=
-\frac{1}{d}\theta^2
-\sigma_{ij}\sigma^{ij}
+\omega_{ij}\omega^{ij}
\nonumber\\
&\quad
+
q\,\mathcal M^{ij}\epsilon_{j\ell m}u^\ell\partial_iB^m
+
q\,\mathcal M^{ij}\epsilon_{j\ell m}\mathcal B^\ell{}_{i}B^m .
\label{eq:ray_parabolic_B}
\end{align}
The first magnetic term is controlled by gradients of the magnetic field. The
second one couples the magnetic field to the deformation tensor of the beam.
Thus a nonuniform magnetic field can act as a local focusing source, while a
uniform magnetic field can still affect the expansion if the congruence is
already deforming.

For a magnetic field directed along the $z$-axis,
\begin{equation}
\mathbf{B}=\left(0,0,B_z(x,y,z)\right),
\end{equation}
the gradient contribution is
\begin{equation}
q\,\mathcal M^{ij}\epsilon_{j\ell 3}u^\ell\partial_iB_z .
\label{eq:mag_grad_general}
\end{equation}
Using the diagonal Hessian in Eq.~\eqref{eq:parabolic_M}, only the terms with
$i=j$ contribute:
\begin{align}
q\,\mathcal M^{ij}\epsilon_{j\ell 3}u^\ell\partial_iB_z
&=
q\left[
\frac{1}{m_x}\epsilon_{1\ell 3}u^\ell\partial_xB_z
+
\frac{1}{m_y}\epsilon_{2\ell 3}u^\ell\partial_yB_z
+
\frac{1}{m_z}\epsilon_{3\ell 3}u^\ell\partial_zB_z
\right].
\end{align}
Since $\epsilon_{3\ell3}=0$, the derivative along $z$ does not contribute to
this term. Using
\begin{equation}
\epsilon_{123}=+1,
\qquad
\epsilon_{213}=-1,
\end{equation}
one obtains
\begin{equation}
q\,\mathcal M^{ij}\epsilon_{j\ell 3}u^\ell\partial_iB_z
=
q\left(
\frac{u^y}{m_x}\partial_xB_z
-
\frac{u^x}{m_y}\partial_yB_z
\right).
\label{eq:mag_source_parabolic}
\end{equation}
This result shows that the magnetic focusing source is transverse: it depends
on the gradients of $B_z$ in the directions perpendicular to the field and on
the corresponding velocity components. The anisotropy enters asymmetrically
through $m_x$ and $m_y$, so the same magnetic profile can produce different
focusing strengths depending on the direction in which the beam crosses it.

\subsection{Interpretation and isotropic limit}

For the anisotropic parabolic band, the Hessian is simply
\begin{equation}
\mathcal M^{ij}
=
\left(m^{-1}\right)^{ij}.
\end{equation}
The Raychaudhuri-type equation therefore becomes an envelope equation in which
electric and magnetic sources are weighted by the inverse effective masses.
Lighter directions react more strongly to field gradients, while heavier
directions reduce the deformation of the beam. This gives a direct geometric
meaning to the effective mass tensor: it determines not only the acceleration
of a single wave packet, but also the focusing properties of a congruence of
nearby wave packets.

In the isotropic limit,
\begin{equation}
m_x=m_y=m_z=m^\ast,
\end{equation}
one has
\begin{equation}
\mathcal M^{ij}
=
\frac{1}{m^\ast}\delta^{ij}.
\end{equation}
The electric contribution reduces to
\begin{equation}
q\,\mathcal M^{ij}\partial_iE_j
=
\frac{q}{m^\ast}\nabla\cdot\mathbf{E},
\end{equation}
while the magnetic terms combine as
\begin{equation}
q\,\mathcal M^{ij}\epsilon_{j\ell m}u^\ell\partial_iB^m
+
q\,\mathcal M^{ij}\epsilon_{j\ell m}\mathcal B^\ell{}_{i}B^m
=
\frac{q}{m^\ast}\nabla\cdot
\left(
\mathbf{u}\times\mathbf{B}
\right).
\end{equation}
Thus the focusing equation becomes
\begin{equation}
\frac{D\theta}{Dt}
=
-\frac{1}{d}\theta^2
-\sigma_{ij}\sigma^{ij}
+\omega_{ij}\omega^{ij}
+
\frac{q}{m^\ast}
\nabla\cdot
\left(
\mathbf{E}
+
\mathbf{u}\times\mathbf{B}
\right).
\label{eq:ray_isotropic}
\end{equation}
This is the expected effective--mass result. The anisotropic case is obtained
by replacing the single factor $1/m^\ast$ with the tensor
$\mathcal M^{ij}$, which weights each field gradient according to the local
principal directions of the band.


\section{Dirac-like dispersion}
\label{sec:dirac}

We now consider a Dirac-like dispersion, which is the simplest example of a
nonparabolic band structure. This case is relevant for systems such as
graphene, topological insulators, and Dirac semimetals. In contrast with the
parabolic band, the quasiparticle speed is fixed by the Fermi velocity, while
the band Hessian becomes momentum dependent and singular near the band-touching
point.

\subsection{Dispersion relation and group velocity}

Consider the two-branch isotropic Dirac dispersion
\begin{equation}
\varepsilon_\pm(\mathbf{k})
=
\pm v_F |\mathbf{k}|,
\qquad
|\mathbf{k}|:=\sqrt{k_i k_i},
\label{eq:dirac_dispersion}
\end{equation}
where $v_F$ is the Fermi velocity. The signs $(\pm)$ label the upper and lower
cones. The group velocity is
\begin{equation}
v^i_{(\pm)}(\mathbf{k})
=
\frac{\partial\varepsilon_\pm}{\partial k_i}.
\end{equation}
Using
\begin{equation}
\frac{\partial|\mathbf{k}|}{\partial k_i}
=
\frac{k_i}{|\mathbf{k}|},
\end{equation}
we obtain
\begin{equation}
v^i_{(\pm)}(\mathbf{k})
=
\pm v_F\,\frac{k_i}{|\mathbf{k}|}.
\label{eq:dirac_velocity}
\end{equation}
Thus the velocity is radial in momentum space and has fixed magnitude,
\begin{equation}
|\mathbf{v}_{(\pm)}|=v_F .
\end{equation}
Changing $|\mathbf{k}|$ does not change the speed of the quasiparticle; it
changes only the direction of propagation.

The band Hessian is
\begin{equation}
\mathcal M^{ij}_{(\pm)}
=
\frac{\partial v^i_{(\pm)}}{\partial k_j}.
\label{eq:def_dirac_M}
\end{equation}
From Eq.~\eqref{eq:dirac_velocity},
\begin{equation}
v^i_{(\pm)}
=
\pm v_F\frac{k_i}{|\mathbf{k}|}.
\end{equation}
Therefore,
\begin{equation}
\frac{\partial v^i_{(\pm)}}{\partial k_j}
=
\pm v_F
\frac{\partial}{\partial k_j}
\left(
\frac{k_i}{|\mathbf{k}|}
\right).
\end{equation}
Now,
\begin{align}
\frac{\partial}{\partial k_j}
\left(
\frac{k_i}{|\mathbf{k}|}
\right)
&=
\frac{\delta_{ij}}{|\mathbf{k}|}
+
k_i\frac{\partial}{\partial k_j}
\left(
\frac{1}{|\mathbf{k}|}
\right)
\nonumber\\
&=
\frac{\delta_{ij}}{|\mathbf{k}|}
-
\frac{k_i k_j}{|\mathbf{k}|^3}.
\end{align}
Hence
\begin{equation}
\mathcal M^{ij}_{(\pm)}
=
\pm v_F
\left(
\frac{\delta_{ij}}{|\mathbf{k}|}
-
\frac{k_i k_j}{|\mathbf{k}|^3}
\right).
\label{eq:dirac_M}
\end{equation}
Introducing
\begin{equation}
\hat{k}_i:=\frac{k_i}{|\mathbf{k}|},
\end{equation}
this result becomes
\begin{equation}
\mathcal M^{ij}_{(\pm)}
=
\pm\frac{v_F}{|\mathbf{k}|}
\left(
\delta_{ij}-\hat{k}_i\hat{k}_j
\right).
\label{eq:dirac_M_projector}
\end{equation}
Thus the Dirac Hessian is proportional to the projector transverse to
$\hat{\mathbf{k}}$,
\begin{equation}
P^\perp_{ij}:=\delta_{ij}-\hat{k}_i\hat{k}_j,
\qquad
\mathcal M^{ij}_{(\pm)}
=
\pm\frac{v_F}{|\mathbf{k}|}P^\perp_{ij}.
\end{equation}
Consequently,
\begin{equation}
\mathcal M^{ij}_{(\pm)}\hat{k}_j=0.
\end{equation}
The radial direction in momentum space is a zero mode of the Hessian, whereas
the transverse directions have eigenvalues
$\pm v_F/|\mathbf{k}|$. This is the main difference from the parabolic case:
the Dirac susceptibility is not an inverse mass tensor with a finite
longitudinal component. It is a purely transverse response tensor.

\begin{figure}[h]
\centering
\begin{tikzpicture}[scale=1.2]

\draw[thick] (0,0) -- (2,2);
\draw[thick] (0,0) -- (-2,2);
\draw[thick] (2,2) arc (0:180:2 and 0.6);

\node at (0,3) {\small Dirac cone};

\draw[->,thick,red] (0,0) -- (1.2,1.2);
\node[red] at (1.4,1.4) {\small $\mathbf{k}$};

\draw[dashed] (1.2,1.2) ellipse (0.8 and 0.3);
\node at (2.3,1.2) {\small transverse};

\end{tikzpicture}

\caption{
Momentum-space geometry of the Dirac dispersion. The tensor
$\mathcal M^{ij}_{(\pm)}$ projects the response onto the subspace transverse to
$\hat{\mathbf{k}}$. Hence longitudinal forcing changes the magnitude of
$\mathbf{k}$, but does not change the quasiparticle velocity at leading order.
}
\label{fig:dirac_geometry}
\end{figure}
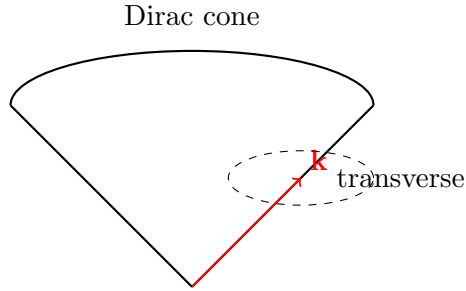

Since the speed is fixed, an electric force parallel to $\mathbf{k}$ cannot
increase the magnitude of the velocity. It only changes the energy measured
along the cone. By contrast, transverse forcing changes the direction of
$\mathbf{k}$ and bends the trajectory. The factor $1/|\mathbf{k}|$ also shows
that the response becomes stronger near the Dirac point, where the cone is
singular and the semiclassical approximation must be used with care.

The acceleration of the wave packet is
\begin{equation}
a^i_{(\pm)}
=
q\,\mathcal M^{ij}_{(\pm)}E_j
+
q\,\mathcal M^{ij}_{(\pm)}
\epsilon_{j\ell m}u^\ell_{(\pm)}B^m .
\label{eq:dirac_accel_start}
\end{equation}
Substituting Eq.~\eqref{eq:dirac_M_projector}, we find
\begin{equation}
a^i_{(\pm)}
=
\pm\frac{qv_F}{|\mathbf{k}|}
P^\perp_{ij}
\left(
E_j+\epsilon_{j\ell m}u^\ell_{(\pm)}B^m
\right).
\label{eq:dirac_accel_final}
\end{equation}
Therefore only the part of the electromagnetic force transverse to
$\hat{\mathbf{k}}$ contributes to the acceleration of the velocity field.

We now insert the Dirac Hessian into the general Raychaudhuri-type equation,
\begin{align}
\frac{D\theta_{(\pm)}}{Dt}
&=
-\frac{1}{d}\theta_{(\pm)}^2
-\sigma^{(\pm)}_{ij}\sigma_{(\pm)}^{ij}
+\omega^{(\pm)}_{ij}\omega_{(\pm)}^{ij}
\nonumber\\
&\quad
+
q\,(\partial_i\mathcal M^{ij}_{(\pm)})E_j
+
q\,\mathcal M^{ij}_{(\pm)}\partial_iE_j
\nonumber\\
&\quad
+
q\,(\partial_i\mathcal M^{ij}_{(\pm)})
\epsilon_{j\ell m}u^\ell_{(\pm)}B^m
\nonumber\\
&\quad
+
q\,\mathcal M^{ij}_{(\pm)}
\epsilon_{j\ell m}\mathcal B^\ell{}_{i\,(\pm)}B^m
\nonumber\\
&\quad
+
q\,\mathcal M^{ij}_{(\pm)}
\epsilon_{j\ell m}u^\ell_{(\pm)}\partial_iB^m .
\label{eq:dirac_ray_start}
\end{align}
Here $\mathcal B^\ell{}_{i\,(\pm)}=\partial_i u^\ell_{(\pm)}$ is the
deformation tensor of the branch congruence.

\subsection{Derivative of the Dirac Hessian}

Because $\mathcal M^{ij}_{(\pm)}$ depends on $\mathbf{k}$, its spatial
derivative contributes whenever $\mathbf{k}$ varies across the beam. From
Eq.~\eqref{eq:dirac_M},
\begin{equation}
\mathcal M^{ij}_{(\pm)}
=
\pm v_F
\left(
\frac{\delta_{ij}}{|\mathbf{k}|}
-
\frac{k_i k_j}{|\mathbf{k}|^3}
\right).
\end{equation}
Differentiating with respect to $x^m$, we have
\begin{equation}
\partial_m\mathcal M^{ij}_{(\pm)}
=
\pm v_F\,\partial_m
\left(
\frac{\delta_{ij}}{|\mathbf{k}|}
-
\frac{k_i k_j}{|\mathbf{k}|^3}
\right).
\label{eq:spatial_derivative_M}
\end{equation}
Since $\mathbf{k}=\mathbf{k}(x,t)$, the chain rule gives
\begin{equation}
\partial_m
\left(
\frac{1}{|\mathbf{k}|}
\right)
=
-\frac{k_r\partial_m k_r}{|\mathbf{k}|^3},
\label{eq:d1overk}
\end{equation}
and
\begin{align}
\partial_m
\left(
\frac{k_i k_j}{|\mathbf{k}|^3}
\right)
&=
\frac{(\partial_m k_i)k_j+k_i(\partial_m k_j)}{|\mathbf{k}|^3}
-
\frac{3k_i k_j}{|\mathbf{k}|^5}
k_r\partial_m k_r .
\label{eq:dkikj}
\end{align}
Combining these expressions, we obtain
\begin{align}
\partial_m\mathcal M^{ij}_{(\pm)}
&=
\pm v_F
\left[
-\delta_{ij}
\frac{k_r\partial_m k_r}{|\mathbf{k}|^3}
-
\frac{(\partial_m k_i)k_j+k_i(\partial_m k_j)}{|\mathbf{k}|^3}
+
\frac{3k_i k_j}{|\mathbf{k}|^5}
k_r\partial_m k_r
\right].
\label{eq:dM_dirac_final}
\end{align}
The powers of $1/|\mathbf{k}|$ show that spatial variations of the Dirac
susceptibility are enhanced near the cone apex. Away from this region, these
terms can be suppressed if the beam samples only a narrow interval of
$\mathbf{k}$.

\subsection{Homogeneous-momentum approximation}

To isolate the effect of external field gradients, let us consider the regime
in which $\mathbf{k}$ varies slowly across the beam. Then
\begin{equation}
\partial_i\mathcal M^{ij}_{(\pm)}\simeq0.
\end{equation}
Equation \eqref{eq:dirac_ray_start} reduces to
\begin{align}
\frac{D\theta_{(\pm)}}{Dt}
&=
-\frac{1}{d}\theta_{(\pm)}^2
-\sigma^{(\pm)}_{ij}\sigma_{(\pm)}^{ij}
+\omega^{(\pm)}_{ij}\omega_{(\pm)}^{ij}
\nonumber\\
&\quad
+
q\,\mathcal M^{ij}_{(\pm)}\partial_iE_j
+
q\,\mathcal M^{ij}_{(\pm)}
\epsilon_{j\ell m}u^\ell_{(\pm)}\partial_iB^m
\nonumber\\
&\quad
+
q\,\mathcal M^{ij}_{(\pm)}
\epsilon_{j\ell m}\mathcal B^\ell{}_{i\,(\pm)}B^m .
\label{eq:ray_dirac_simple}
\end{align}
Using Eq.~\eqref{eq:dirac_M_projector}, this becomes
\begin{align}
\frac{D\theta_{(\pm)}}{Dt}
&=
-\frac{1}{d}\theta_{(\pm)}^2
-\sigma^{(\pm)}_{ij}\sigma_{(\pm)}^{ij}
+\omega^{(\pm)}_{ij}\omega_{(\pm)}^{ij}
\nonumber\\
&\quad
\pm\frac{qv_F}{|\mathbf{k}|}
P^\perp_{ij}\partial_iE_j
\pm\frac{qv_F}{|\mathbf{k}|}
P^\perp_{ij}\epsilon_{j\ell m}u^\ell_{(\pm)}\partial_iB^m
\nonumber\\
&\quad
\pm\frac{qv_F}{|\mathbf{k}|}
P^\perp_{ij}
\epsilon_{j\ell m}\mathcal B^\ell{}_{i\,(\pm)}B^m .
\label{eq:ray_dirac_projected}
\end{align}
This equation contains three characteristic features of the Dirac cone. The
two branches enter with opposite overall sign, the electromagnetic forcing is
projected transversely to $\hat{\mathbf{k}}$, and the strength of the response
is scaled by $1/|\mathbf{k}|$.

\subsection{Pure electric focusing}

For a purely electric configuration, $\mathbf{B}=0$, one obtains
\begin{equation}
\frac{D\theta_{(\pm)}}{Dt}
=
-\frac{1}{d}\theta_{(\pm)}^2
-\sigma^{(\pm)}_{ij}\sigma_{(\pm)}^{ij}
+\omega^{(\pm)}_{ij}\omega_{(\pm)}^{ij}
\pm
\frac{qv_F}{|\mathbf{k}|}
P^\perp_{ij}\partial_iE_j .
\label{eq:ray_dirac_E}
\end{equation}
The electric focusing source is therefore
\begin{equation}
\mathcal S^{(E)}_{(\pm)}
=
\pm
\frac{qv_F}{|\mathbf{k}|}
P^\perp_{ij}\partial_iE_j .
\label{eq:dirac_electric_source}
\end{equation}
Only the transverse part of the electric-field gradient contributes. Hence an
electric texture that is longitudinal with respect to $\hat{\mathbf{k}}$ does
not focus the Dirac beam at this order. The opposite signs of the two branches
show that the same electric profile can focus one cone while defocusing the
other, depending on the sign of $qP^\perp_{ij}\partial_iE_j$.

\subsection{Pure magnetic focusing}

For a purely magnetic configuration, $\mathbf{E}=0$, Eq.
\eqref{eq:ray_dirac_projected} gives
\begin{align}
\frac{D\theta_{(\pm)}}{Dt}
&=
-\frac{1}{d}\theta_{(\pm)}^2
-\sigma^{(\pm)}_{ij}\sigma_{(\pm)}^{ij}
+\omega^{(\pm)}_{ij}\omega_{(\pm)}^{ij}
\nonumber\\
&\quad
\pm\frac{qv_F}{|\mathbf{k}|}
P^\perp_{ij}\epsilon_{j\ell m}u^\ell_{(\pm)}\partial_iB^m
\nonumber\\
&\quad
\pm\frac{qv_F}{|\mathbf{k}|}
P^\perp_{ij}
\epsilon_{j\ell m}\mathcal B^\ell{}_{i\,(\pm)}B^m .
\label{eq:ray_dirac_B}
\end{align}
The first magnetic term is produced by magnetic-field gradients. The second
one couples the magnetic field to the internal deformation of the congruence.
Thus a uniform magnetic field has no gradient source, but it can still affect
the expansion when the beam already carries shear, rotation, or expansion. In
both terms the projector $P^\perp_{ij}$ keeps only the part of the response
transverse to the direction of propagation.

The Dirac cone is therefore qualitatively different from the parabolic band.
The susceptibility tensor is momentum dependent,
\begin{equation}
\mathcal M^{ij}_{(\pm)}
=
\pm\frac{v_F}{|\mathbf{k}|}P^\perp_{ij},
\end{equation}
and becomes large near $|\mathbf{k}|=0$. As a result, focusing and defocusing
effects are expected to be more sensitive close to the Dirac point, while the
semiclassical description becomes less reliable exactly at the singular point.
The two cones also respond with opposite signs, which gives a simple
quasiparticle analogue of branch-dependent beam evolution.

\subsection{Special two-dimensional reduction}

For systems such as graphene, it is useful to specialize to two spatial
dimensions. Let
\begin{equation}
\varepsilon_\pm(k_x,k_y)
=
\pm v_F\sqrt{k_x^2+k_y^2}.
\end{equation}
Then
\begin{equation}
\mathcal M^{ij}_{(\pm)}
=
\pm\frac{v_F}{|\mathbf{k}|}
\left(
\delta_{ij}-\hat{k}_i\hat{k}_j
\right),
\qquad
i,j=1,2.
\end{equation}
The Raychaudhuri-type equation becomes
\begin{equation}
\frac{D\theta_{(\pm)}}{Dt}
=
-\frac12\theta_{(\pm)}^2
-\sigma^{(\pm)}_{ij}\sigma_{(\pm)}^{ij}
+\omega^{(\pm)}_{ij}\omega_{(\pm)}^{ij}
+
\partial_i a^i_{(\pm)},
\label{eq:ray_2D_dirac}
\end{equation}
with
\begin{equation}
a^i_{(\pm)}
=
\pm\frac{qv_F}{|\mathbf{k}|}
P^\perp_{ij}
\left(
E_j+\epsilon_{j\ell}u^\ell_{(\pm)}B
\right),
\end{equation}
where $\epsilon_{j\ell}$ is the two-dimensional antisymmetric symbol and $B$
is the out-of-plane magnetic field. This form is suited to electron-optics
setups in graphene, where electrostatic gates and magnetic textures can
collimate, bend, or focus quasiparticle beams. The transverse projector shows
which part of those external profiles actually changes the direction of the
Dirac velocity field.


\section{Weyl-type two-branch systems and quasiparticle birefringence}
\label{sec:weyl}

We now consider multibranch systems in which different dispersion branches are
available at the same momentum. This situation appears in conical band
structures, Weyl semimetals, and systems with branch-resolved quasiparticle
dynamics. These systems provide a condensed--matter realization of the same
geometric mechanism found in the spin--nondegenerate Lorentz--violating
sectors: each branch defines its own tangent field, its own response tensor,
and therefore its own congruence evolution.

\subsection{Weyl-type dispersion relations}

A minimal isotropic conical dispersion is
\begin{equation}
\varepsilon_\pm(\mathbf{k})
=
\pm v_F|\mathbf{k}|,
\label{eq:weyl_dispersion}
\end{equation}
where the signs label the two branches of the cone. In a Weyl semimetal, an
additional chirality label may be attached to each Weyl node, but the
branch-resolved structure studied here is already visible in
Eq.~\eqref{eq:weyl_dispersion}.

More generally, an anisotropic Weyl-type cone can be written as
\begin{equation}
\varepsilon_\pm(\mathbf{k})
=
\pm
\sqrt{v_i v_j k_i k_j},
\label{eq:weyl_anisotropic}
\end{equation}
where the parameters $v_i$ determine the directional slopes of the cone. The
isotropic case is recovered when $v_i=v_F$ in all directions. In both cases,
the two branches occupy the same momentum space but have opposite energy
slopes, which is the origin of the opposite response tensors obtained below.

\subsection{Branch-dependent velocities and Hessians}

For the isotropic dispersion \eqref{eq:weyl_dispersion}, the group velocities
are
\begin{equation}
u^i_{(\pm)}
=
\frac{\partial\varepsilon_\pm}{\partial k_i}
=
\pm v_F\frac{k_i}{|\mathbf{k}|}.
\end{equation}
The corresponding Hessians are
\begin{equation}
\mathcal M^{ij}_{(\pm)}
=
\frac{\partial u^i_{(\pm)}}{\partial k_j}
=
\pm\frac{v_F}{|\mathbf{k}|}
\left(
\delta_{ij}-\hat{k}_i\hat{k}_j
\right),
\qquad
\hat{k}_i:=\frac{k_i}{|\mathbf{k}|}.
\label{eq:weyl_M}
\end{equation}
Thus
\begin{equation}
\mathcal M^{ij}_{(+)}
=
-\mathcal M^{ij}_{(-)}.
\end{equation}
Although this is only an overall sign, it changes the sign of the
electromagnetic contribution to the branch acceleration and to the
Raychaudhuri equation. The projector
$\delta_{ij}-\hat{k}_i\hat{k}_j$ also shows that the response is transverse to
the propagation direction, as expected for a conical dispersion with fixed
speed.

\subsection{Branch-dependent acceleration}

The semiclassical acceleration of each branch is
\begin{equation}
a^i_{(\pm)}
=
q\,\mathcal M^{ij}_{(\pm)}E_j
+
q\,\mathcal M^{ij}_{(\pm)}
\epsilon_{j\ell m}u^\ell_{(\pm)}B^m .
\end{equation}
Using Eq.~\eqref{eq:weyl_M}, this becomes
\begin{equation}
a^i_{(\pm)}
=
\pm\frac{qv_F}{|\mathbf{k}|}
P^\perp_{ij}
\left(
E_j+\epsilon_{j\ell m}u^\ell_{(\pm)}B^m
\right),
\qquad
P^\perp_{ij}:=\delta_{ij}-\hat{k}_i\hat{k}_j .
\label{eq:weyl_accel}
\end{equation}
The same electromagnetic field is therefore converted into opposite branch
responses. This does not mean that the Lorentz force itself changes sign; the
sign reversal comes from the Hessian of the dispersion, which converts
momentum-space forcing into real-space acceleration.

The expansion scalars $\theta_{(\pm)}$ obey
\begin{align}
\frac{D\theta_{(\pm)}}{Dt}
&=
-\frac{1}{d}\theta_{(\pm)}^2
-\sigma^{(\pm)}_{ij}\sigma_{(\pm)}^{ij}
+\omega^{(\pm)}_{ij}\omega_{(\pm)}^{ij}
+
\partial_i a^i_{(\pm)} .
\label{eq:weyl_ray_general}
\end{align}
Substitution of Eq.~\eqref{eq:weyl_accel} gives
\begin{align}
\partial_i a^i_{(\pm)}
&=
\pm
\partial_i
\left[
\frac{qv_F}{|\mathbf{k}|}
P^\perp_{ij}
\left(
E_j+\epsilon_{j\ell m}u^\ell_{(\pm)}B^m
\right)
\right].
\label{eq:weyl_forcing}
\end{align}
If the beam samples a narrow region of momentum space, so that the spatial
variation of the Hessian can be neglected, the Raychaudhuri equation reduces to
\begin{align}
\frac{D\theta_{(\pm)}}{Dt}
&=
-\frac{1}{d}\theta_{(\pm)}^2
-\sigma^{(\pm)}_{ij}\sigma_{(\pm)}^{ij}
+\omega^{(\pm)}_{ij}\omega_{(\pm)}^{ij}
\nonumber\\
&\quad
\pm\frac{qv_F}{|\mathbf{k}|}
P^\perp_{ij}\partial_iE_j
\nonumber\\
&\quad
\pm\frac{qv_F}{|\mathbf{k}|}
P^\perp_{ij}
\epsilon_{j\ell m}u^\ell_{(\pm)}\partial_iB^m
\nonumber\\
&\quad
\pm\frac{qv_F}{|\mathbf{k}|}
P^\perp_{ij}
\epsilon_{j\ell m}
\mathcal B^\ell{}_{i\,(\pm)}B^m .
\label{eq:weyl_ray_final}
\end{align}
Here
\begin{equation}
\mathcal B^\ell{}_{i\,(\pm)}:=\partial_i u^\ell_{(\pm)}
\end{equation}
is the deformation tensor of the branch congruence. The first two
electromagnetic terms are controlled by gradients of the electric and magnetic
fields, while the last term couples the magnetic field to an already deforming
beam.

\subsection{Branch-dependent focusing and quasiparticle birefringence}

The main feature of Eq.~\eqref{eq:weyl_ray_final} is the overall branch sign
multiplying the electromagnetic source terms. For a purely electric field, the
source reduces to
\begin{equation}
\mathcal S^{(E)}_{(\pm)}
=
\pm\frac{qv_F}{|\mathbf{k}|}
P^\perp_{ij}\partial_iE_j .
\label{eq:weyl_electric_source}
\end{equation}
Thus an electric-field gradient that decreases $\theta_{(+)}$ may increase
$\theta_{(-)}$, and vice versa. The effect is controlled only by the transverse
part of the field gradient, since the projector $P^\perp_{ij}$ removes the
longitudinal component along $\hat{\mathbf{k}}$.

This behavior is analogous to birefringence in optics. The two quasiparticle
branches play the role of two propagation channels, while the band Hessians
play the role of branch-dependent response tensors. In the present case, an
initially collimated beam entering a region with nonuniform electromagnetic
fields can split into components with different expansion rates. We refer to
this effect as \emph{quasiparticle birefringence of congruences}.

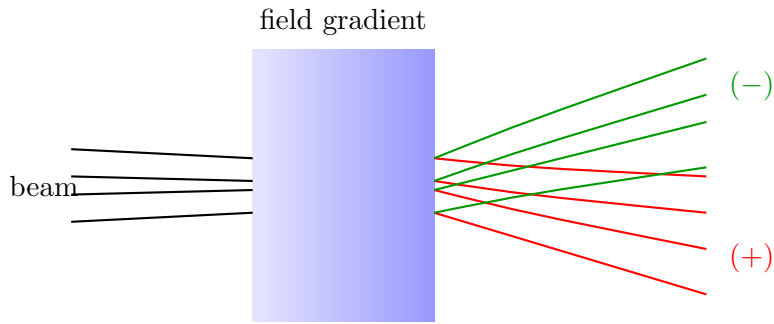
\begin{figure}[h]
\centering
\begin{tikzpicture}[scale=1.2]

\shade[left color=blue!10,right color=blue!40] (-2,-1.5) rectangle (0,1.5);

\node at (-1,1.8) {\small field gradient};

\draw[thick] (-4,-0.4) -- (-2,-0.3);
\draw[thick] (-4,-0.1) -- (-2,-0.05);
\draw[thick] (-4,0.1) -- (-2,0.05);
\draw[thick] (-4,0.4) -- (-2,0.3);

\node at (-4.3,0) {\small beam};

\draw[thick,red] (0,-0.3) .. controls (1,-0.6) .. (3,-1.2);
\draw[thick,red] (0,-0.05) .. controls (1,-0.3) .. (3,-0.7);
\draw[thick,red] (0,0.05) .. controls (1,-0.1) .. (3,-0.3);
\draw[thick,red] (0,0.3) .. controls (1,0.2) .. (3,0.1);

\node[red] at (3.5,-0.8) {\small $(+)$};

\draw[thick,green!60!black] (0,-0.3) .. controls (1,-0.1) .. (3,0.2);
\draw[thick,green!60!black] (0,-0.05) .. controls (1,0.2) .. (3,0.7);
\draw[thick,green!60!black] (0,0.05) .. controls (1,0.4) .. (3,1.0);
\draw[thick,green!60!black] (0,0.3) .. controls (1,0.7) .. (3,1.4);

\node[green!60!black] at (3.5,1.1) {\small $(-)$};

\end{tikzpicture}

\caption{
Schematic illustration of quasiparticle birefringence. An initially collimated
beam enters a region with spatially varying electromagnetic field and separates
into two branch components. Since
$\mathcal M^{ij}_{(+)}=-\mathcal M^{ij}_{(-)}$, the electromagnetic forcing
term in the Raychaudhuri equation has opposite signs for the two branches,
leading to different focusing behavior.
}
\label{fig:birefringence}
\end{figure}

The mechanism can be stated in a compact way. The band dispersion fixes
$u^i_{(\pm)}$, the Hessian $\mathcal M^{ij}_{(\pm)}$ fixes the branch response,
and the Raychaudhuri equation converts this response into the evolution of
$\theta_{(\pm)}$. Therefore, branch splitting at the level of the dispersion
becomes branch splitting at the level of beam deformation.

The branch dependence is also visible in the reduced phase-space flow. The two
branches define different local vector fields in the $(x,k)$ plane, and these
flows deform nearby trajectories in different ways. This phase-space picture is
shown schematically in Fig.~\ref{fig:phase_space_branches}.

\begin{figure}[h]
\centering
\begin{tikzpicture}[scale=1.0]

\begin{scope}[shift={(-5.2,0)}]
\draw[->] (-2.8,0) -- (2.8,0) node[right] {\small $x$};
\draw[->] (0,-2.2) -- (0,2.5) node[above] {\small $k$};

\draw[->,gray] (-2.3,1.8) -- (-1.8,1.4);
\draw[->,gray] (-1.5,1.5) -- (-0.9,1.0);
\draw[->,gray] (-0.7,1.1) -- (-0.1,0.6);
\draw[->,gray] (0.4,0.7) -- (1.0,0.3);
\draw[->,gray] (1.3,0.4) -- (1.9,0.1);

\draw[->,gray] (-2.3,0.8) -- (-1.6,0.6);
\draw[->,gray] (-1.4,0.6) -- (-0.7,0.4);
\draw[->,gray] (-0.5,0.4) -- (0.2,0.2);
\draw[->,gray] (0.5,0.2) -- (1.2,0.0);
\draw[->,gray] (1.5,0.0) -- (2.1,-0.2);

\draw[->,gray] (-2.3,-0.5) -- (-1.6,-0.4);
\draw[->,gray] (-1.4,-0.4) -- (-0.7,-0.3);
\draw[->,gray] (-0.5,-0.3) -- (0.2,-0.3);
\draw[->,gray] (0.5,-0.3) -- (1.2,-0.4);
\draw[->,gray] (1.5,-0.5) -- (2.1,-0.7);

\draw[very thick,red]
(-2.3,1.3) .. controls (-1.4,0.9) and (-0.6,0.5) .. (0.2,0.1)
             .. controls (1.0,-0.2) and (1.7,-0.5) .. (2.3,-0.8);

\draw[very thick,red]
(-2.3,0.9) .. controls (-1.4,0.6) and (-0.6,0.3) .. (0.2,-0.1)
             .. controls (1.0,-0.4) and (1.7,-0.8) .. (2.3,-1.2);

\draw[dashed] (-1.7,1.15) -- (-1.7,0.85);
\draw[dashed] (1.7,-0.65) -- (1.7,-1.05);

\node[red] at (-1.5,2.1) {\small $(+)$ branch};
\end{scope}

\begin{scope}[shift={(5.2,0)}]
\draw[->] (-2.8,0) -- (2.8,0) node[right] {\small $x$};
\draw[->] (0,-2.2) -- (0,2.5) node[above] {\small $k$};

\draw[->,gray] (-2.3,1.3) -- (-1.7,1.5);
\draw[->,gray] (-1.5,1.0) -- (-0.8,1.2);
\draw[->,gray] (-0.6,0.7) -- (0.1,1.0);
\draw[->,gray] (0.5,0.4) -- (1.2,0.9);
\draw[->,gray] (1.4,0.2) -- (2.0,0.7);

\draw[->,gray] (-2.3,0.2) -- (-1.6,0.5);
\draw[->,gray] (-1.4,0.0) -- (-0.7,0.4);
\draw[->,gray] (-0.5,-0.1) -- (0.2,0.3);
\draw[->,gray] (0.5,-0.2) -- (1.2,0.2);
\draw[->,gray] (1.5,-0.3) -- (2.1,0.1);

\draw[->,gray] (-2.3,-0.8) -- (-1.7,-0.4);
\draw[->,gray] (-1.5,-1.0) -- (-0.8,-0.5);
\draw[->,gray] (-0.6,-1.2) -- (0.1,-0.7);
\draw[->,gray] (0.5,-1.3) -- (1.2,-0.8);
\draw[->,gray] (1.4,-1.4) -- (2.0,-1.0);

\draw[very thick,green!60!black]
(-2.3,0.8) .. controls (-1.4,0.9) and (-0.6,1.0) .. (0.2,1.1)
             .. controls (1.0,1.2) and (1.7,1.3) .. (2.3,1.5);

\draw[very thick,green!60!black]
(-2.3,0.4) .. controls (-1.4,0.6) and (-0.6,0.8) .. (0.2,1.0)
             .. controls (1.0,1.2) and (1.7,1.5) .. (2.3,1.8);

\draw[dashed] (-1.7,0.7) -- (-1.7,0.45);
\draw[dashed] (1.7,1.38) -- (1.7,1.68);

\node[green!60!black] at (-1.5,2.1) {\small $(-)$ branch};
\end{scope}

\node at (0,-2.8) {\small Branch-resolved phase-space flow in a reduced $(x,k)$ description};

\end{tikzpicture}

\caption{
Branch-resolved phase-space flow for a two-branch quasiparticle system in a
reduced $(x,k)$ plane. The arrows represent the local semiclassical flow in
each branch. Nearby trajectories define a congruence, and the different local
flows deform the $(+)$ and $(-)$ congruences in different ways. This provides
the phase-space origin of the branch-dependent focusing shown in
Fig.~\ref{fig:birefringence}.
}
\label{fig:phase_space_branches}
\end{figure}

From the experimental point of view, the effect should be most visible when a
collimated quasiparticle beam crosses an electrostatic or magnetic texture with
nonzero spatial gradients. The separation between the two branch components is
controlled by the factor $v_F/|\mathbf{k}|$, so beams prepared closer to the
node are more sensitive to the external profile, within the regime where the
semiclassical approximation remains valid. In Weyl materials, this suggests
that branch-dependent focusing can appear as different beam divergence,
chirality- or band-sensitive transmission patterns, or asymmetric collimation
in electron-optics geometries.

The comparison with the relativistic Lorentz--violating construction is then
straightforward. In both cases, the dispersion relation splits the motion into
branches, each branch carries its own response tensor, and the Raychaudhuri
equation evolves each congruence separately. The condensed--matter system has
the advantage that the external electromagnetic texture and the position of
the beam in momentum space can be controlled experimentally, making the
branch-resolved deformation of quasiparticle beams a potentially accessible
signature of the same geometric mechanism.


\section{Electric and magnetic focusing mechanisms}
\label{sec:fields}

In this section we collect the results obtained for the different band
structures and identify the mechanisms responsible for focusing and
defocusing of quasiparticle congruences. The goal is to isolate the common
structure of the Raychaudhuri-type equation and to clarify the roles played by
electric fields, magnetic fields, band curvature, and branch splitting.

\subsection{General structure of the focusing equation}

From Sec.~\ref{sec:raychaudhuri}, the expansion scalar obeys
\begin{equation}
\frac{D\theta}{Dt}
=
-\frac{1}{d}\theta^2
-\sigma_{ij}\sigma^{ij}
+\omega_{ij}\omega^{ij}
+
\partial_i a^i,
\label{eq:ray_general_fields}
\end{equation}
where the semiclassical acceleration is
\begin{equation}
a^i
=
q\,\mathcal M^{ij}E_j
+
q\,\mathcal M^{ij}\epsilon_{j\ell m}u^\ell B^m.
\label{eq:accel_fields_repeat}
\end{equation}
Thus the external contribution to the expansion is determined by
\begin{equation}
\mathcal S:=\partial_i a^i.
\label{eq:source_def}
\end{equation}
Using Eq.~\eqref{eq:accel_fields_repeat}, this source can be written as
\begin{align}
\mathcal S
&=
q\,(\partial_i\mathcal M^{ij})E_j
+
q\,\mathcal M^{ij}\partial_iE_j
\nonumber\\
&\quad
+
q\,(\partial_i\mathcal M^{ij})\epsilon_{j\ell m}u^\ell B^m
+
q\,\mathcal M^{ij}\epsilon_{j\ell m}\mathcal B^\ell{}_{i}B^m
+
q\,\mathcal M^{ij}\epsilon_{j\ell m}u^\ell\partial_iB^m,
\label{eq:source_general}
\end{align}
with
\begin{equation}
\mathcal B^\ell{}_{i}:=\partial_i u^\ell .
\end{equation}
Equation \eqref{eq:source_general} displays the two ingredients that control
the focusing source: the electromagnetic texture and the band Hessian. If the
Hessian is constant across the beam, the terms involving
$\partial_i\mathcal M^{ij}$ vanish and the remaining terms are controlled by
field gradients and by the deformation already present in the congruence.

\subsection{Electric focusing}

The electric part of the source is
\begin{equation}
\mathcal S_E
=
q\,(\partial_i\mathcal M^{ij})E_j
+
q\,\mathcal M^{ij}\partial_iE_j .
\label{eq:source_electric_general}
\end{equation}
In the homogeneous Hessian approximation, this reduces to
\begin{equation}
\mathcal S_E
=
q\,\mathcal M^{ij}\partial_iE_j .
\label{eq:source_electric_homogeneous}
\end{equation}
Therefore, a uniform electric field does not generate local focusing in this
limit, since $\partial_iE_j=0$. It can still bend individual trajectories, but
it does not change the local expansion of an initially uniform beam.

For anisotropic bands, the tensor $\mathcal M^{ij}$ weights the electric
gradient differently in different directions. Hence two electric textures with
the same ordinary divergence may produce different focusing strengths if their
gradients are oriented along different principal directions of the Hessian. In
a multibranch system, the replacement
$\mathcal M^{ij}\rightarrow\mathcal M^{ij}_{(\pm)}$ makes the electric source
branch dependent,
\begin{equation}
\mathcal S_E^{(\pm)}
=
q\,\mathcal M^{ij}_{(\pm)}\partial_iE_j ,
\end{equation}
so that the same electric profile can drive distinct evolutions for the two
congruences.

\subsection{Magnetic focusing}

The magnetic part of the source is
\begin{align}
\mathcal S_B
&=
q\,(\partial_i\mathcal M^{ij})\epsilon_{j\ell m}u^\ell B^m
+
q\,\mathcal M^{ij}\epsilon_{j\ell m}\mathcal B^\ell{}_{i}B^m
+
q\,\mathcal M^{ij}\epsilon_{j\ell m}u^\ell\partial_iB^m .
\label{eq:source_magnetic_general}
\end{align}
This term differs from the electric contribution because it depends explicitly
on the velocity field. The magnetic gradient is sampled according to the
direction of propagation, and the result is again weighted by the band Hessian.

If $\mathcal M^{ij}$ is constant and the beam has no initial deformation, the
leading magnetic focusing source comes from
\begin{equation}
q\,\mathcal M^{ij}\epsilon_{j\ell m}u^\ell\partial_iB^m .
\end{equation}
A uniform magnetic field has no contribution of this type. However, the term
\begin{equation}
q\,\mathcal M^{ij}\epsilon_{j\ell m}\mathcal B^\ell{}_{i}B^m
\end{equation}
shows that a uniform magnetic field can still affect the expansion if the
congruence already carries shear, rotation, or expansion. This point separates
magnetic focusing from the purely electric case in the homogeneous Hessian
approximation.

\subsection{Comparison between band structures}

The role of $\mathcal M^{ij}$ becomes clearer when the different examples are
compared. For an anisotropic parabolic band,
\begin{equation}
\mathcal M^{ij}
=
(m^{-1})^{ij}
=
{\rm const}.
\end{equation}
All Hessian-gradient terms vanish, and the focusing sources are determined by
the electromagnetic gradients together with the inverse effective masses. The
anisotropy is fixed once the principal masses $m_x$, $m_y$, and $m_z$ are
specified.

For a Dirac cone,
\begin{equation}
\mathcal M^{ij}_{(\pm)}
=
\pm\frac{v_F}{|\mathbf{k}|}P^\perp_{ij},
\qquad
P^\perp_{ij}:=\delta_{ij}-\hat{k}_i\hat{k}_j .
\end{equation}
The response is transverse to the propagation direction and grows as the beam
approaches the Dirac point. The sign difference between the two cones makes
the focusing source branch dependent. This is why the Dirac case is not just a
parabolic band with a different effective mass; its Hessian projects out the
longitudinal direction and changes strongly with momentum.

For a Weyl-type two-branch system,
\begin{equation}
\mathcal M^{ij}_{(+)}
=
-\mathcal M^{ij}_{(-)}.
\end{equation}
Consequently, the electromagnetic source terms enter the two Raychaudhuri
equations with opposite signs when the homogeneous-momentum approximation is
valid. A single electric or magnetic texture can then produce two different
beam evolutions. This is the quasiparticle birefringence mechanism discussed
above.

\subsection{Focusing versus trajectory bending}

The Raychaudhuri equation separates trajectory bending from congruence
focusing. The Lorentz force can bend the path of each quasiparticle even in a
uniform electromagnetic field. By contrast, focusing is governed by
$\partial_i a^i$, which measures whether neighboring trajectories converge or
separate locally.

This distinction is important for interpreting electron-optics setups. A
uniform electric field can accelerate the beam, and a uniform magnetic field
can curve it. However, local focusing requires a nonzero divergence of the
acceleration field, or a coupling between the magnetic field and an already
deforming congruence. The observable signal is therefore not only the bending
of the central trajectory, but the change in the beam width encoded in
$\theta$.

\subsection{Conditions for focusing}

From Eq.~\eqref{eq:ray_general_fields}, a congruence focuses when
$\theta$ decreases along the flow. If shear and vorticity are neglected and
one considers an initially weakly expanding beam, the condition is controlled
by the sign of the source,
\begin{equation}
\mathcal S<0.
\end{equation}
When shear is present, it already contributes negatively to
$D\theta/Dt$, while vorticity contributes with the opposite sign. The
electromagnetic source can then either reinforce or compensate these
kinematical effects.

In a two-branch Weyl-type system satisfying
$\mathcal M^{ij}_{(+)}=-\mathcal M^{ij}_{(-)}$, the source obeys
\begin{equation}
\mathcal S_{(+)}
=
-\mathcal S_{(-)}
\end{equation}
under the same approximations. Thus the field profile that contributes to
focusing in one branch contributes to defocusing in the other. In this way,
the band Hessian determines not only the strength of the response, but also
which branch of the quasiparticle beam is selected by a given electromagnetic
texture.


\section{{Observational and analogue implications}}
\label{sec:observational_implications}

{
The formalism developed in this work is an envelope description of charged branch congruences. The most direct observable is not the bending of a single trajectory, but the change in the transverse profile of a family of neighboring trajectories. This distinction is standard in charged-beam and accelerator optics, where the physically relevant quantities are beam envelopes, focusing lengths, angular spreads, and transverse widths \cite{Wiedemann2015}. In the present case, the relevant signatures are branch-dependent beam narrowing or broadening, shifts in the focusing parameter, changes in the caustic position, and relative asymmetries between the two spin--nondegenerate components. These quantities are controlled by the same response tensor $M^{\mu\nu}_{(\pm)}$ that appears in the electromagnetic source of the Raychaudhuri equation.

\paragraph{Charged-particle propagation.}
In a relativistic setting, the effect may be relevant for charged particles crossing inhomogeneous electromagnetic regions. Examples include magnetized plasma environments, compact-object magnetospheres, jet-like configurations, shock fronts, and laboratory beam lines \cite{Amato:2017,Lazarian:2023,Chen:2018,Wiedemann2015}. In such systems, a nonuniform electromagnetic field can modify the local expansion of a charged beam through the divergence of the effective branch force. The distinctive Lorentz--violating feature is that two spin--nondegenerate branches can acquire different expansion histories under the same external field configuration. In practice, this would appear as a branch-dependent angular spread, branch-dependent collimation, or a small displacement of the effective focusing length.

\paragraph{Astrophysical interpretation.}
Astrophysical environments provide long baselines and strong electromagnetic gradients, but they also introduce significant uncertainties associated with plasma composition, turbulence, magnetic-field modeling, source geometry, and propagation effects \cite{Mattingly:2005,Liberati:2013,Amato:2017,Lazarian:2023}. For this reason, the present framework should not be interpreted as giving immediate model-independent astrophysical bounds. Its main use in this context is to identify which observables would be sensitive to the branch-dependent response: angular broadening of charged-particle beams, spin- or branch-dependent collimation, orientation-dependent transport through magnetic gradients, and possible differences between focusing and defocusing regions along the same path. A quantitative astrophysical constraint would require embedding the present local congruence equation into a concrete propagation model for a specified source and electromagnetic background.

\paragraph{Lorentz-violation searches.}
The connection with Lorentz-violation phenomenology is through the dependence of $M^{\mu\nu}_{(\pm)}$ and $k^\mu_{(\pm)}$ on the SME coefficients. Existing bounds are strongly sector-, species-, frame-, and component-dependent, so no universal numerical estimate should be assigned to all coefficients \cite{Kostelecky:2011,Mattingly:2005,Liberati:2013}. Instead, the practical procedure is to choose a particle species and sector, insert the corresponding allowed SME coefficients from the data tables, and compute the induced shift in the focusing length or in the branch asymmetry. Within this interpretation, the present formalism converts SME coefficients into beam-envelope observables. Possible laboratory signatures include orientation-dependent focusing, branch-dependent beam-width evolution, charge-conjugation comparisons when appropriate, and sidereal variations if the laboratory orientation changes with respect to the background coefficients.

\paragraph{Analogue condensed-matter signatures.}
The condensed-matter realization is experimentally more direct because the dispersion relation, the electromagnetic texture, and the beam position in momentum space can be controlled. In graphene, gate-defined $p$--$n$ junctions and electrostatic profiles can act as electron-optics elements, including Veselago-type lenses and caustic-forming structures \cite{Cheianov:2007,Brun:2018}. In Dirac- and Weyl-type materials, the two branches can carry different or even opposite Hessians, and Weyl/Dirac semimetals provide solid-state platforms in which relativistic quasiparticle kinematics and branch-sensitive transport can be investigated \cite{Armitage:2018,Jia:2016}. A gate-defined electrostatic lens or a magnetic texture can then produce different expansion rates for the two quasiparticle branches. Observable signatures include different beam widths after crossing the lens, branch- or chirality-dependent transmission, asymmetric collimation, and spatial separation of components that experience focusing and defocusing. In this sense, the quasiparticle system provides a tunable analogue platform for visualizing the same response-tensor mechanism that appears in the relativistic Lorentz--violating theory.

The main observational channels can be summarized as follows:
\begin{table}[h]
\centering
{
\hspace*{-0.5cm}%
\begin{tabular}{@{}c c c@{}}
\hline\hline
Setting & Controlled quantity & Possible signature \\
\hline
Particle beams &
EM gradient/profile &
shifted focus or beam width \\
Astrophysical media &
magnetized gradients &
branch-dependent broadening/collimation \\
SME tests &
$b_\mu$, $H_{\mu\nu}$, $d_{\mu\nu}$ &
orientation/branch focusing asymmetry \\
Condensed-matter analogues &
band Hessian and external textures &
chirality-dependent transmission/divergence \\
\hline\hline
\end{tabular}
}
\caption{{
Representative observational and analogue signatures of branch-dependent electromagnetic focusing. The entries are qualitative, since the effect depends on the particle species, SME sector, electromagnetic profile, and validity of the branch-resolved approximation.
}}
\label{tab:observational_signatures}
\end{table}

These implications also clarify the physical status of the focusing and defocusing terminology used throughout the paper. Focusing refers to the contraction of a congruence cross section, while defocusing refers to its local broadening. The branch-dependent effect is measurable only when the two components can be prepared, selected, or inferred separately. This condition is difficult in generic astrophysical propagation, but it is more realistic in controlled laboratory beam settings and in analogue condensed-matter platforms.
}

\section{Conclusion}
\label{sec:discussion}

In this work, we investigated the electromagnetic forcing of spin--nondegenerate Lorentz--violating congruences through a gauge--covariant Raychaudhuri formulation. The minimal coupling was introduced by replacing the canonical momentum with the gauge--covariant momentum $P_{\mu}=\pi_{\mu}-qA_{\mu}$. This replacement preserved the branch structure of the free dispersion relation and made the electromagnetic field act through the standard field strength $F_{\mu\nu}$. The Lorentz--violating information remained contained in the branch dispersion function $D^{(\pm)}(P)$.

For a generic branch, the tangent vector and the momentum Hessian were written as
$k^{\mu}_{(\pm)}=(e/2)\partial D^{(\pm)}/\partial P_{\mu}$ and
$M^{\mu\nu}_{(\pm)}=-\partial k^{\mu}_{(\pm)}/\partial P_{\nu}$. These quantities fixed the non-geodesic part of the motion,
$a^{\mu}_{(\pm)}=-qM^{\mu\nu}_{(\pm)}F_{\nu\rho}k^{\rho}_{(\pm)}$.
Substitution into the Raychaudhuri equation gave the universal branch-resolved form
\[
\frac{d\theta_{(\pm)}}{d\lambda}
=
-\frac{1}{2}\theta^{2}_{(\pm)}
-\sigma^{(\pm)}_{\mu\nu}\sigma_{(\pm)}^{\mu\nu}
+\omega^{(\pm)}_{\mu\nu}\omega_{(\pm)}^{\mu\nu}
-R_{\mu\nu}k^{\mu}_{(\pm)}k^{\nu}_{(\pm)}
-q\nabla_{\mu}
\left(
M^{\mu\nu}_{(\pm)}F_{\nu\rho}k^{\rho}_{(\pm)}
\right).
\]
This equation displayed the separation between geometry, electromagnetic forcing, and Lorentz--violating branch response. Since the background coefficients were kept constant, they did not generate independent force terms. Their effect was to change the tangent and the Hessian through which the same electromagnetic field was converted into acceleration.

We applied the construction to the $b_{\mu}$, $H_{\mu\nu}$, and $d_{\mu\nu}$ sectors. In each case, the minimally coupled Hamiltonian kept the same branch dispersion relation after the replacement $p_{\mu}\rightarrow P_{\mu}$. The explicit Hessians showed how the two spin--nondegenerate branches respond differently to the same electromagnetic background. The inverse square-root structures appearing in these tensors also showed that the response can become stronger near the degeneracy regions of the dispersion relation, where the branch description has to be treated with care.

In flat spacetime, the curvature term was absent and the expansion was controlled by the divergence of the effective electromagnetic branch force. For slowly varying beam data, the leading local source was governed by field gradients. A uniform electric field can accelerate or bend the trajectories, but it does not produce a leading focusing term for an initially uniform congruence with nearly constant Hessian. The magnetic sector is less restrictive: a magnetic gradient can focus the beam, while a uniform magnetic field may still contribute when the congruence already carries shear, rotation, or expansion. In other words, the Raychaudhuri equation separated trajectory bending from true local focusing.

We also formulated the corresponding Raychaudhuri-type equation for semiclassical quasiparticle beams. Starting from the wave-packet equations of motion, the acceleration was written in terms of the band Hessian $M_{ij}=\partial^{2}\varepsilon/\partial k_i\partial k_j$. The expansion equation then contained electric and magnetic sources weighted by $M_{ij}$ and by its spatial variation across the beam. This gave a direct condensed-matter analogue of the relativistic branch response: the band Hessian controls not only the acceleration of a single quasiparticle, but also the deformation of a beam of nearby trajectories.

The examples made this structure explicit. For an anisotropic parabolic band, the Hessian reduced to the inverse effective mass tensor, so the focusing strength was weighted by the principal effective masses. For a Dirac-like dispersion, the Hessian became transverse to the propagation direction and momentum dependent, with a stronger response near the band-touching point. For Weyl-type two-branch systems, the two Hessians had opposite signs in the homogeneous-momentum approximation. As a result, the same electromagnetic texture could decrease the expansion of one branch while increasing the expansion of the other. 

The present analysis was performed for constant Lorentz--violating
backgrounds and for semiclassical band dynamics in the absence of
Berry-curvature corrections. Within this regime, the same geometric
quantity governed the electromagnetic response in both relativistic and
semiclassical systems: the Hessian of the underlying branch or band
dispersion relation. This tensor determines how electromagnetic fields
are converted into effective acceleration and how such acceleration
modifies the local evolution of the congruence expansion.

The framework developed here reveals that congruence dynamics in
Lorentz-violating systems possesses an intrinsically branch-resolved
electromagnetic geometry. In this picture, focusing is no longer a
purely curvature-induced phenomenon, but instead becomes controlled by
the interplay between external electromagnetic fields and the modified
dispersion geometry encoded in the branch response tensor.

The central physical result of this work is therefore the emergence of branch-dependent electromagnetic focusing. Although the external field is common to all branches, the effective response tensor converting the field into acceleration is branch resolved. As a consequence, different branches may undergo distinct focusing or defocusing behaviors under the same electromagnetic environment, providing a geometric realization of electromagnetic birefringence at the level of congruence evolution.

As a further perspective, it would be interesting to investigate the modified dispersion relations associated with the particular systems discussed in this paper by employing the ensemble approach developed in Refs.~\cite{araujo2023thermodynamics,araujo2022thermal,araujo2022does,AraujoFilho:2025rwr,araujo2022particles,araujo2023thermodynamical}. In addition, the analysis of gravitational wave polarization and related phenomenological aspects could also be explored along the lines of Refs.~\cite{araujo2026gravitssational,AraujoFilho:2026vcf,AraujoFilho:2026oiy,amarissslo2024gravitational}.

\appendix
\section{Derivation of the evolution equation for the deformation tensor}
\label{app:deformation}

In this appendix we present the derivation of the evolution equation for the
deformation tensor used in Sec.~\ref{sec:raychaudhuri}. The calculation follows
the usual kinematics of congruences, but it is written in the language of the
semiclassical quasiparticle flow. To avoid confusion with the magnetic field,
we denote the deformation tensor by $\mathcal B_{ij}$.

\subsection{Definitions}

Let $u^i(x,t)$ be the velocity field of the quasiparticle congruence. The
deformation tensor is defined as
\begin{equation}
\mathcal B_{ij}:=\partial_j u_i .
\end{equation}
The derivative along the flow is the convective derivative
\begin{equation}
\frac{D}{Dt}
=
\partial_t+u^k\partial_k ,
\end{equation}
and the acceleration field is
\begin{equation}
a_i
=
\frac{D u_i}{Dt}.
\end{equation}

\subsection{Evolution of the deformation tensor}

We start from
\begin{align}
\frac{D}{Dt}\mathcal B_{ij}
&=
\frac{D}{Dt}\left(\partial_j u_i\right)
\nonumber\\
&=
\partial_t\left(\partial_j u_i\right)
+
u^k\partial_k\left(\partial_j u_i\right).
\label{eq:app_start}
\end{align}
The first term can be written as
\begin{equation}
\partial_t\left(\partial_j u_i\right)
=
\partial_j\left(\partial_t u_i\right).
\end{equation}
For the convective part, the product rule gives
\begin{equation}
u^k\partial_k\left(\partial_j u_i\right)
=
\partial_j\left(u^k\partial_k u_i\right)
-
\left(\partial_j u^k\right)
\left(\partial_k u_i\right).
\end{equation}
Therefore,
\begin{align}
\frac{D}{Dt}\mathcal B_{ij}
&=
\partial_j
\left(
\partial_t u_i+u^k\partial_k u_i
\right)
-
\left(\partial_j u^k\right)
\left(\partial_k u_i\right).
\end{align}
Using
\begin{equation}
\partial_t u_i+u^k\partial_k u_i=a_i,
\end{equation}
and the definition of $\mathcal B_{ij}$, we obtain
\begin{equation}
\frac{D}{Dt}\mathcal B_{ij}
=
\partial_j a_i
-
\mathcal B_{ik}\mathcal B^k{}_{j}.
\label{eq:app_deformation_evolution}
\end{equation}
This identity is the basic kinematical equation behind the
Raychaudhuri-type description. The first term measures the inhomogeneity of
the acceleration field, while the quadratic term accounts for the deformation
already present in the congruence.

\subsection{Trace and Raychaudhuri-type equation}

The expansion scalar is the trace of the deformation tensor,
\begin{equation}
\theta
=
\mathcal B^i{}_{i}.
\end{equation}
Taking the trace of Eq.~\eqref{eq:app_deformation_evolution}, we find
\begin{equation}
\frac{D\theta}{Dt}
=
\partial_i a^i
-
\mathcal B_{ik}\mathcal B^{ki}.
\label{eq:app_theta_pre}
\end{equation}
We now decompose $\mathcal B_{ij}$ into trace, symmetric traceless, and
antisymmetric parts,
\begin{equation}
\mathcal B_{ij}
=
\frac{1}{d}\theta\,\delta_{ij}
+
\sigma_{ij}
+
\omega_{ij},
\end{equation}
where $\sigma_{ij}$ is symmetric and traceless, while $\omega_{ij}$ is
antisymmetric. The contraction becomes
\begin{equation}
\mathcal B_{ik}\mathcal B^{ki}
=
\frac{1}{d}\theta^2
+
\sigma_{ij}\sigma^{ij}
-
\omega_{ij}\omega^{ij}.
\label{eq:app_B_contraction}
\end{equation}
The minus sign in the vorticity contribution follows from the contraction
$\omega_{ik}\omega^{ki}=-\omega_{ij}\omega^{ij}$, whereas the mixed terms
vanish because they involve either the trace of $\sigma_{ij}$ or contractions
between symmetric and antisymmetric tensors.

Substituting Eq.~\eqref{eq:app_B_contraction} into
Eq.~\eqref{eq:app_theta_pre}, we obtain
\begin{equation}
\frac{D\theta}{Dt}
=
-\frac{1}{d}\theta^2
-
\sigma_{ij}\sigma^{ij}
+
\omega_{ij}\omega^{ij}
+
\partial_i a^i .
\label{eq:app_raychaudhuri}
\end{equation}
This is the Raychaudhuri-type equation for the real-space quasiparticle
congruence. The terms proportional to $\theta^2$ and
$\sigma_{ij}\sigma^{ij}$ drive contraction of the beam, the vorticity term
acts with the opposite sign, and the divergence of the acceleration contains
the external forcing produced by the semiclassical dynamics.


\section*{Acknowledgments}

A. A. Araújo Filho is supported by Conselho Nacional de Desenvolvimento Cient\'{\i}fico e Tecnol\'{o}gico (CNPq) with project number 150223/2025-0. JAASR acknowledges partial financial support from UESB through Grant AuxPPI (Edital No. 267/2024) and also gratefully recognizes support from FAPESB–CNPq/Produtividade (Grant No. 12243/2025, TOB-BOL2798/2025). A. F. Santos is partially supported by National Council for Scientific and Technological Development - CNPq project No. 312406/2023-1.
V. B. B. is partially supported by the National Council for
Scientific and Technological Development—CNPq, Grant
No. 307211/2020-7.

\section*{Data Availability Statement}

No Data associated in the manuscript.

	\bibliography{main}

\end{document}